\documentclass[twocolumn,rmp,aps]{revtex4}

\usepackage{graphicx}
\usepackage{amsmath,amssymb,bm}
\usepackage{fancyhdr}

\def\Z{Z_{\rm eff}}
\def\Zr{\Z^{(\rm res)}}
\def\Zth{\Z^{\rm th}}
\def\eps{\varepsilon }
\def\eb{\varepsilon _b}

\begin{document}
\title{Positron-molecule interactions: resonant attachment, annihilation,\\
and bound states}

\author{G. F. Gribakin}
\email{g.gribakin@qub.ac.uk}
\affiliation{Department of Applied Mathematics and Theoretical Physics,
Queen's University Belfast, Belfast BT7 1NN, Northern Ireland, UK}
\author{J. A. Young}
\email{jyoung@physics.ucsd.edu}
\affiliation{Jet Propulsion Laboratory, California Institute of Technology,
4800 Oak Grove Drive, Pasadena, California 91109, USA}
\author{C. M. Surko}
\email{csurko@ucsd.edu}
\affiliation{Department of Physics, University of California, San Diego,
9500 Gilman Drive, La Jolla, California 92093-0319, USA}

\begin{abstract}
This article presents an overview of current understanding of the
interaction of low-energy positrons with molecules with emphasis on
resonances, positron attachment and annihilation.
%
Measurements of annihilation rates resolved as a function of positron energy
reveal the presence of vibrational Feshbach resonances (VFR) for many
polyatomic molecules. These resonances lead to strong enhancement of
the annihilation rates. They also provide evidence that positrons bind
to many molecular species.
A quantitative
theory of VFR-mediated attachment to small molecules is presented. It is
tested successfully for selected molecules (e.g., methyl halides and methanol)
where all modes couple to the positron continuum. Combination and overtone
resonances are observed and their role is elucidated. Molecules that do not
bind positrons, and hence do not exhibit such resonances, are discussed. In
larger molecules, annihilation rates from VFR far exceed those explicable on
the basis of single-mode resonances. These enhancements increase rapidly with
the number of vibrational degrees of freedom, approximately as the fourth
power of the number of atoms in the molecule. While the details are as yet
unclear, intramolecular vibrational energy redistribution (IVR) to states
that do not couple directly to the positron continuum appears to be
responsible for these enhanced annihilation rates. In connection with IVR,
experimental evidence indicates that inelastic positron escape channels are
relatively
rare. Downshifts of the VFR  from the vibrational mode energies, obtained by
measuring annihilate rates as a function of incident positron energy, have
provided binding energies for thirty species. Their dependence upon molecular
parameters and their relationship to positron-atom and positron-molecule
binding energy calculations are discussed. Feshbach resonances and positron
binding to molecules are
compared with the analogous electron-molecule (negative ion) cases. The
relationship of VFR-mediated annihilation to other phenomena such as
Doppler-broadening of the gamma-ray annihilation spectra, annihilation
of thermalized positrons in gases, and annihilation-induced
fragmentation of molecules is discussed. Possible areas for future theoretical
and experimental investigation are also discussed.
\end{abstract}                                                                 

\maketitle

\thispagestyle{fancy}
\renewcommand{\headrulewidth}{0pt}
\rhead{\small REVIEWS OF MODERN PHYSICS, VOLUME 82, 2557--2607 (2010)
\hspace{3.5cm}}

\tableofcontents

\section{Introduction and overview}\label{sec:intro}

The subject of this review is the interaction of low-energy positrons with
molecules. Positrons, the antiparticles of electrons, are
important in many areas of science and technology. Much of their utility
relies on the fact that, when an electron and positron interact, they can
annihilate, producing a characteristic burst of gamma-rays. The lowest order
process results in two back-to-back photons, each with the energy of the rest
mass of the electron (or positron), 511 keV. 

The annihilation of low-energy (e.g., $\leq 50$~eV) positrons on atoms and
molecules plays a particularly important role in many fields. 
In medicine, positron emission
tomography (PET) exploits two-gamma annihilation to study human metabolic
processes \cite{Wah02}. In materials science, there are numerous
positron-based techniques to study the properties of matter
\cite{SL88,PN94,DM95,Col00}, including the Fermi surfaces in metals
\cite{MDW04}, microscopic pores in solids \cite{GDF00,GPV06}, the free
volume in polymers \cite{DSF98}, and the composition and structure of
surfaces \cite{DKS01}.
In astronomy, 511 keV annihilation radiation, the strongest gamma ray
line of extraterrestrial origin, has proven useful in elucidating
astrophysical processes \cite{RLC92,CSS05,GJG10}.
A current research goal is the
creation of a Bose condensate of positronium (Ps) atoms (i.e., the
electron-positron analog of the hydrogen atom) that offers promise for the
development of an annihilation gamma-ray laser \cite{CM07,MCG04,Mil02,Mil07}.

Typically, positrons from conventional sources (e.g., radioisotopes or
electron accelerators) slow down from energies of kilovolts to hundreds of
kilovolts to $\lesssim 50$~eV before annihilating.
In the case of atoms or molecules,
if the incident positron energy $\eps $ is greater than the Ps-formation
threshold
$E_{\rm th} = E_i - E_{\rm Ps} $, where $E_i$ is the ionization energy of the
target and $E_{\rm Ps} =6.8$~eV is the binding energy of the ground-state Ps
atom, then the
dominant annihilation process is through Ps formation. The resulting Ps atom
subsequently annihilates by emitting two or three gamma-ray quanta.

In this review, attention
is restricted to positron energies below the
Ps-formation threshold, $0<\eps < E_{\rm th}$, where the Ps channel is
closed. Here annihilation occurs as a result of the overlap of the positron
and electron densities during the collision. The basic rate in this case
is the Dirac rate $\lambda_D$ for two-gamma annihilation in a free electron
gas \cite{Dir30}
\begin{equation}\label{eq:Dirac}
\lambda_D=\pi r_0^2cn_e,
\end{equation}
where $r_0$ is the classical electron radius, $r_0= e^2/mc^2$ in cgs units,
$e$ and $m$ are the electron charge and mass,
$c$ is the speed of light, and $n_e$ is the electron density.

In his seminal discovery of the positronium atom,
\textcite{Deu51a,Deu51b} found a curious effect. Although the annihilation
rate for thermal positrons at 300~K in atomic and molecular gases was
approximately in accord with Eq.~(\ref{eq:Dirac}) for some species
(e.g., argon and nitrogen), the rate for dichlorofluoromethane
CCl$_2$F$_2$ (``Freon-12'') was much
larger. Deutsch insightfully ascribed this effect to some type of resonant
positron-molecule attachment process. A decade later, \textcite{PS63} measured
annihilation rates in gases of alkane molecules C$_n$H$_{2n+2}$,
from methane to butane, $n=1$--4. They found that
the rate $\lambda $ was much greater than $\lambda _D$
and that the ratio $\lambda /\lambda _D$ increased
exponentially with molecular size.

Annihilation rates in gases are conventionally normalized to the Dirac
rate. The corresponding dimensionless quantity
is the ``effective number of electrons''\footnote{In
chemical kinetics, $\Z $ corresponds to the (normalized) {\em rate constant}
of the annihilation reaction. In positron physics this quantity is commonly
referred to as the ``annihilation rate''.}
\begin{equation}\label{eq:Zefflam}
\Z =\frac{\lambda }{\pi r_0^2cn},
\end{equation}
where $n$ is the density of atoms or molecules \cite{Pom49,Fra68}.
For a simple collision, and neglecting electron-positron correlations, one
might expect that $\lambda \sim \lambda _D$, so that $\Z$ is comparable to
$Z=n_e/n$, the total number of electrons per target atom or molecule.
However, values of $\Z$ are often much larger (e.g., for butane,
$\Z /Z = 600$).

Positron annihilation in atoms and molecules was subsequently studied
for a wide range of species \cite{Osm65a,Osm65b,Tao65,Tao70,%
MSR75,SM78,CGH80,HCG85,SHW85,HCD86,ACB00,CWA02,CWL06}. Early experiments
were done with thermal positrons in gases at atmospheric densities,
$n\sim 1$~amagat \cite{Deu53,PS63,GH78}.\footnote{$1~\mbox{amagat}~=2.69\times
10^{19}$ cm$^{-3}$ is the density of an ideal gas at standard 
temperature and pressure, 273.15~K and 101.3~kPa, respectively.}
Later, experiments were done at much lower densities
using positrons trapped and cooled to 300~K \cite{SPL88,MS91,IGM95,IGS97}.
The experiments showed that the annihilation rates for many molecular species
exceeded greatly the naive benchmark rate, $\Z \sim Z$, and a
number of chemical trends were identified.

Since Deutsch's first results, these large annihilation rates were associated
with some kind of resonance phenomenon or attachment process. \textcite{GS64}
discussed the possibility of resonance-enhanced annihilation due to
a bound or virtual positron state close to zero energy. \textcite{SP70}
considered the possibility that the large annihilation rates in molecules were
due to a vibrational resonance, and several other explanations were
proposed \cite{SPL88,DFG96,LW97,Gri00}. However, progress
was hampered greatly by the lack of data other than for positrons
with thermal energy distributions at 300~K. The summary statement in 1982 by
Sir Harrie Massey was that annihilation studies were ``completely mysterious at
present in almost all substances'' \cite{FBC82,Mas82}, and this remained
more or less correct for another twenty years.

In the broader view, processes that are commonplace in physics involving
matter, such as low-energy, two-body scattering events, have frequently been
found to be frustratingly difficult to study when antiparticles are
involved \cite{SL88,Col00,CH01,EH99}. The advent of efficient
positron traps marked a turning point \cite{MS92,SLP88,SGB05},
enabling a new generation of studies \cite{SPL88,MS91,IGM95,KGS96,IGS97}.
Experiments with trapped positrons cooled to 300~K
permitted studies of test species at low densities (e.g.,
$\leq 10^{-6}$~amagat). This ensured that
annihilation was strictly due to binary collisions, rather than many-particle
effects \cite{IK82}, and it enabled study of a broader range of chemical
species, including low-vapor-pressure liquids and solids. Gamma-ray spectra
were measured for many molecules \cite{IGS97}.
The $\Z /Z$ ratios were found to increase rapidly with
molecular size up to species as large as naphthalene and hexadecane
(C$_{16}$H$_{34}$), reaching values
$\gtrsim 10^4$ \cite{SPL88,MS91}.\footnote{The theoretical maximum for the
magnitude of $\Z$ is given by the unitarity limit of the inelastic cross
section \cite{QM}, $\Z \lesssim 10^7$ for room-temperature positrons.}

A key to further progress was the development of a trap-based positron
beam with a narrow energy spread ($\sim 40$~meV) \cite{KGG98,GKG97}.
Using this beam, annihilation rates for atoms and molecules were measured
as a function of incident positron energy from 50 meV to
many electron volts. The result was the discovery of {\em resonances}
associated with the molecular vibrational modes, namely vibrational Feshbach
resonances (VFR) \cite{GBS02}.

A crucial point is that VFRs generally require the existence of a bound state
of the positron and the molecule. They occur when the incident positron
excites a vibrational mode and simultaneously makes a transition from
the continuum into the bound state. The existence of
both low-lying vibrational excitations and a positron bound state
thus enables the formation of long-lived positron-molecule resonant complexes
in a two-body collision. The lifetime of these quasibound states is limited
by positron autodetachment accompanied by vibrational de-excitation.
The upper limit on the lifetime is $\lesssim 0.1$~ns, set by the positron
annihilation rate in the presence of atomic-density electrons.

\begin{figure}[ht]
\includegraphics*[width=7cm]{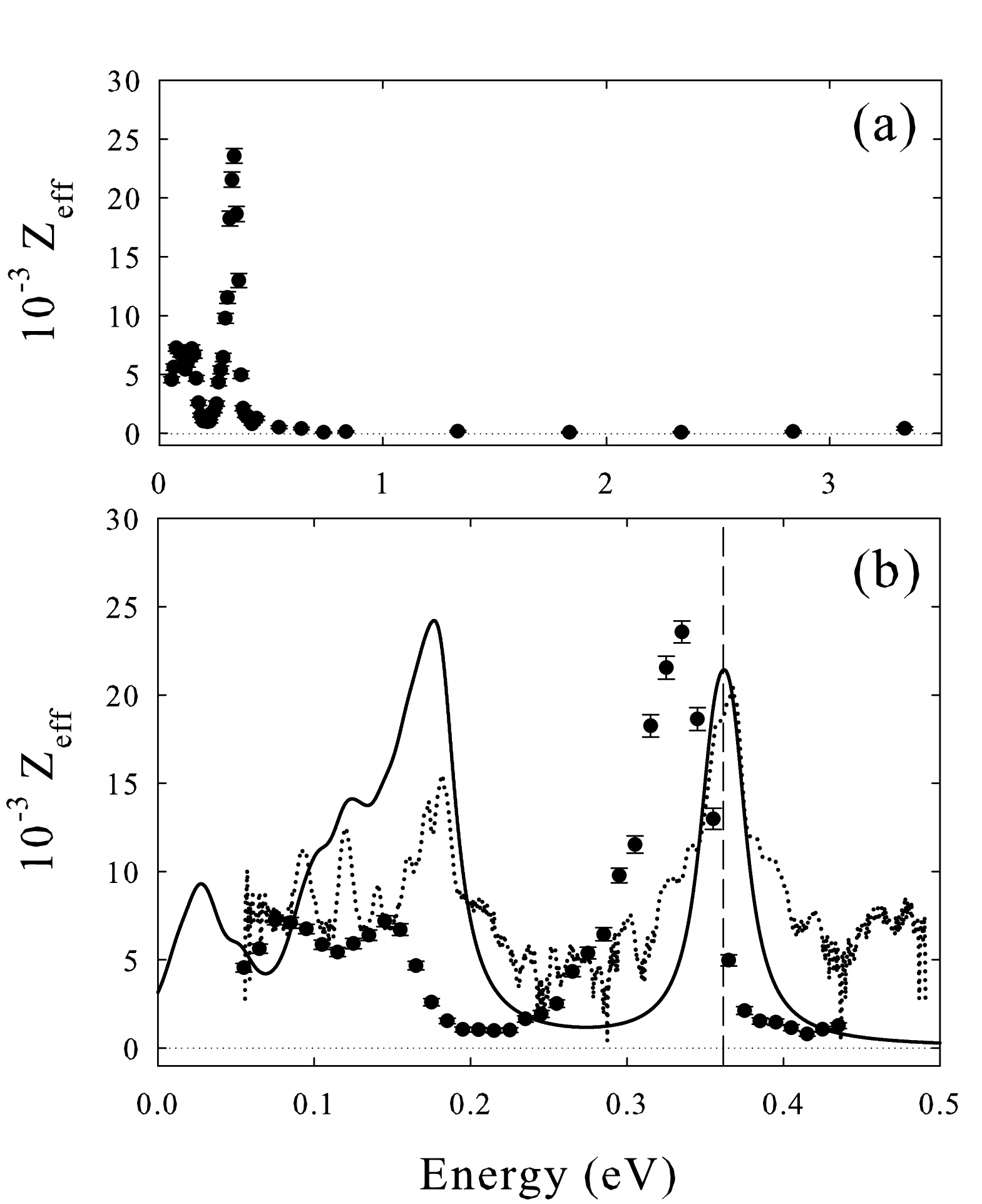}
\caption{The normalized annihilation rate $\Z (\eps )$
for butane C$_4$H$_{10}$ ($\bullet $) as a function of the total incident
positron energy $\eps $: (a) up to the Ps formation threshold,
$E_{\rm th}=3.8$~eV, and (b) in the region of the molecular vibrations;
dotted curve, the infrared absorption spectrum \cite{NIST} (logarithmic
vertical scale, arbitrary units); solid curve, the vibrational mode density
(in arbitrary units), with the modes represented by Lorentzians with an
arbitrary FWHM of 10~meV; dashed line, mean energy of the C-H stretch
fundamentals.}
\label{fig:butane}
\end{figure}

The annihilation rate as a function of positron energy, $\Z (\eps )$ (i.e.,
the ``annihilation spectrum'') for the four-carbon alkane, butane, is shown in
Fig. \ref{fig:butane} \cite{GBS02}. While there is some qualitative
correspondence between the $\Z (\eps )$ and the infrared (IR) absorption
spectrum of the molecule, the shapes of the spectral
features are quite different \cite{BGS03}.

These positron VFRs can be compared to resonances that play an important role
in {\em electron} attachment to molecules and clusters \cite{Chr84,HRA03}.
The electron collision results in the production of
long-lived (metastable) parent anions, or molecular fragment negative ions via
dissociative attachment. A dominant mechanism of electron capture by molecules
is via negative-ion resonant states \cite{Chr84}. Dissociative attachment
usually proceeds via electron shape resonances of ground or
electronically-excited molecules. Such resonances are quite common in diatomic,
triatomic and polyatomic species at energies in the range $\sim 0$--4~eV. The
theoretical description of them involves (complex) Born-Oppenheimer
potential-energy surfaces \cite{Bar68a,OMa66,Dom81}.
All the data indicate that positrons generally do not form
shape resonances or electronic Feshbach resonances in low-energy collisions
with molecules. Instead, energy-resolved annihilation studies point
to the important role played by the VFR.

These vibrational (or ``nuclear-excited'') Feshbach resonances involve
coupling of the electronic and the nuclear motion beyond
the Born-Oppenheimer approximation. It cannot be described
by potential-energy surfaces. This type of resonances was originally
introduced by \cite{Bar68b} as an ``indirect'' mechanism for dissociative
electron recombination and described using Breit-Wigner theory. In the case
of electrons, these VFRs
lead to large attachment cross sections which typically reach their maximum
values at thermal electron energies \cite{Chr84}. They are also responsible
for the formation of long-lived parent negative ions for many complex
polyatomic molecules.

Referring to Fig.~\ref{fig:butane}, the energy of the VFR corresponding to
mode $\nu $ is given by energy conservation,
\begin{equation}\label{eq:enVFR}
\eps _\nu =\omega _\nu -\eb ,
\end{equation}
where $\eb$ is the positron-molecule binding energy, and $\omega _\nu $
is the vibrational mode energy. The positron binding energy
(i.e., the positron affinity) can be measured by the
downshift of the resonances from the energies of the vibrational
modes.
In Fig. \ref{fig:butane} for butane, this is most easily seen as the shift
in the C-H stretch vibrational resonance. The corresponding peak in $\Z$
occurs at 330~meV, as compared with the vibrational mode frequency of 365~meV,
indicating that $\eb = 35$~meV.
The resonances at lower energies are due to C-C modes and C-H bend modes
and exhibit the same downshift.

There are a number of important chemical trends associated with resonant
annihilation on molecules \cite{MS91,IGM95,YS08a,YS08b}. Examples are shown in
Table \ref{tab:Zeff}. Very small molecules, such as CO$_2$, CH$_4$, or H$_2$O,
have relatively small values of $\Z$ (e.g., $\Z /Z\lesssim 10$), and
typically they do not exhibit resonant annihilation peaks.
Positrons either do not bind to these species
(i.e., $\eb<0$), or they bind extremely weakly. With the exception of
methane, all of the alkanes exhibit VFRs, with values of $\eb$ increasing
linearly with the number of carbon atoms $n$, and the magnitudes of $\Z$
increasing approximately exponentially with $n$. Most hydrocarbons, including
aromatic molecules, alkenes and alcohols, exhibit similar resonant annihilation
spectra.

\begin{table}[ht]
\caption{Annihilation rates $\Z$, and binding energies $\eb$ for
selected molecules.}
\label{tab:Zeff}
\begin{ruledtabular}
\begin{tabular}{ccrrr}
Class & Molecule & $Z$ & $\eb$ (meV)\footnotemark[1] & $\Z$\footnotemark[1] \\
\hline
Small inorganics & H$_2$O & 10 & $<0$ & 170\footnotemark[2] \\
& NH$_3$ & 9 & $>0$ & 300\footnotemark[2] \\
\hline
Methyl halides & CH$_3$F & 18 & $>0$ & 250\footnotemark[2] \\
& CH$_3$Br & 44 & $40$ & 2000\footnotemark[2] \\
\hline
Alkanes & CH$_4$ & 10& $<0$ & 70\footnotemark[2] \\
& C$_2$H$_6$ & 18 & $>0$ & 900\footnotemark[3] \\
& C$_3$H$_8$ & 26 & 10 & 10\,500\footnotemark[3] \\
& C$_6$H$_{14}$ & 50 & 80 & 184\,000\footnotemark[3] \\
& C$_{12}$H$_{26}$ & 98 & 220 & 9\,800\,000\footnotemark[3] \\
\hline
Alcohols & CH$_3$OH & 18 & $>0$ & 750\footnotemark[2]  \\
& C$_2$H$_5$OH & 26 & 45 & 4500\footnotemark[2]  \\
\hline
Aromatics & C$_6$H$_6$ & 42 & 150 & 47\,000\footnotemark[3]  \\
& C$_{10}$H$_8$ & 68 & 300 & 1\,240\,000\footnotemark[3]
\end{tabular}
\end{ruledtabular}
\footnotetext[1]{Values from energy-resolved measurements \cite{YS08a,YS08b};
typical uncertainties in $\Z$ and $\eb $ are $\pm $20\% and $\pm $10~meV,
respectively.}
\footnotetext[2]{Maximum values for positron energies $\eps \geq 50$~meV.}
\footnotetext[3]{Values of $\Z $ at the C-H resonance peak.}
\end{table}

Much progress has been made in the theoretical understanding of
resonant positron annihilation in molecules
\cite{Gri00,GG04,Gri01}. A quantitative theory has been developed
for the case of isolated resonances of IR-active vibrational modes,
such as those observed in experiments for selected small molecules.
The prototypical example is that of the methyl halides,
CH$_3$X, where X is a F, Cl or Br atom. Positron coupling to the IR-active
modes is evaluated in the dipole approximation
using data from IR absorption measurements.
The only free parameter in the theory is the positron binding energy,
which can be taken from experiment. This yields theoretical annihilation
spectra for methyl halides that are in good agreement with the measurements
\cite{GL06a}.

A more stringent test of the theory relies on the fact that positron binding
energies are expected to change little upon isotope substitution.
For deuteration this was confirmed experimentally. The binding energies
measured for CH$_3$Cl and CH$_3$Br were used to predict $\Z$ for their
deuterated analogs. The result is excellent agreement between theory and
experiment with no adjustable parameters \cite{YGL08}. In other small
molecules, such as ethylene and methanol, IR-inactive modes and multimode
vibrations are prominent, and must be included to explain the
observations \cite{YGL08}.

This theoretical approach explains $\Z$ for small polyatomics in which the
positron coupling to the mode-based VFR, and possibly a few overtones, can be
estimated (e.g., when they have
dipole coupling). Their $\Z$ values are between a few hundred and a few
thousand.\footnote{The heights of resonant peaks in the measured $\Z$ spectra
are related to the energy spread $\delta \eps $ of the incident positron beam.
The value of $\Z \sim 10^3$ corresponds to the typical
$\delta \eps \sim 40$~meV used to date (see Sec. \ref{sec:exp}).}
However, larger molecules with more than one or two carbons, have values
of $\Z$ that
cannot be explained by this theory (cf. Fig.~\ref{fig:butane} for butane).
The current physical picture ascribes their large annihilation
rates to large densities of vibrational resonances, known as ``dark states''
\cite{Gri00,Gri01}, that are not coupled directly to the positron continuum.
The positron first attaches to the molecule via a vibrational
``doorway state'' (e.g., a dipole-allowed mode-based VFR) \cite{GG04}. The
vibrational energy is then transferred to the ``dark states'' in
a process known as intramolecular vibrational energy redistribution (IVR).
Such IVR is important for many physical and chemical processes in molecules,
including dissociative attachment \cite{UM91,NF96}.

The magnitudes of resonant contributions to $\Z$ exhibit a relatively weak
dependence on $\eb$ and on the incident positron energy $\eps $. It is of the
form $g = \sqrt{\eb/\eps }$ and follows from rather general theoretical
considerations. When this dependence is factored out, it is found
experimentally that the resulting quantity, $\Z /g$, scales as $\sim N^4$,
where $N$ is the number of atoms in the molecule. This dependence
on $N$ is thought to reflect the rapid increase in the density of the molecular
vibrational spectrum with the number of vibrational modes. This
dependence is interpreted as evidence that IVR does indeed play an important
role in the annihilation process.

Estimates of $\Z$ in large molecules, which assume that the IVR process is
complete and all modes are populated statistically,
predict $\Z$ values far in excess of those that are observed. Such estimates
also fail to reproduce the energy dependence of $\Z$,
which is largely determined by the mode-based
vibrational doorways. One hypothesis, as yet unconfirmed, is that the
IVR process does not run to ``completion''. It appears that selective coupling
of multimode vibrations leaves a large portion of them inactive. The
calculation of $\Z$ then requires a detailed
knowledge of the vibrational mode couplings, and this has not yet been done. 

The energy-resolved annihilation experiments provide measures of
positron-molecule binding energies, either directly using Eq.~(\ref{eq:enVFR}),
or for very weakly bound states, indirectly through the dependence of $\Z$
on $g$.
To date, binding energies for about thirty molecules have been
measured. They range from $\sim 1$~meV in small molecules such as CH$_3$F,
to $\sim 300$~meV for large alkanes \cite{YS08a,YS08b}.
A recent analysis indicates that these binding energies increase approximately
linearly with the molecular dipole polarizability and dipole moment, and for
aromatic molecules, the number of $\pi $ bonds \cite{DYS09}.

For atoms, comparatively accurate positron binding energies have been
predicted theoretically for about ten species \cite{MBR02}, but there are no
measurements.
There have been a number of calculations for positron binding to molecules
\cite{SW76,KJ78,KJ81,DT88,SM01,CS06,Str04,Str99,BMM98,GFB06,TBK03,BLM05,%
BL08,CLG08}. Most of these molecules have large dipole moments which
facilitate binding. In contrast, the molecules for which the binding energies
are known from experiment are either nonpolar or only weakly polar. Thus at
present, there are no species for which experiment and theory can be
compared, and so this is a critical area for future research.

Presented here is a review of theoretical and experimental results for positron
annihilation on molecules in the range of energies below the positronium
formation threshold. Emphasis is placed upon the case in which positrons bind
to the target and annihilation proceeds via the formation of vibrational
Feshbach
resonances. Current knowledge of positron-molecule binding energies, obtained
from both experiment and theoretical calculations, is summarized. These
results are related to studies of positron-induced fragmentation of molecules,
annihilation gamma ray spectra, annihilation in dense gases where nonlinear
effects are observed, and to analogous electron interactions with molecules
and clusters.

\section{Theory}\label{sec:theory}

\subsection{Annihilation basics}\label{subsec:basics}

The process of electron-positron annihilation is described by quantum
electrodynamics (QED). In the nonrelativistic Born approximation, the
cross section for annihilation into two photons
averaged over the electron and positron spins is \cite{QED}
\begin{equation}\label{eq:sig2gam}
\overline{\sigma }_{2\gamma }=\pi r_0^2\frac{c}{v},
\end{equation}
where $v$ is the relative velocity of the two particles. This cross section
obeys the $1/v$ threshold law which describes
inelastic collisions with fast particles in the final state \cite{QM}.

The two-photon annihilation described by Eq.~(\ref{eq:sig2gam})
is allowed only when the total spin $S$ of the electron-positron pair
is zero. For $S=1$ the smallest possible number of annihilation photons is
three. The corresponding spin-averaged cross section is \cite{QED}
\begin{equation}\label{eq:sig3gam}
\overline{\sigma }_{3\gamma }=\frac{4}{3}(\pi ^2-9)\alpha r_0^2\frac{c}{v},
\end{equation}
where $\alpha =e^2/\hbar c $ (in cgs units) is the fine structure constant,
$\alpha \approx 1/137$. Since $\overline{\sigma }_{3\gamma }$ is 400 times
smaller than $\overline{\sigma }_{2\gamma }$, positron annihilation in
many-electron systems is dominated by the two-gamma process.

Numerically, the cross section in Eq.~(\ref{eq:sig2gam}) is
$\overline{\sigma }_{2\gamma }\sim 10^{-8}c/v$ a.u.\footnote{We make
use of atomic units (a.u.), in which $m=|e|=\hbar =1$,
$c=\alpha ^{-1}\approx 137$~a.u., and the Bohr radius $a_0=\hbar^2/me^2$
(in cgs units) also equals unity.}
Hence
the annihilation rate is usually much smaller than the rates for other atomic
collision processes, even at low positron velocities (e.g., thermal,
$v\sim 0.05$~a.u. at 300~K).
When a fast positron, such as that emitted in a $\beta ^+$ decay, moves
through matter, it loses energy quickly through collisions,
first by direct ionization, positronium formation and electronic excitation,
and then by vibrational excitation and elastic collisions. As a result,
the positrons typically slow to thermal energies
(i.e., $\sim 25$~meV for $T=300$~K) before annihilation.

At small velocities, $v\lesssim 1$ a.u., Eq.~(\ref{eq:sig2gam}) must be
modified to take into account the Coulomb interaction between
the electron and positron. The typical momenta exchanged in the annihilation
process are $p\sim mc$. The corresponding separation, $r\sim \hbar /mc $, is
small compared to $a_0$, and in the nonrelativistic
limit, the annihilation takes
place when the electron and positron are at the same point. The cross
section in Eq.~(\ref{eq:sig2gam}) must then be multiplied by the probability
density at the origin \cite{QM},
\begin{equation}\label{eq:Gamow}
|\psi (0)|^2=\frac{2\pi }{v(1-e^{-2\pi /v})},
\end{equation}
where the wave function $\psi $ is normalized by
$\psi ({\bf r})\simeq e^{i{\bf k}\cdot {\bf r}}$ at $r\gg a_0$.
This increases the annihilation cross section.

The annihilation cross section for many-electron targets is traditionally
written as \cite{Pom49,Fra68}
\begin{equation}\label{eq:sigma_a}
\sigma _a=\overline{\sigma }_{2\gamma }\Z=\pi r_0^2\frac{c}{v}\Z ,
\end{equation}
where $\Z$
represents the {\em effective number of electrons} that contribute to the
annihilation. In the Born approximation,
$\Z=Z$, the total number of target electrons.

However, at small positron energies (e.g., $\eps \lesssim 1$ eV), $\Z$ can be
different from $Z$.
First, there is a strong repulsion between the positron and the atomic nuclei.
This prevents the positron from penetrating deep into the atoms, so that
the annihilation involves predominantly electrons in the valence and
near-valence subshells, thereby reducing $\Z$. On the other hand, the
positron is attracted to the target by a long-range polarization potential
$-\alpha _d /2r^4$, where $\alpha _d$ is the target dipole polarizability,
which enhances $\Z$. There is also a
short-range enhancement of $\Z$ due to the Coulomb interaction between
the annihilating electron and positron, which has the
same origin as the expression in Eq.~(\ref{eq:Gamow}).
Finally, if the target binds the positron, the annihilation can
be enhanced by positron capture into this bound state. The cross section
for radiative capture (i.e., by emission of a photon) is small, namely
$\sigma _c\sim \alpha ^3 a_0^2$ \cite{QED}.
In collisions with molecules, the positron can transfer its
energy to vibrations, forming a positron-molecule complex. This process
is effective in enhancing the annihilation rate. It is the
principal focus of the present review.

As follows from its definition by Eq.~(\ref{eq:sigma_a}), $\Z$ is equal to the
electron density at the positron,
\begin{equation}\label{eq:Zeff}
\Z =\int \sum _{i=1}^Z\delta ({\bf r}-{\bf r}_i)|\Psi _{\bf k}
({\bf r}_1,\dots ,{\bf r}_Z,{\bf r})|^2d{\bf r}_1\dots d{\bf r}_Zd{\bf r}~,
\end{equation}
where $\Psi _{\bf k}({\bf r}_1,\dots ,{\bf r}_Z,{\bf r})$ 
is the total wave function of the system, with electron
coordinates ${\bf r}_i$ and positron coordinate ${\bf r}$. This
wave function describes the scattering of the positron with initial
momentum ${\bf k}$ by the atomic or molecular target, and is
normalized to the positron plane wave at large separations,
\begin{equation}\label{eq:asymp}
\Psi _{\bf k}({\bf r}_1,\dots ,{\bf r}_Z,{\bf r})\simeq \Phi _0
({\bf r}_1,\dots ,{\bf r}_Z)e^{i{\bf k}\cdot {\bf r}},
\end{equation}
where $\Phi _0$ is the wave function of the initial (e.g., ground) state
of the target. For molecules, both $\Psi _{\bf k}$ and $\Phi _0$ also depend
on the nuclear coordinates, which must
be integrated over in Eq.~(\ref{eq:Zeff}).

Equations (\ref{eq:sigma_a}) and (\ref{eq:Zeff}) determine the annihilation
rate in binary positron-molecule collisions,
\begin{equation}\label{eq:rate}
\lambda =\sigma _a v n=\pi r_0^2c\Z n,
\end{equation}
where $n$ is the gas number density. To compare with experiment, this rate
is averaged over the positron energy distribution. For thermal positrons
this distribution is a Maxwellian, while in beam experiments, it
is determined by the parameters of the beam. Empirically
Eq.~(\ref{eq:rate}) is also used to describe experiments at high densities
where $\Z$ becomes density dependent (see Sec.~\ref{subsec:density}).

\subsection{Gamma-ray spectra and annihilation rates}\label{subsec:anspec}

In the nonrelativistic limit, the two-photon QED annihilation amplitude
can be expressed in terms of an effective annihilation operator
\begin{equation}\label{eq:annop1}
\hat O_a({\bf P})\equiv \int e^{-i{\bf P}\cdot {\bf r}}\hat \psi ({\bf r})
\hat \varphi ({\bf r})d{\bf r},
\end{equation}
where $\hat \psi ({\bf r})$ and $\hat \varphi ({\bf r})$ are the electron and
positron destruction operators,\footnote{In Eq.~(\ref{eq:annop1}) the spin
indices in $\hat \psi ({\bf r})$ and $\hat \varphi ({\bf r})$
are suppressed, and summation over them is assumed. This form can
be used in systems with paired electron spins or when averaging over the
positron spin. The modulus-squared amplitude is then multiplied
by the spin-averaged QED factor $\pi r_0^2c$. In general, one should
use the spin-singlet combination of the annihilation operators
in Eq.~(\ref{eq:annop1}), $\frac{1}{\sqrt{2}}
(\hat \psi _{\uparrow }\hat \varphi _{\downarrow }
-\hat \psi _{\downarrow }\hat \varphi _{\uparrow })$,
together with the two-photon annihilation factor $4\pi r_0^2c$.}
and ${\bf P}$ is the total momentum of the photons \cite{Fer56,Lee57,DG06a}.
The probability distribution of ${\bf P}$ in an annihilation event is given by
\begin{equation}\label{eq:wP}
W_f({\bf P})=\pi r_0^2c\left|\langle f|\hat O_a({\bf P})|i\rangle \right|^2,
\end{equation}
where $|i\rangle $ is the initial state with $Z$ electrons and the positron
(e.g., that with the wave function $\Psi _{\bf k}$), and $|f\rangle $ is
the state of $Z-1$ electrons after the annihilation.

For ${\bf P}=0$, the two photons are emitted in opposite directions
and have equal energies,
$E_{\gamma 1}=E_{\gamma 2}\equiv E_\gamma \approx mc^2$.
For ${\bf P}\neq 0$ the photon energy is Doppler-shifted, e.g.,
\begin{equation}\label{eq:Dop}
E_{\gamma 1}=E_\gamma +mc|{\bf V}|\cos \theta ,
\end{equation}
where 
${\bf V}={\bf P}/2m$ is the center-of-mass velocity of the electron-positron
pair, and $\theta $ is the angle between ${\bf V}$ and the direction of the
photon. Averaging the distribution of the Doppler shifts
$\epsilon = E_{\gamma 1}-E_\gamma  =(Pc/2)\cos \theta $, over the direction
of emission of the photons, gives the photon energy spectrum,
\begin{equation}\label{eq:wepsP}
w_f(\epsilon )=\frac{1}{c}\int \!\! \int _{2|\epsilon|/c}^{\infty}
W_f({\bf P}) \frac{Pd Pd\Omega _{\bf P}}{(2\pi )^3}.
\end{equation}
In Cartesian coordinates,
\begin{equation}\label{eq:wepsC}
w_f(\epsilon )=\frac{2}{c}\int \!\!\! \int W_f(P_x,P_y,2\epsilon /c)
\frac{d P_xd P_y}{(2\pi )^3}.
\end{equation}
This form shows that the energy spectrum is proportional to the probability
density for a component of ${\bf P}$. This quantity can be measured 
either by sampling the Doppler spectrum of the gamma rays or by measuring
the angular deviation of the two photons (see Sec.~\ref{subsec:gam_meas}).

When a low-energy positron annihilates with a bound electron with energy
$\eps _n$, the mean photon energy $E_\gamma $ is shifted by $\eps _n/2$
relative to $mc^2$. This shift is much smaller than the typical Doppler shift
$\epsilon $ due to the momentum of the bound electron,
$P\sim \sqrt{2m|\eps _n|}$, which corresponds to
$\epsilon \sim Pc\sim \sqrt{|\eps _n|mc^2}\gg|\eps _n|$. The resulting width
and shape of the gamma spectrum contain important information about the bound
electrons.

In most experiments, the annihilation photons are not detected in
coincidence with the final state $f$, and the observed spectrum is the sum
over all final states,
$w(\epsilon )=\sum _f w_f(\epsilon )$. However, this spectrum still reveals
contributions of different final states. For example, in partially
fluorinated alkanes, annihilation with the tightly bound fluorine $2p$ electrons
results in a broader spectral component than annihilation with the more
diffuse C-H bond electrons. This allows one to deduce the
relative fraction of the corresponding annihilation events \cite{IGG97}
(see Sec.~\ref{subsec:gamma}).

The total annihilation rate in the state $i$ leading to the final state
$f$ is obtained by integration over the momenta,
\begin{eqnarray}
\lambda _f=\pi r_0^2c \int |\langle f|\hat O_a({\bf P})|i\rangle |^2
\frac{d^3P}{(2\pi )^3},\label{eq:lambda}
\end{eqnarray}
and the total annihilation rate in state $i$ is
\begin{equation}\label{eq:lambda1}
\lambda =\sum _f\lambda _f=
\pi r_0^2c \int \langle i|\hat n_-({\bf r})\hat n_+({\bf r})|i\rangle d{\bf r},
\end{equation}
where $\hat n_-({\bf r})=\hat \psi ^\dagger ({\bf r})\hat \psi ({\bf r})$ and
$\hat n_+({\bf r})=\hat \varphi ^\dagger ({\bf r})\hat \varphi ({\bf r})$
are the electron and positron density operators. The annihilation rate
is thus given by the expectation value of the electron density at the
positron. Equation (\ref{eq:lambda1}) gives the two-photon annihilation rate
in a system of one positron and one target atom or molecule. For a positron
moving through a gas
of density $n$, the annihilation rate takes the form of
Eq.~(\ref{eq:rate}). Normalizing the initial state $i$ to one positron
per unit volume, as $\Psi _{\bf k}$ in Eq.~(\ref{eq:asymp}), one obtains
\begin{equation}\label{eq:Zeff1}
\Z =\int \langle i|\hat n_-({\bf r})\hat n_+({\bf r})|i\rangle d{\bf r}.
\end{equation}
In the coordinate representation, this yields Eq.~(\ref{eq:Zeff}).

In the independent-particle approximation, the electronic parts of the
initial and final states are Slater determinants constructed
from the electron orbitals (e.g., in the Hartree-Fock scheme). The
incident positron is described by its own wave function
$\varphi _{\bf k}({\bf r})$, and the annihilation amplitude
$\langle f|\hat O_a({\bf P})|i\rangle $ takes the form
\begin{equation}\label{eq:Amp0}
A_{n{\bf k}}({\bf P})=
\int e^{-i{\bf P}\cdot {\bf r}} \psi_n({\bf r})\varphi_{{\bf k}}({\bf r}) 
d {\bf r},
\end{equation}
where $\psi _n({\bf r})$ is the orbital of the annihilated electron.
In this approximation
\begin{equation}\label{eq:Zeff0}
\Z =\sum _{n=1}^Z\int |\psi _n({\bf r})|^2|\varphi_{\bf k}({\bf r})|^2 d{\bf r},
\end{equation}
i.e., the average {\em product} of the electron and positron densities.

\subsection{Positron-molecule wave function}
\label{subsec:anmech}

The wave function $\Psi _{\bf k}$ for the positron colliding with a molecule,
can be written as \cite{Gri00,Gri01}
\begin{equation}\label{eq:totpsi}
\Psi _{\bf k}=\Psi ^{(0)}_{\bf k}+\sum _\nu
\frac{\Psi _\nu \langle \Psi _\nu |V|\Psi ^{(0)}_{\bf k}\rangle }
{\eps -\eps _\nu +\frac{i}{2}\Gamma _\nu }~.
\end{equation}
The first term on the right-hand side describes
direct, or potential scattering of the positron by the target.
The corresponding wave function $\Psi ^{(0)}_{\bf k}$ is determined by the
positron interaction with the charge distribution of the ground-state
target and electron-positron correlation effects (e.g., target polarization
and virtual Ps formation). It neglects the coupling $V$ between
the electron-positron and nuclear (vibrational) degrees of freedom.

The second term describes positron capture into the vibrational Feshbach
resonances. It is present for molecules that can bind the positron.
These resonances correspond to vibrationally-excited states $\Psi _\nu $
of the positron-molecule complex, embedded in the positron continuum.
They occur when the positron energy $\eps =k^2/2$ is close to
$\eps_{\nu} = E_{\nu} -\eb $, where $\eb $ is the positron binding energy, and
$E_\nu $ is the vibrational excitation energy of the positron-molecule complex.
Equation (\ref{eq:totpsi}) has the appearance of a
standard perturbation-theory formula, but the energies of the
positron-molecule quasibound states $\Psi _\nu $ in the denominator are
complex, $\eps _\nu -\frac{i}{2}\Gamma _\nu $, where
$\Gamma _\nu = \Gamma _\nu ^a+\Gamma _\nu^e+\Gamma _\nu^i$ is the total
{\em width} of the resonance. In atomic units $\Gamma _\nu $ is equal to the
decay rate of the resonant state.  It contains contributions of positron
annihilation and elastic escape, $\Gamma _\nu ^a$ and $\Gamma _\nu^e$,
respectively, and possibly also the inelastic escape rate $\Gamma _\nu^i$.
The latter describes positron autodetachment accompanied by
vibrational transitions to the states other than the initial state.

Molecular rotations are, in general, not expected to affect positron
annihilation. The rotational motion is slow compared to the motion of the
positron or the vibrational motion. Accordingly, direct scattering can
be considered for fixed molecular orientation, and the results averaged over
the orientations. Positron capture in VFRs at low energies is dominated by
the $s$ wave, or at most a few lower partial waves. Hence in the capture
process, the angular momentum of the molecule remains unchanged or changes
little.

The positron capture amplitude
$\langle \Psi _\nu |V|\Psi ^{(0)}_{\bf k}\rangle $ determines
the elastic width in state $\nu $,
\begin{equation}\label{eq:gamc}
\Gamma _\nu ^e =2\pi \int |\langle \Psi _\nu |V|\Psi ^{(0)}_{\bf k}
\rangle |^2 \frac{ k d\Omega _{\bf k}}{(2\pi )^3}.
\end{equation}
If the positron interaction with the vibrations cannot be described by
perturbation theory, Eqs.~(\ref{eq:totpsi}) and (\ref{eq:gamc}) remain
valid provided the amplitudes
$\langle \Psi _\nu |V|\Psi ^{(0)}_{\bf k}\rangle $ are replaced by
their nonperturbative values.

According to Eq.~(\ref{eq:lambda1}), the annihilation rate of the
positron-molecule state $\Psi _\nu $ is given by
\begin{equation}\label{eq:gama}
\Gamma _\nu ^a=\pi r_0^2c \rho _{ep}~,
\end{equation}
where $\rho _{ep}$ is the average electron density at the positron,
\begin{equation}\label{eq:rho_ep}
\rho _{ep}=\int \sum _{i=1}^Z\delta ({\bf r}-{\bf r}_i)
|\Psi _\nu ({\bf r}_1,\dots ,{\bf r}_Z,{\bf r})|^2
d{\bf r}_1\dots d{\bf r}_Zd{\bf r},
\end{equation}
with the integration extending to the nuclear coordinates in
the wave function $\Psi _\nu$. The amplitude of the nuclear motion
is small, and $\rho _{ep}$ is expected to depend weakly on the degree of
vibrational excitation in state $\nu $.

To calculate $Z_{\rm eff}$, the wave function from Eq.~(\ref{eq:totpsi}) is
substituted into Eq.~(\ref{eq:Zeff}), which yields
\begin{eqnarray}\label{eq:zeff1}
\Z &=&\langle \Psi _{\bf k}| \sum _{i=1}^Z
\delta ({\bf r}-{\bf r}_i)| \Psi _{\bf k}\rangle \nonumber \\
&=&\langle \Psi _{\bf k}^{(0)}| \sum _{i=1}^Z
\delta ({\bf r}-{\bf r}_i)| \Psi _{\bf k}^{(0)}\rangle +
\left\{ {{\rm interference}\atop {\rm terms}} \right\} \nonumber \\
&+&\frac{2\pi ^2}{k}\sum _{\mu \nu}\frac{A_\mu ^*\langle \Psi _\mu |
\sum _{i=1}^Z \delta ({\bf r}-{\bf r}_i)| \Psi _\nu \rangle A_\nu }
{(\eps -\eps _\mu -\frac{i}{2}\Gamma _\mu )
(\eps -\eps _\nu +\frac {i}{2}\Gamma _\nu )}~,
\end{eqnarray}
where the capture amplitude $A_\nu $ is related to the elastic width by
$\Gamma _\nu ^e=2\pi |A_\nu |^2$.
The terms on the right-hand side describe the contributions of
direct and resonant annihilation and the interference
between the two. We will now examine the two main contributions in
detail.

The separation of the wave function into the direct and resonant parts
in Eq.~(\ref{eq:totpsi}) is valid because the positron VFRs are narrow.
This is a consequence of the weakness of coupling between the
positron and the vibrational motion (i.e., small capture widths
$\Gamma _\nu^e$, see Sec.~\ref{subsec:IR}). In spite of this, the resonant
contribution to the annihilation rate for complex polyatomics exceeds
the direct contributions by orders of magnitude.

\subsection{Direct annihilation: virtual and weakly bound positron states}
\label{subsec:dir}

The potential scattering wave function $\Psi _{\bf k}^{(0)}$
satisfies the Schr\"odinger equation
\begin{equation}\label{eq:Schr}
(T+U-E_0)\Psi _{\bf k}^{(0)}=\varepsilon \Psi _{\bf k}^{(0)},
\end{equation}
where $T$ is the kinetic energy operator for the electrons and positron,
$U$ is the sum of all Coulomb interactions between the particles (with the
nuclei at their equilibrium positions), and $E_0$ is the target ground-state
energy.

For positron energies below the Ps-formation threshold, annihilation
occurs when the positron is within the range of the target ground-state
electron cloud. At such distances, the interaction $U$ between the
particles is much greater than the
positron energy $\eps $. Therefore, the $\varepsilon \Psi _{\bf k}^{(0)}$
term in Eq. (\ref{eq:Schr}) can be neglected, and the solution
$\Psi _{\bf k}^{(0)}$ at these small separations does not depend on
$\eps $, except through a normalization factor.

When the positron is outside the target, $\Psi _{\bf k}^{(0)}$
contains contributions of the incident and scattered positron waves,
\begin{equation}\label{eq:out}
\Psi _{\bf k}^{(0)}({\bf r}_1,\dots ,{\bf r}_Z,{\bf r})\simeq \Phi _0
({\bf r}_1,\dots ,{\bf r}_Z)\left[e^{i{\bf k}\cdot {\bf r}}+
f_{{\bf kk}'}\frac{e^{ikr}}{r} \right] ,
\end{equation}
where $f_{{\bf kk}'}$ is the scattering amplitude, and ${\bf k}'=k{\bf r}/r$.
Inside the target, $\Psi _{\bf k}^{(0)}$ is determined
by matching it with Eq.~(\ref{eq:out}) at the target boundary $r=R$, where
$R$ is the characteristic radius of the target. For small positron momenta,
$kR\ll 1$, the scattering is dominated by the $s$ wave, and the amplitude
$f_{{\bf kk}'}$ can be replaced by the $s$-wave amplitude $f_0$. As a result,
the integrand in Eq.~(\ref{eq:Zeff}) for $\Z$ is proportional to
$| 1+f_0/R|^2$ \cite{DFK93}. This gives the following estimate for $\Z$ due
to direct annihilation \cite{Gri00},
\begin{equation}\label{eq:Zdir}
Z_{\rm eff}^{\rm (dir)}\simeq 4\pi \rho _e\delta R \left( R^2+
2R\,{\rm Re}f_0 +\frac{\sigma _{\rm el}}{4 \pi }\right)~,
\end{equation}
where $\rho _e$ is the effective electron density in the region of annihilation,
$\delta R$ is the range of distances where the positron annihilates, and
$\sigma _{\rm el}$ is the elastic cross section. At small positron energies,
$\sigma _{\rm el}\simeq 4\pi |f_0|^2$, and in the zero-energy limit
$\sigma _{\rm el}=4\pi a^2$, where $a$ is the positron scattering length,
$a=-f_0$ at $k=0$.\footnote{If the target molecule has a permanent dipole
moment $\mu $,
the long-range dipole potential $\bm{\mu }\cdot {\bf r}/r^3$ dominates the
low-energy scattering \cite{Fab77}. This makes $\sigma _{\rm el}$
infinite, while $\Z$ remains finite, making Eq.~(\ref{eq:Zdir})
invalid.}

A simple estimate of the factor $4\pi \rho _e\delta R\equiv F$ in
Eq.~(\ref{eq:Zdir}) is obtained using the
Ps density at the origin, $\rho _e\sim \rho _{\rm Ps}=1/8\pi $, and
$\delta R\sim 1$, which yields $F \sim 0.5$. Equation (\ref{eq:Zdir}) then
shows that the magnitude of $Z_{\rm eff}^{\rm (dir)}$ is comparable to
the geometrical cross section of the target (in atomic units), unless
$\sigma _{\rm el}$ is much greater than $R^2$.

When the scattering cross section is large, the annihilation rate is greatly
enhanced. This occurs when the positron has a virtual or a bound state close
to zero energy \cite{GS64}. Such states are characterized by a small parameter
$\kappa =1/a$, $|\kappa |\ll R^{-1}$. It is related to the energy of the
bound state $\varepsilon _0=-\kappa ^2/2$ (for $\kappa >0$), or virtual state,
$\varepsilon _0=\kappa ^2/2$ (for $\kappa <0$). This parameter determines
the low-energy $s$-wave scattering amplitude $f_0=-(\kappa +ik)^{-1}$
and cross section $\sigma _{\rm el}\simeq 4\pi /(\kappa ^2 +k^2)$ \cite{QM}.
For small $\kappa $, this cross section can be much greater than the
geometrical size of the target. The last term in brackets
in Eq.~(\ref{eq:Zdir}) then dominates, and $\Z^{\rm (dir)}$ shows a similar
enhancement \cite{DFK93,MI02},\footnote{The long-range
polarization potential $-\alpha _d/2r^4$ modifies the near-threshold
form of $\sigma_{\rm el}$ and $\Z ^{\rm (dir)}$ \cite{Gri00,Mit02},
but Eq.~(\ref{eq:Z_virt}) can still be used as an estimate.}
\begin{equation}\label{eq:Z_virt}
\Z^{\rm (dir)}\simeq \frac{F}{\kappa ^2+k^2}.
\end{equation}

\begin{figure}[ht]
\includegraphics*[width=7cm]{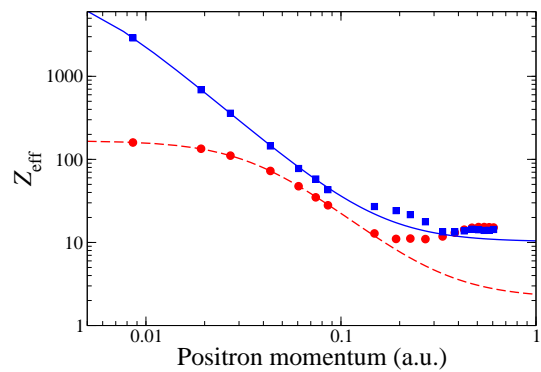}
\caption{Comparison of the $\Z$ values calculated using the
SMC method for C$_2$H$_2$ (squares) and C$_2$H$_4$
(circles) by \textcite{VCL02} (see Sec.~\ref{subsubsec:ann}) with the fit
using Eq.~(\ref{eq:Z_virt})
with a constant vertical offset. Parameters of the fit:
C$_2$H$_2$ (solid curve), $F=0.261$, $\kappa =0.0041$; 
C$_2$H$_4$ (dashed curve), $F=0.230$, $\kappa =0.0372$.}
\label{fig:Zeff_Lima}
\end{figure}

The applicability of Eq.~(\ref{eq:Z_virt}) is shown in
Fig.~\ref{fig:Zeff_Lima}. It shows the $\Z$ values from the Schwinger
multichannel (SMC) calculation for C$_2$H$_2$ and C$_2$H$_4$ \cite{VCL02},
fitted using Eq.~(\ref{eq:Z_virt}) with a small vertical offset
to account for the nonresonant $\Z$ background. According to the SMC
calculation, both molecules possess virtual positron states.
This results in the characteristic rise of $\Z$ at
small positron momenta described by Eq.~(\ref{eq:Z_virt}). The virtual
level in C$_2$H$_2$ (fitted value $\kappa =0.0041$) lies closer to zero
energy than in C$_2$H$_4$ ($\kappa =0.0372$), which manifests in the large
$\Z$ values for acetylene. The fitted factor $F\approx 0.25$ for the
two molecules is smaller than the estimate obtained from high-quality atomic
calculations (see below). This is likely an
indication of the lack of short-range correlation terms in the SMC
calculation, which would enhance the electron density at the
positron (see Sec.~\ref{subsubsec:ann}).

Positron virtual states explain the large thermal $\Z$ values observed at
room temperatures in heavier noble gases \cite{DFK93,DFG96}. The
value of $\Z=400$ observed for Xe \cite{MS90} is close to the maximum direct
annihilation rate for thermal positrons at 300~K. It is estimated from
Eq.~(\ref{eq:Z_virt}) to be
\begin{equation}\label{eq:Zdir_lim}
\Z^{\rm (dir)}\lesssim 10^3.
\end{equation}
Higher $\Z$ values observed in many polyatomics (see, e.g.,
Table \ref{tab:Zeff}) can be understood only by considering positron-molecule
binding and resonances.

The annihilation rate for the positron bound to an atom or molecule is
\begin{equation}\label{eq:Gamma_a}
\Gamma ^a=\pi r_0^2c\int \sum _{i=1}^Z\delta ({\bf r}-{\bf r}_i)|\Psi _0
({\bf r}_1,\dots ,{\bf r}_Z,{\bf r})|^2d{\bf r}_1\dots d{\bf r}_Zd{\bf r}~,
\end{equation}
where $\Psi _0$ is the wave function of the bound state. For a weakly
bound state (e.g., $\eb \ll 1$ eV) $\Gamma ^a$ can be estimated in a way
similar to that used for $\Z^{\rm (dir)}$ above. When the positron is outside the
target ($r>R$), $\Psi_0$ takes the form
\begin{equation}\label{eq:Psi_out}
\Psi _0({\bf r}_1,\dots ,{\bf r}_Z,{\bf r})\simeq \Phi _0
({\bf r}_1,\dots ,{\bf r}_Z)\frac{A}{r}e^{-\kappa r},
\end{equation}
where $A$ is the asymptotic normalization constant.\footnote{Equation
(\ref{eq:Psi_out}) assumes that the ionization potential of the atomic system
satisfies $E_i>E_{\rm Ps}$. For $E_i<E_{\rm Ps}$ the asymptotic form is
that of Ps bound to the positive ion \cite{MBR02}.}
For weak binding ($\kappa \ll R^{-1}$) the main
contribution to the normalization integral
\begin{equation}
\int |\Psi _0({\bf r}_1,\dots ,{\bf r}_Z,{\bf r})|^2d{\bf r}_1\dots
d{\bf r}_Zd{\bf r}=1,
\end{equation}
comes from large positron separations where Eq.~(\ref{eq:Psi_out}) is valid.
This yields
\begin{equation}\label{eq:A}
A=\sqrt{\kappa /2\pi }.
\end{equation}
By matching the wave function $\Psi_0$ in Eq.~(\ref{eq:Gamma_a}) at $r=R$
with the asymptotic form in Eq. (\ref{eq:Psi_out}), one obtains
\begin{equation}\label{eq:Gam_est}
\Gamma ^a\simeq \pi r_0^2c\,4\pi \rho _e\delta R |A|^2=
\pi r_0^2cF\frac{\kappa }{2\pi },
\end{equation}
\cite{Gri01}. Hence the electron-positron contact density from
Eq. (\ref{eq:rho_ep}) is estimated by
\begin{equation}\label{eq:rhoep}
\rho_{ep}\simeq (F/2\pi )\kappa .
\end{equation}

\begin{figure}[ht]
\includegraphics*[width=7cm]{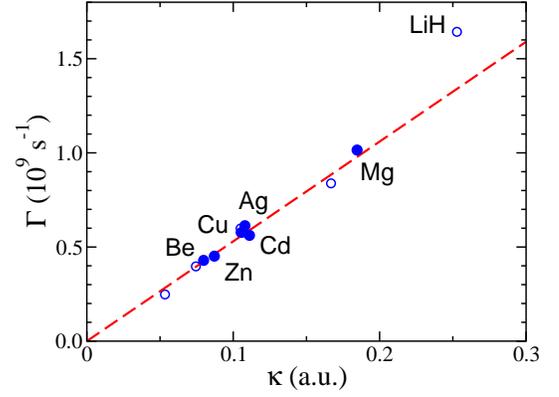}
\caption{Dependence of the annihilation rate $\Gamma ^a$ for positron
bound states on the parameter $\kappa =\sqrt{2\eb }$: solid circles,
recent results for six atoms
\cite{MBR02,MZB08,BM02,BM06,BM10}; open circles, earlier
results for these atoms \cite{MBR02} and LiH molecule \cite{MR00};
dashed line is the fit $\Gamma ^a=5.3\kappa $~(in $10^9$~s$^{-1}$) which
corresponds to $F=0.66$ a.u.}
\label{fig:rate_kap}
\end{figure}

Equation (\ref{eq:Gam_est}) shows that $\Gamma ^a$ is proportional to
$\kappa =\sqrt{2\eb}$ (i.e., to the square root of the binding energy;
see \textcite{MI02} for an alternative derivation).\footnote{Equation
(\ref{eq:Psi_out}) is valid if the positron-target interaction is
short-range. It must be modified if the molecule has a dipole moment
[see, e.g., \textcite{Fab77}]. However, Eq.~(\ref{eq:Gam_est}) can be
used as an estimate if the dipole force does not play a dominant role in the
binding.}
This relationship between $\Gamma ^a$ and $\kappa $ is confirmed by
positron-atom bound state calculations \cite{MBR02}. Figure~\ref{fig:rate_kap}
shows values for six atoms with $E_i>E_{\rm Ps}$, namely Be, Mg, Cd, Cu, Zn and
Ag, obtained using high-quality configuration interaction and stochastic
variational methods (see Sec.~\ref{subsubsec:bind} for details). Note that
the datum for the LiH molecule also follows this trend,
in spite of its large dipole moment [$\mu = 5.9$~D \cite{CRC}] and
relatively strong binding. A linear fit through the atomic data points gives
a value for the factor $F=4\pi \rho _e\delta R$ in Eq.~(\ref{eq:Gam_est}),
namely $F\approx 0.66$~a.u., which is close to the rough estimate given above. 
One can use this value to evaluate the annihilation rates for positron-molecule
bound states from Eq.~(\ref{eq:Gam_est}), provided their binding energies
are known.

\subsection{Resonant annihilation}\label{subsec:res}

The effect of resonances on $\Z$ is described by the second and third terms in
Eq.~(\ref{eq:zeff1}). It is dominated by the diagonal part of the double
sum in the last term. The off-diagonal and interference terms vanish upon
averaging over the positron energy and can usually be neglected.
The resonant contribution to the annihilation cross section is described
by the Breit-Wigner formula~\cite{QM},
\begin{equation}\label{eq:siga}
\sigma _a=\frac{\pi}{k^2}\sum _\nu \frac{b_\nu \Gamma_\nu^a \Gamma_\nu ^e}
{(\eps - \eps _\nu)^2+\frac 14 \Gamma_\nu^2},
\end{equation}
where $b_\nu$ is the degeneracy of the $\nu$th resonance.
Equation (\ref{eq:sigma_a}), (\ref{eq:gama}), and (\ref{eq:siga}) then give
the resonant $\Z$,
\begin{equation}\label{eq:Zeffres}
\Zr =\frac{\pi }{k}\rho_{ep}\sum_\nu \frac{b_\nu \Gamma_\nu^e}
{(\eps - \eps _\nu )^2+\frac 14 \Gamma_\nu^2}.
\end{equation}
The contact density $\rho _{ep}$ can be estimated from Eq.~(\ref{eq:rhoep})
if the positron binding energy is known. To calculate $\Zr$, one also
needs the energies and widths of the resonances. The former are
determined by the positron binding energy and the vibrational excitation
energies of the positron-molecule complex. The elastic and total rates depend
on the strength of coupling between the positron and the vibrational
motion, and for overtones and combination excitations, on the strength
of anharmonic terms in the vibrational Hamiltonian. This makes an
{\it ab initio} calculation of resonant $\Z$ a multifaceted problem.

\subsection{Resonances due to infrared-active modes}\label{subsec:IR}

One case in which such a calculation is possible is that of isolated vibrational
resonances of IR-active fundamentals \cite{GL06a}.
Consider a small polyatomic molecule which supports a bound
positron state with a small binding energy $\eb=\kappa^2/2 \ll $~1.
The wave function of the bound positron is very diffuse. Outside the
molecule it behaves as $\varphi _0({\bf r})=Ar^{-1}e^{-\kappa r}$, with $A$
given by Eq.~(\ref{eq:A}).

Suppose that the vibrational modes in this molecule are not mixed with
overtones or combination vibrations. Due to the weakness of the positron
binding, the vibrational excitation energies of the positron-molecule
complex are close to the vibrational fundamentals
$\omega _\nu $ of the neutral molecule,
$E_\nu \approx \omega _\nu $.\footnote{There is extensive experimental
evidence that $E_\nu \approx \omega _\nu $ for most resonances observed
(cf. Secs. \ref{sec:small} and \ref{sec:large}). Apparent exceptions, where
shifts $\sim $\,10--20 meV are observed are the C-H stretch mode
of CH$_3$F and the O-H stretch in methanol (cf. Sec.~\ref{sec:small}).}
In this case the sum in Eq.~(\ref{eq:Zeffres}) is over the modes $\nu $, and
the resonant energies are $\eps _\nu =\omega _\nu -\eb$.
Some (or all) of these modes can be IR active. Positron capture into
the corresponding VFR is mediated by the long-range dipole coupling, and one
can readily evaluate this contribution to $\Zr$.

Consider a positron with momentum ${\bf k}$ incident upon a molecule in the
vibrational ground state $\Phi_0 ({\bf R})$, where ${\bf R}$ represents all
of the molecular coordinates. If $k^2/2\approx \eps _\nu $, the
positron can be captured into a VFR, and thus be bound to the molecule in a
vibrationally excited state $\Phi _\nu ({\bf R})$. The corresponding
rate $\Gamma_\nu^e $, given by Eq.~(\ref{eq:gamc}), can be found using a method
similar to the Born dipole approximation \cite{Lan80} with the coupling
$V=\hat{\bf d}\cdot {\bf r} /r^3$, where $\hat{\bf d}$ is the dipole moment
operator of the molecule.\footnote{Experiments show that the Born dipole
approximation provides good estimates, or lower bounds, for the excitation
cross sections of IR-active modes by low-energy
positrons \cite{MS05a,MGS06}.} This gives the amplitude
\begin{eqnarray}
\langle \Phi _\nu |V|\Psi ^{(0)}_{\bf k} \rangle &=&
\int \varphi _0({\bf r})\Phi_\nu ^\ast ({\bf R})
\frac{\hat{\bf d}\cdot {\bf r} }{r^3}e^{i{\bf k}\cdot{\bf r}}\Phi_0 ({\bf R})
\,d{\bf r}d{\bf R} \nonumber \\
&=& \frac{4\pi i}{3}\,\frac{{\bf d}_\nu \cdot {\bf k}}{\sqrt{2\pi\kappa}}
\, _2F_1\left(\frac{1}{2},1;\frac{5}{2};-\frac{k ^2}{\kappa^2 }\right),
\label{eq:Anuk}
\end{eqnarray}
where ${\bf d}_\nu =\langle \Phi _\nu |\hat{\bf d}|\Phi _0\rangle $, and
$_2F_1$ is the hypergeometric function.\footnote{Here,
$_2F_1\left(\frac{1}{2},1;\frac{5}{2};-z^2\right )
=\frac32 z^{-2}[(1+z^2)z^{-1}\arctan z -1]$.} The corresponding elastic rate is
\begin{equation}\label{eq:Gam_eh}
\Gamma _\nu ^e =\frac{16\omega _\nu d_\nu ^2}{27}\, h(\xi ),
\end{equation}
where $h(\xi )=\xi ^{3/2}(1-\xi )^{-1/2}
\left[ _2F_1\left(\frac{1}{2},1;\frac{5}{2};-\frac{\xi }{1-\xi }\right)
\right]^2 $
is a dimensionless function of $\xi =1-\eb/\omega _\nu$, with a
maximum $h\approx 0.75$ at $\xi \approx 0.89$.

Equation (\ref{eq:Gam_eh}) shows that the elastic rate for the resonance
of an IR-active mode is determined largely by its frequency
$\omega _\nu $ and transition dipole amplitude $d_\nu $. Their values
are known for many species from IR absorption measurements [e.g., see
\textcite{BC82}].

This theory has been successfully applied to the methyl halides \cite{GL06a}.
Energy-resolved measurements of $\Z$ for CH$_3$F \cite{BGS03}, CH$_3$Cl, and
CH$_3$Br \cite{BYS06} show peaks close to the vibrational mode energies,
pointing to a sizeable contribution of resonant annihilation in these molecules.
The methyl halides have $C_{3v}$ symmetry, and all six vibrational modes are
IR active (see Table \ref{tab:CH3Cl} for CH$_3$Cl). Methyl halides are also
relatively small, so that IVR effects may not be important in the energy range
of the fundamentals (see Sec. \ref{subsec:large}).
Thus for these molecules, Eqs. (\ref{eq:rhoep}), (\ref{eq:Zeffres}), and
(\ref{eq:Gam_eh}) allow one to calculate the contribution of all
VFR to $\Zr$.

\begin{table}[ht]
\caption{Characteristics of the vibrational modes of CH$_3$Cl.}
\label{tab:CH3Cl}
\begin{ruledtabular}
\begin{tabular}{cccrcc}
Mode & Symmetry & $b _\nu $ & $\omega_\nu $\footnotemark[1]
& $d_\nu$\footnotemark[1] & $\Gamma _\nu ^e$\footnotemark[1]\\
 & & & (meV) & (a.u.) &  ($\mu $eV) \\
\hline
$\nu_1$ & $A_1$ & 1 & 363 & $0.0191$ & 57.2 \\
$\nu_2$ & $A_1$ & 1 & 168 & $0.0176$ & 22.9 \\
$\nu_3$ & $A_1$ & 1 &  91 & $0.0442$ & 65.7 \\
$\nu_4$ & $E$ & 2 & 373 & $0.0099$ & 15.9 \\
$\nu_5$ & $E$ & 2 & 180 & $0.0162$ & 20.9 \\
$\nu_6$ & $E$ & 2 & 126 & $0.0111$ & 6.4
\end{tabular}
\end{ruledtabular}
\footnotetext[1]{Mode energies and transition amplitudes from \textcite{BC82};
elastic widths from Eq.(\ref{eq:Gam_eh}) for $\eb =25$~meV.}
\end{table}

The only free parameter in the theory is the positron binding energy.
It can be chosen by comparison with the experimental $\Z$ spectrum. To do
this, the theoretical $\Z$ must be averaged over the energy distribution of the
positron beam $f_{\rm b}(\eps -\bar \eps )$ from Eq.~(\ref{eq:posen}),
\begin{equation}\label{eq:Zeffresav}
\bar Z_{\rm eff}^{\rm (res)}(\bar \eps)=\int \Zr (\eps )
f_{\rm b}(\eps -\bar \eps) d\eps .
\end{equation}
This integral is simplified by the fact that the resonances are very narrow,
since the total width $\Gamma _\nu =\Gamma _\nu^e+\Gamma _\nu^a$ is small
compared to the energy spread of the beam. For example, the values in
Table \ref{tab:CH3Cl} show that $\Gamma _\nu ^e<0.1$~meV.
The annihilation width is even smaller; for $\eb=25$~meV,
Eq.~(\ref{eq:Gam_est}) yields $\Gamma _\nu ^a = 0.15~\mu {\rm eV}$.
Hence,
\begin{equation}\label{eq:Zeff_fin}
\bar Z _{\rm eff}^{(\rm res)}(\bar \eps )=
2\pi^2\rho_{ep}\sum_\nu \frac{b_\nu \Gamma _\nu^e}{k_\nu \Gamma _\nu}
\Delta (\bar \eps -\eps _\nu),
\end{equation}
where $k_\nu =\sqrt{2\eps _\nu }$, and $\Delta (E)\equiv f_{\rm b}(-E)$
describes the shape of the resonances as measured in the positron-beam
experiment (see Sec. \ref{subsec:beam}).

The above estimates show that $\Gamma _\nu ^a \ll\Gamma _\nu^e$. In this case
the total decay rate is dominated by the elastic rate,
$\Gamma _\nu \approx \Gamma _\nu^e$, and the contributions of individual
resonances to the sum in Eq.~(\ref{eq:Zeff_fin}) are not sensitive to the
precise values of $\Gamma _\nu^e$. Therefore
even relatively weak positron-vibrational coupling is sufficient to
fully ``turn on'' the resonance contribution. This explains why the $\Z$
and IR absorption spectra of a molecule can be quite different, even
when the resonant $\Z$ is determined by dipole coupling.

Application of this theory to methyl halides and other small molecules
is discussed in Sec. \ref{sec:small}.

\subsection{Resonant annihilation in large molecules}\label{subsec:large}

\subsubsection{Vibrational level densities}\label{subsubsec:dens}

In general, a molecule with $N$ atoms has $3N-6$ vibrational degrees
of freedom. If positron attachment proceeds only via excitation
of the single-mode VFRs, the resonant $\Z$ values will grow
linearly with the size of the molecule.
However the experimental $\Z$ data show a much faster increase (cf. the data
for alkanes in Table~\ref{tab:Zeff}). These large $\Z$ values can only be
explained if the positrons can couple, at least indirectly, with
multiquantum vibrations.

In large polyatomic species, the total vibrational level density increases
rapidly with the excitation energy and is quite high, even in the energy
range of the fundamentals. The spacing between the multimode VFRs
in such spectra is small compared with the energy spread of the incident
positrons. Averaging $\Zr$ from Eq.~(\ref{eq:Zeffres}) over
an energy interval which contains many resonances near positron
energy $\eps $, one obtains
\begin{equation}\label{eq:Zeffav}
\Z (\eps )=\frac{2 \pi ^2\rho _{ep}}{k}
\left\langle \frac{\Gamma ^{e}(\varepsilon )}
{\Gamma (\varepsilon)}\right\rangle \rho (\varepsilon +\eb),
\end{equation}
where $\rho (\varepsilon +\eb)$ is the vibrational level density
of the positron-molecule complex. From now on we omit the superscript
in $\Zr $, since $\Z$ in large molecules is almost entirely
due to resonant annihilation. In Eq.~(\ref{eq:Zeffav}) it is assumed that
the positron collides with a molecule in the ground vibrational state.
Larger molecules have a significant thermal energy content at room
temperature, which can be taken into account (see below).

The $\Z$ in Eqs.~(\ref{eq:Zeffres}) and (\ref{eq:Zeffav}),
and the contributions of individual resonances in Eq.~(\ref{eq:Zeff_fin}),
are proportional to $\rho _{ep}/k$. Since $\rho _{ep}\propto \sqrt{\eb}$
[cf. Eq.~(\ref{eq:rhoep})], the resonant contribution to $\Z$ is proportional
to the dimensionless ``kinematic'' factor $g=\sqrt{\eb/\eps }$.
The magnitude and energy dependence of $\Z$ beyond this factor are due to
the dynamics of molecular vibrations and positron interaction with them, etc.
Normalizing $\Z$ by $g$ has proven to be useful in
analyzing various trends in resonant annihilation \cite{YS07}
(see Secs.~\ref{sec:small} and \ref{sec:large}).

\begin{figure}[ht]
\includegraphics*[width=7cm]{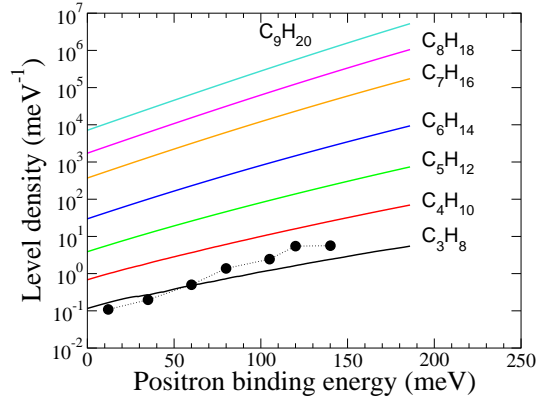}
\caption{Vibrational level densities in alkanes as functions of the positron
binding energy: solid curves, calculated from Eq.~(\ref{eq:vdensT}) for
room-temperature incident positrons; circles, estimated from
experimental $Z_{\rm eff}$ values at 300~K for C$_3$H$_8$ to
C$_9$H$_{20}$, using Eqs.~(\ref{eq:rhoep}) and (\ref{eq:Zeffav}) and
assuming $\Gamma ^{e}/\Gamma =1$, and plotted against the
experimental binding energies (Table \ref{tab:Zeff_N}).}
\label{fig:dens}
\end{figure}

According to Eq. (\ref{eq:Zeffav}), the resonant $\Z$ is proportional
to the vibrational spectrum density $\rho $. This density can be evaluated
easily in the harmonic approximation in which $E_\nu =\sum _kn_k\omega _k$,where
$n_k$ are non-negative integers, and $\omega _k$ are the mode frequencies.
For weakly bound positron-molecule complexes, these frequencies are close to
those of the neutral molecule, and can be taken from experiment \cite{NIST} or
quantum chemistry calculations (e.g., using Q-Chem, \textcite{QChem}).
To apply Eq.~(\ref{eq:Zeffav}) to the
annihilation of thermal positrons at temperature $T$, the density must be
averaged over the Maxwellian positron energy distribution,
\begin{equation}\label{eq:vdensT}
\bar \rho (\eb)=\int _0^\infty \rho (\eb+k^2/2)
\frac{e^{-k^2/2k_BT}}{(2\pi k_BT)^{3/2}}\,4\pi k^2dk,
\end{equation}
where $k_B$ is the Boltzmann constant. Figure~\ref{fig:dens} shows
$\bar \rho (\eb )$ for alkanes with $n=3$--9 carbons.
These densities increase rapidly with the positron binding energy and 
with the size of the alkane molecule.

The estimates made in Sec.~\ref{subsec:IR} show that
$\Gamma ^e/\Gamma \approx 1$ for the VFRs of IR-active modes. If the same were
true for the multimode resonances, the increase of $\Z$ values along the
alkane series would match the growth of their vibrational level densities,
as per Eq.~(\ref{eq:Zeffav}). However, experimental room-temperature $\Z$
values increase at a much slower rate. As a result, the {\em effective}
vibrational densities $\rho $ estimated using Eqs.~(\ref{eq:Zeffav}) and
(\ref{eq:rhoep}) from experimental thermal $\Z$ values and binding energies
(Table \ref{tab:Zeff_N}), and assuming $\Gamma ^e/\Gamma = 1$,
are much lower than the calculated densities for all alkanes larger than
propane (see Fig.~\ref{fig:dens}). Hence the above assumption is generally
incorrect, and instead, $\Gamma ^e/\Gamma \ll 1$ \cite{GG04}.

What is the physical reason for the suppressed $\Gamma ^e/\Gamma $ ratio?
First, the positron coupling to multimode VFR is likely much weaker than
that of single-mode resonances. For example, the dipole coupling analyzed
in Sec. \ref{subsec:IR} can cause only single-quantum vibrational transitions
(in the harmonic approximation). Combination vibrations and overtones
can be excited due to anharmonic or Coriolis terms in the vibrational
Hamiltonian. In this case, the coupling strength of a single-mode excitation
is divided between many multiquantum excitations. The value of $\Gamma ^e$
for a multimode VFR is then only a small {\em fraction} of the typical
single-mode $\Gamma ^e$. This
is discussed in more detail in Sec. \ref{subsubsec:door}.
In addition, some vibrational excitations may be completely inaccessible
to the incident positron (e.g., due to symmetry). This will reduce
the average $\langle \Gamma ^{e}/\Gamma \rangle$, or, effectively, reduce
the vibrational density in Eq.~(\ref{eq:Zeffav}).

Another reason for the reduction of $\Gamma ^{e}/\Gamma $ could be the
contribution of vibrationally {\em inelastic} escape to the total width
$\Gamma $.
At present there is little direct experimental evidence of inelastic positron
escape (see Sec. \ref{sec:large}). However, if the system was in the regime
of strong mixing between the single-mode and multimode vibrational excitations,
then one cannot see why such channels would be closed.\footnote{Vibrationally
inelastic scattering following capture has been studied in electron-molecule
collisions via energy-loss spectra. In addition to
the excitation of modes and overtones, vibrational state
``quasicontinua'' have been observed in larger polyatomic molecules
at electron-volt excitation energies \cite{All84}.}

A simple estimate of this effect can be made assuming complete,
statistical mixing of all vibrational excitations \cite{GL09}.
When a positron with energy $\eps $ collides with a molecule with 
vibrational energy $E_v$ then all final vibrational states with energies
$E_v'<E_v+\eps $ can be populated. Assuming that the positron coupling
strengths to all vibrational excitations are similar, one has
$\Gamma ^e/\Gamma \approx 1/N(\eps +E_v)$, where 
$N(\eps +E_v)=\int _0^{\eps +E_v}\rho (E_v')dE_v'$ is the number of open
vibrational escape channels. Equation~(\ref{eq:Zeffav}) then
becomes 
\begin{equation}\label{eq:Zeff_N}
\Z\approx \frac{2 \pi ^2\rho _{ep}}{k}
\frac{\rho (\varepsilon +E_v+\eb)}{N(\eps +E_v)}.
\end{equation}
For a given molecule this expression contains only one free parameter, namely,
the positron binding energy.

\begin{figure}[ht]
\includegraphics*[width=7cm]{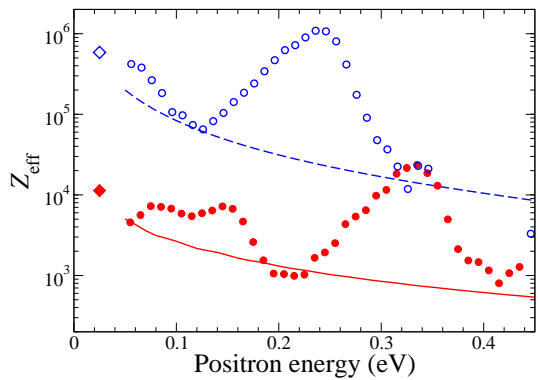}
\caption{Comparison of experimental $\Z$ for butane (solid symbols) and
octane (open symbols) with the predictions of the statistical
model: diamonds, experimental $\Zth$ values for
thermal positrons at 300~K (cf. Table \ref{tab:Zeff_N}); circles, $\Z$
measured as a function of positron energy \cite{BGS03}; curves, $\Z$
calculated from Eq.~(\ref{eq:Zeff_N}), using $\eb$ fitted to reproduce
thermal $\Z$.}
\label{fig:alk}
\end{figure}

The binding energies can be chosen by comparison with
experimental room-temperature data for thermal positrons
(Table~\ref{tab:Zeff_N}), averaging $\Z$ from Eq.~(\ref{eq:Zeff_N})
over the initial target states using the Boltzmann factors $\exp (-E_v/k_BT)$,
and the Maxwellian positron energy distribution. For alkanes with
3--8 carbons such a fit gives the binding energies $\eb=22$, 42, 65, 90, 103
and 122~meV, respectively. These values are in good agreement with those
measured in the positron-beam experiments (cf. Table~\ref{tab:Zeff_N}). 
However, the dependence of $\Z$ on the positron energy predicted by
Eq.~(\ref{eq:Zeff_N}) is in striking disagreement with the measured
energy-resolved $\Z$, as shown in Fig.~\ref{fig:alk} for butane and octane.
The $\Z$ values from Eq.~(\ref{eq:Zeff_N}) decrease monotonically,
the increase in $N(\eps +E_v)$ being faster than that of the density
$\rho $. In contrast, the experimental $\Z$ spectra show resonant peaks
corresponding to the vibrational fundamentals (downshifted by $\eb $,
cf. Fig.~\ref{fig:butane}), and the peak $\Z$ values exceed the
predictions of Eq.~(\ref{eq:Zeff_N}) by a factor of $\sim 50$.
Thus the model that assumes complete statistical mixing with all
vibrational excitations coupled to the positron continuum, does not explain
either the resonant structure or the large values of $\Z$ that are observed.

\subsubsection{Mode-based resonant doorway states}\label{subsubsec:door}

The energy dependence of the measured $\Z$ in alkanes shown in
Fig.~\ref{fig:alk} is of the type expected for {\em mode-based} VFRs
described in Sec. \ref{subsec:IR}. On the other hand, the magnitude
of $\Z$ and its increase with the size of the molecule, are much greater than
the contribution of mode-based resonances. This suggests a two-step model of
positron capture that involves mode-based vibrational doorway
resonances\footnote{The term ``doorway resonance'' originates in
nuclear
physics, where it means ``a metastable state formed in the initial state of the
reaction'', which ``may decay partly into the open channels (direct reactions),
and partly through the coupling to the internal degrees of
freedom'' \cite{BM98}.} \cite{GG04}.

In this model, the incident positron first forms a quasibound state with the
molecule by transferring its excess energy to a single mode with near-resonant
energy $\omega _n\approx \eps +\eb$. This simple {\em doorway} state
of the positron-molecule complex [also termed a zeroth-order
``bright'' state \cite{NF96}] is embedded in the dense spectrum of multimode
vibrations (or ``dark'' states, as they are not coupled directly to the
positron continuum). Due to vibrational state mixing caused by anharmonic
or rotationally-induced coupling terms in the vibrational Hamiltonian, the
doorway state can then ``spread'' into multimode vibrational
states. This process of vibrational energy redistribution
takes place on a time scale $\tau \sim 1/\Gamma _{\rm spr}$,
where $\Gamma _{\rm spr}$ is known as the spreading width in nuclear physics,
or the IVR rate in molecular physics.

To link the multimode-VFR and the doorway-state-resonance pictures,
consider a perturbative expression for the elastic rate,
\begin{equation}\label{eq:Gamma_e}
\Gamma _\nu ^{e}=2\pi |\langle \Psi _\nu |V|0,\varepsilon \rangle |^2,
\end{equation}
where $|0,\varepsilon \rangle $ describes the positron incident on the
ground-state molecule.\footnote{See Eq. (\ref{eq:gamc}); for
simplicity, in Eq.~(\ref{eq:Gamma_e}) we assume that only one positron partial
wave contributes to the rate and that its wave function is normalized to a
$\delta $-function in energy rather than to a plane wave.}
The multimode eigenstate of the positron-molecule complex
$|\Psi _\nu \rangle $ can be expanded in the basis of the noninteracting
(e.g., harmonic approximation) multimode vibrational states $|\Phi _i \rangle $,
\begin{equation}\label{eq:C}
|\Psi _\nu \rangle =\sum _iC_i^{(\nu )} |\Phi _i \rangle ,
\end{equation}
where the coefficients $C_i^{(\nu )}$ are obtained by diagonalizing
the vibrational Hamiltonian in the basis of $|\Phi _i \rangle $.

Let us assume that of all $|\Phi _i \rangle $, only those which
describe a bound positron and a single-mode excitation (i.e.,
$|n,\varepsilon _0\rangle $, where $n$ indicates the mode) are coupled
to $|0,\varepsilon \rangle $. These states $|n,\varepsilon _0\rangle $
are the doorway states introduced above. The coefficients $C_i^{(\nu )}$
describe the mixing of the doorway states with the multimode eigenstates
$\nu $  (i.e., ``spreading''). The corresponding probabilities can be
approximated by a Breit-Wigner line shape,
\begin{equation}\label{eq:C2}
|C_i^{(\nu )}|^2 \propto \frac{\Gamma _{\rm spr} ^2/4}
{(E_\nu -E_i)^2+\Gamma _{\rm spr}^2/4},
\end{equation}
subject to normalization $\sum _i| C_i^{(\nu )}|^2 =1$. Here $E_\nu $ and
$E_i$ are the energies of the eigen- and basis states of the positron-molecule
complex, respectively, so that $E_\nu \approx \eps -\eps _0=\eps +\eb $ and
$E_i= \omega _n $ for $|\Phi _i \rangle = |n,\varepsilon _0\rangle $.

Using Eqs.~(\ref{eq:Zeffav}), (\ref{eq:Gamma_e}),
(\ref{eq:C}) and (\ref{eq:C2}), one obtains $Z_{\rm eff}$,
averaged over energy on the scale of closely spaced VFRs, as
\begin{equation}\label{eq:Zefffun}
Z_{\rm eff}=\frac{2 \pi ^2\rho _{ep}}{k}\frac{\Gamma _{\rm spr}}
{2\pi \Gamma (\varepsilon )}\sum _n\frac{\Gamma _n^e}
{(\varepsilon -\omega _n+\eb)^2+\frac{1}{4}\Gamma _{\rm spr}^2},
\end{equation}
where $\Gamma _n^{e}=2\pi |\langle n,\varepsilon _0|V
|0,\varepsilon \rangle |^2$ is the elastic rate of the doorway of mode $n$.
Equation (\ref{eq:Zefffun}) has the same form as Eq.~(\ref{eq:Zeffres}),
except that it contains the elastic rates of the doorways,
$\Gamma _n^{e}$, and the sum is over the {\em modes}.

Comparison with Eq. (\ref{eq:Zeffav}) shows that the resonant energy
dependence
of $\Z$ in Eq. (\ref{eq:Zefffun}) is due to the modulation of the elastic rate
by the mode-based doorways. Evaluating $\Z$ from Eq.~(\ref{eq:Zefffun})
at the doorway resonance energy (i.e., $\varepsilon =\omega _n-\eb$) and
comparing it with Eq.~(\ref{eq:Zeffav}) gives an estimate of the elastic rate
of the VFR:
\begin{equation}\label{eq:GamGam_n}
\Gamma ^{e}(\varepsilon )\sim \frac{2\Gamma _n^e}
{\pi \rho (\eps +\eb)\Gamma _{\rm spr}}.
\end{equation}
The product $\rho (\eps +\eb)\Gamma _{\rm spr}$ is the number of vibrational
eigenstates within the energy interval $\Gamma _{\rm spr}$. In the regime
of strong level mixing this number is large,
$\rho (\eps +\eb)\Gamma _{\rm spr}\gg 1$. Equation (\ref{eq:GamGam_n}) shows
that the elastic rates of the multimode VFRs are much smaller than those of
the doorways, as each VFR carries a small fraction of the positron coupling
strength of the doorway.

Equation (\ref{eq:Zefffun}) shows that the enhancement of $\Z$ for
larger molecules is caused by (i)~larger binding energies and
$\rho _{ep}\propto \sqrt{\eb }$, (ii)~larger number of modes/doorways in the
sum, and most importantly (iii)~smaller total decay rates $\Gamma (\eps )$
(i.e., longer lifetimes) of the VFRs. This decrease in $\Gamma (\eps )$ must
be related to the suppression of $\Gamma ^{e}$ due to higher level densities.
The total rate is, however, always bounded from below by the
annihilation contribution, $\Gamma (\eps )>\Gamma ^a$. The annihilation rates
are typically smaller than the elastic rates of mode-based doorways by 2--3
orders of magnitude (see Sec. \ref{subsec:IR}). Hence, one can expect
a similar-sized increase in the contribution to $\Z$ from each mode-based
doorway resonance in a large molecule, compared with that in a small molecule.

\begin{figure}[t]
\includegraphics*[width=7cm]{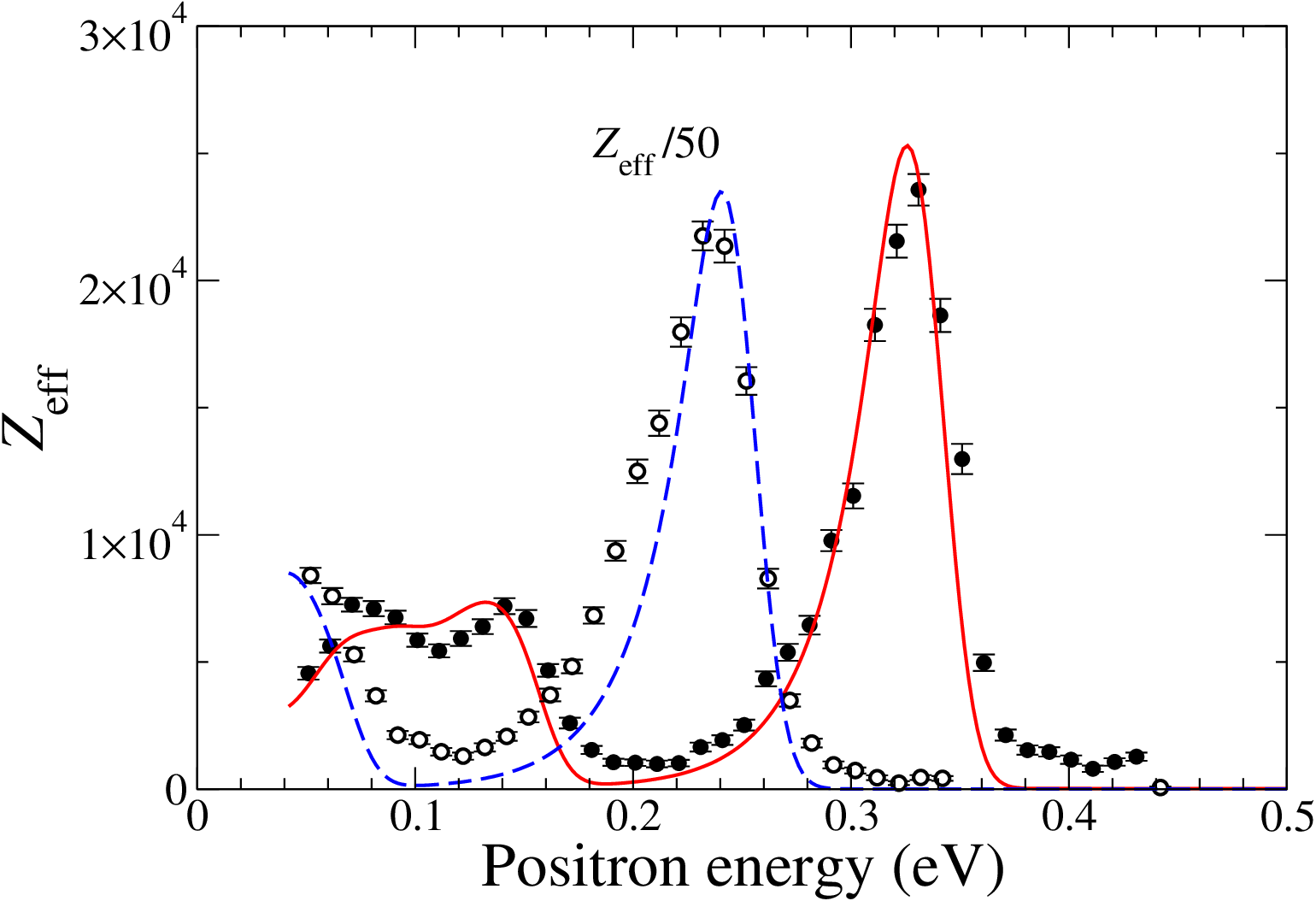}
\caption{Comparison of the experimental $\Z$ for ($\bullet $) butane,
C$_4$H$_{10}$ and ($\circ $) octane, C$_8$H$_{18}$ (scaled by a factor $1/50$)
\cite{BGS03}, with $\Z$ from Eq.~(\ref{eq:Zefffun}). The theoretical
$\Z$ is averaged over the positron beam energy distribution. The fit for
butane uses $\eb=35$~meV, $\Gamma _n^e/\Gamma =7.2$ for C-H stretch
modes and $\Gamma _n^e/\Gamma =1.2$ for the rest; for octane,
$\eb=122$~meV, $\Gamma _n^e/\Gamma =84$ for C-H stretch modes and
$\Gamma _n^e/\Gamma =14$ for the rest.}
\label{fig:c4c8}
\end{figure}

Figure \ref{fig:c4c8} shows the applicability of Eq.~(\ref{eq:Zefffun}).
Here it is used to fit the experimental energy-resolved $\Z$ data for butane
and octane \cite{BGS03}. For this comparison, $\Z$ from Eq.~(\ref{eq:Zefffun})
is averaged
over the positron energy distribution (Sec. \ref{subsec:beam}). The result
does not depend on the spreading width, as long as it is much smaller than
the typical energy spread of the beam (i.e., $\Gamma _{\rm spr}\ll 40$ meV).
The binding energies of the two species are chosen to be $\eb=35$ and 122~meV.
The remaining unknown parameter is the ratio
$\Gamma _n^e/\Gamma (\varepsilon )$. Given the differences between
the C-H stretch and lower-energy resonances in Fig. \ref{fig:c4c8}, two
different values are used for these groups of modes, the former six times
greater than the latter. The ratios $\Gamma _n^e/\Gamma $ for octane are
approximately 12 times larger than for butane. In the context of the model,
this reflects the greater degree of mixing between multimode
VFRs in the larger molecule.

The results shown in Fig.~\ref{fig:c4c8} summarize the extent of our
understanding of annihilation in large molecules. To explain the observed
magnitudes of $\Z$, positron capture into multimode VFR must be invoked,
mediated by a process such as IVR. However, if the enhancement
is indeed due to IVR, it appears to be far from statistically complete.

\subsubsection{Annihilation and the onset of IVR}\label{subsubsec:IVR}

Experimentally, resonant annihilation is observed in relatively small
polyatomics with four or five atoms (Sec.~\ref{sec:small}), as well as in much
larger molecules (Sec.~\ref{sec:large}).
In the first case, the annihilation is due to the single-mode VFRs, possibly
augmented by contributions of combination and overtone resonances
(see Sec.~\ref{subsec:meth}). In large molecules enhanced $\Z$ values
are assumed to be due to the spreading of the vibrational energy into
multimode VFRs. The energy dependence of $\Z$ in large molecules is
typically well represented by the spectrum of the fundamentals,
which act as doorways. Isolated overtone and combination VFRs cannot, in
general, be identified as distinct
doorways in the $\Z$ spectrum, though there are exceptions (e.g.,
benzene, Sec. \ref{sec:large}). This can probably be explained by
the much smaller ratios of $\Gamma _n^e/\Gamma (\varepsilon )$ for
the combination/overtone doorways [cf. Eq.~(\ref{eq:Zefffun})].

The transition from the small- to the large-molecule behavior occurs with the
onset of strong vibrational mixing (i.e., IVR) involving, in particular,
the vibrational fundamentals. The phenomenon of IVR has been studied
widely using a number of techniques \cite{NF96}. In particular, high-resolution
(0.0005~cm$^{-1}$) measurements of vibrational spectra of jet-cooled molecules
allow direct observation of the splitting of single-mode transitions into
clumps of vibrationally mixed multimode levels spread over an energy interval
$\Gamma _{\rm spr}\sim 0.02$~cm$^{-1}\approx 2~\mu$eV \cite{MN90}.

Another technique used to study IVR in the range of the
C-H stretch modes, is IR fluorescence \cite{SM83}. Here, a fundamental
vibration is excited by a short laser pulse. If this vibration
is mixed with multimode excitations, the amount of fluorescence at the
fundamental frequency is reduced. The measured fluorescence, normalized to the
known IR absorption strength, gives the ``dilution factor''. Its reciprocal
characterizes the number of eigenstates that are strongly
coupled to the fundamental. If all vibrational states within the energy
range $\Gamma _{\rm spr}$
are mixed, the dilution factor will be $\sim (\rho \Gamma _{\rm spr})^{-1}$,
where $\rho $ is the vibrational spectrum density at the relevant energy. Note
that a similar factor enters Eq.~(\ref{eq:GamGam_n}). \textcite{SM83}
observed that the dilution factor drops rapidly when the level density is
increased beyond the ``threshold'' value of $\rho =10$--100$/\mbox{cm}^{-1}$.
This is in agreement with the value $\Gamma _{\rm spr}\sim 0.02$~cm$^{-1}$,
as $\rho \Gamma _{\rm spr}\gtrsim 1$ marks the onset of IVR.

In order to see if a similar threshold governs the transition from the
small- to the large-molecule behavior in annihilation, vibrational
densities have been evaluated for a number of molecules using the harmonic
approximation. The densities at the excitation energy of the C-H stretch mode
($E=2900~\mbox{cm}^{-1}=0.36~\mbox{eV}$) are shown in Table \ref{tab:dens}. To
assess the effect of finite molecular temperature, the densities were
evaluated both at zero temperature, denoted $\rho (E)$, and at room temperature,
$\rho _T(E)=\sum _v\rho (E+E_v)e^{-E_v/k_BT}/\sum _ve^{-E_v/k_BT}$.
In Table~\ref{tab:dens} values of the IR fluorescence dilution factor, where
known, are also listed.

\begin{table}[ht]
\caption{Vibrational excitation densities for zero- and
room-temperature molecules at the C-H stretch mode energy.}
\label{tab:dens}
\begin{ruledtabular}
\begin{tabular}{lcccc}
\multicolumn{2}{c}{Molecule} & $\rho $\footnotemark[1] &
$\rho _T$\footnotemark[1] & Dilution\\
& & $1/\mbox{cm}^{-1}$ & $1/\mbox{cm}^{-1}$ & factors\footnotemark[2]\\
\hline
Acetylene & C$_2$H$_2$ & 0.05 & 0.06 & $-$\\
Methyl chloride & CH$_3$Cl & 0.07 & 0.07 & $-$\\
Ethylene & C$_2$H$_4$ & 0.08 & 0.09 & $-$\\
Methanol & CH$_3$OH & 0.12 & 0.13 & $-$\\
Ethane & C$_2$H$_6$ & 0.42 & 0.54 & 0.8,~1\\
Cyclopropane & C$_3$H$_6$ & 0.59 & 0.81 & 0.3,~0.7,~0.8\\
Ethanol & C$_2$H$_5$OH & 2.84 & 6.22 & $-$ \\
Propane & C$_3$H$_8$ & 5.19 & 14.3 & 0.2,~0.4\\
Benzene & C$_6$H$_6$ & 5.76 & 22.3 & 0.6,~0.7,~0.9\\
Butane & C$_4$H$_{10}$ & 106 & 921 & 0.16,~0.05\footnotemark[3] 
\end{tabular}
\end{ruledtabular}
\footnotetext[1]{Densities at $E=2900~\mbox{cm}^{-1}$
for $T=0$ and $T=293$~K, averaged over a Gaussian
with 10 meV, FWHM. Mode frequencies for most
molecules are from \textcite{NIST}.}
\footnotetext[2]{Values for the C-H stretch modes \cite{SM83}.}
\footnotetext[3]{Values for 1-butyne \cite{KKM87} and isobutane \cite{SM83},
respectively.}
\end{table}

Energy-resolved $\Z$ measurements put the top four molecules in
Table~\ref{tab:dens} in the small-molecule category (cf. Sec. \ref{sec:small}).
Propane, benzene and butane at the bottom of the table, behave as large
molecules (Sec. \ref{sec:large}), as they exhibit $\Z$ values that cannot
be explained by the mode-based VFR.
Ethane and cyclopropane appear to be borderline, with characteristics that
place them in both categories (e.g., the C-H stretch peak in both molecules
appears to be strongly enhanced). The values of the densities in
Table~\ref{tab:dens} are broadly in agreement with this classification, with
the threshold density, $\rho \sim 1/\mbox{cm}^{-1}$. This value is
lower than the threshold density in the fluoresecence studies, perhaps due
to differences in the nature of the vibrational energy transfer in the two
cases. Somewhat surprisingly, $\Z$ measurements
for ethanol (cf. Sec.~\ref{subsec:smallsize}) indicate that it is a ``small
molecule'', in contrast with its vibrational density value. This
suggests that the threshold value depends on the details of the vibrational
Hamiltonian.

Table ~\ref{tab:dens} shows that molecular temperature can strongly affect
the density for larger molecules. The only experiments to investigate the
effect of molecular temperature on $\Z$ in similar-sized molecules, heptane
and pentane, show only a small effect at the C-H stretch peak
(Sec.~\ref{subsubsec:temp}). While the analysis here indicates that molecules
with $\rho \sim 1/\mbox{cm}^{-1}$ might change from exhibiting small- to
large-molecule behavior as the temperature is increased, molecules with these
density values exhibit only small changes in $\rho $ with temperature. Thus
the combination of this analysis and the experimental results on large
molecules indicate that temperature likely plays a limited role in
changing the details of the $\Z$ spectra.

When the level densities for larger molecules are examined as a function of the
excitation energy $E$, the densities in the C-H stretch energy range
are an order-of-magnitude greater than those in the range of other
fundamentals, $E\lesssim 1000$~cm$^{-1}$. In principle, this could explain
the stronger enhancement of $\Z$ in the C-H stretch peak,
as compared to that in the low-energy mode peaks (see Fig.~\ref{fig:c4c8} and
Sec.~\ref{sec:large} for further discussion).

To summarize, optics-based IVR studies show that the degree of IVR increases
rapidly as a function of molecular size. The onset of strong IVR occurs for
molecules that are similar in size to those for which mode-based, VFR-mediated
annihilation begins to fail to explain the $\Z$ spectra (e.g., ethane,
cyclopropane and propane). Thus, while indirect, these results fit well
with the physical picture that IVR is responsible for the very large
annihilation rates observed in large molecules.

\subsection{Calculations of annihilation and binding}\label{subsec:calc}

\subsubsection{Annihilation}\label{subsubsec:ann}

Most calculations of $\Z$ for molecules have been done for diatomics or small
polyatomics with fixed nuclei, ignoring the vibrational dynamics. This is a
good approximation for molecules that do not possess VFRs, where direct
annihilation is the dominant mechanism. Such a calculation is still far from
trivial, given the large role of electron-positron correlations.

The H$_2$ molecule is an example. In this case one can construct
a sufficiently flexible trial wave function for the positron and two
electrons and use the generalized Kohn method to solve the scattering problem
\cite{Arm84} and calculate $\Z$ \cite{AB85}. This calculation
provides an accurate description of the positron elastic scattering
cross section below 5 eV \cite{ABP90}. Calculation of $\Z$ requires
the inclusion of terms with explicit dependence on the
electron-positron distance \cite{AB86}.
However, it still yields a thermal room-temperature annihilation rate of
$\Z=10.2$, well below the experimental value of $14.8\pm 0.2$ \cite{MSB79}.
This discrepancy has been finally resolved using the stochastic variational
method \cite{ZMV09}, which is one of the most powerful methods for studying
few-body systems.

Another {\em ab initio} method used to calculate positron-molecule scattering
and annihilation is the Schwinger multichannel (SMC) method \cite{GL93,SGL94}.
In this scheme the ($Z+1$)-particle wave function is expanded in a Cartesian
Gaussian basis set, with the functions centered on the atomic nuclei
and on additional centers outside the molecule. The latter are important for
representing electron-positron correlation effects. This method has been
applied to H$_2$ \cite{LGS98}, N$_2$ \cite{CVL00}, C$_2$H$_4$ \cite{SGL96},
and C$_2$H$_2$ \cite{CVL03}. In all cases the differential and total elastic
scattering cross sections are in good agreement with the experimental
data. In contrast, the
calculated $\Z$ values for room-temperature positron energies \cite{VCL02} are
well below the measured thermal data, namely,
$\Z=7.3$ {\em vs} 14.8 for H$_2$ \cite{MSB79}, 9.3 {\em vs} 30.5 for
N$_2$ \cite{HCG82}, 73 {\em vs} 1200 for C$_2$H$_4$ \cite{IGS94},
and 145 {\em vs} 3160 for C$_2$H$_2$ \cite{IGS97}.\footnote{The SMC computer
code used to calculate the $\Z$ values prior to 2001
contained an extra factor of $Z$ due to a programming mistake \cite{VCL02},
which gave an illusion of agreement with experiment.} The discrepancy for
H$_2$ and N$_2$ is most likely due to a lack of basis functions that
describe short-range electron-positron correlations. The much
larger gap between the theory and experiment for ethylene and acetylene has a
different origin. As discussed in Sec.~\ref{subsec:forbid}, the energy-resolved
$\Z$ data for both molecules show large contributions
of resonant annihilation that cannot be described by a fixed-nuclei
calculation.

In spite of the failure to reproduce the $\Z$ values, the SMC calculations
provide an important clue about the physics of the positron interaction with
ethylene and acetylene. They show that electron-positron correlations in these
systems lead to strong positron-molecule attraction that produces
virtual levels close to zero energy \cite{CVL03,SGL96}. As a result,
the calculated $\Z$ display a characteristic growth at low energies
(cf. Fig.~\ref{fig:Zeff_Lima}). Observation of resonant annihilation
for these molecules (Sec. \ref{subsec:forbid}) indicates that they do, in
fact, support positron {\em bound} states with $\eb \lesssim 10$~meV. This means
that a relatively small (though, computationally challenging)
improvement in the SMC calculations could prove the existence of these
bound states. This would represent an important step towards {\em ab initio}
theoretical description of resonant annihilation.

Positron interaction with acetylene (and methane) was also examined by
\textcite{NG03} using a body-fixed vibrational close-coupling (VCC) method.
In this approach, the positron-molecule interaction is modeled using a
correlation-polarization potential (CPP). These calculations
revealed the existence of a positron
virtual state for methane ($\kappa \approx -0.15$~a.u.) and acetylene
($\kappa \approx -0.01$~a.u.). As in the case of the SMC calculations, it
appears that only a small increase in the potential is required to turn the
virtual state into the observed bound state in acetylene. Such an increase
may well be
within the uncertainty of the CPP method. In fact, a fixed-nuclei
calculation with the existing CPP, showed that the change from a
virtual state to the bound states can be achieved by stretching the C-H bonds
in C$_2$H$_2$, C$_2$H$_4$, and C$_2$H$_6$ by 15--30\% \cite{NG05}.
In acetylene this change can also be induced by symmetric bending by about
$16$ degrees \cite{NG04}. However, this mechanism of bound-state formation
appears to be problematic, since the energy required to distort the molecule
far exceeds the positron binding energy and the thermal energy
of molecules at $\sim 300$~K.

The earliest calculation of positron annihilation on polyatomic molecules
is probably that of \textcite{JT83} for CH$_4$ and NH$_3$. They
described the interaction between the positron and the target by means of
a potential $V_s({\bf r})+V_p({\bf r})$, where $V_s({\bf r})$ is the full
electrostatic potential of the ground-state molecule, and $V_p({\bf r})$
is a CPP. This potential has the correct
form $V_p({\bf r})\simeq -\alpha _d/2r^4$ at large distances. Solving the
Schr\"odinger equation for the positron wave function
$\varphi _{\bf k}({\bf r})$ yields the scattering amplitude and cross section,
and $\varphi _{\bf k}({\bf r})$ is then used in Eq.~(\ref{eq:Zeff0}) to
calculate $\Z$. \textcite{JT83} showed that it is important to go
beyond this approximation and include the effect of distortion of the electron
density by the positron. This led to an increase of
$\Z$ values for CH$_4$ by a factor of two. Their final result for CH$_4$,
$\Z=99.5$ at $\eps =0.025$ eV, is reasonably close to, but still smaller than
the
experimental room-temperature value $\Z=142\pm 1$ \cite{WCC83}. They concluded
that the remaining discrepancy could be removed in a better positron-molecule
calculation. More significantly, they also concluded that the somewhat large,
measured $\Z$ value was not the result of the formation of a positron-molecule
complex. As discussed in Sec. \ref{subsec:dir}, this is in complete agreement
with the current understanding of the way in which positron-molecule virtual
states can enhance low-energy $\Z$ values.

An approach similar to that of \textcite{JT83} has now been tested for
a variety of organic and inorganic polyatomic molecules \cite{GMO01,OG03},
with ethane and benzene being the largest. The short-range part of the CPP
employed in these calculations is based on density-functional treatments
of the electron-positron correlations, while the long-range behavior is
described analytically (e.g., as $-\alpha _d/2r^4$, for the dipole polarization
term) \cite{JG91}. The positron wave function from these calculations is
used to calculate $\Z$ from Eq.~(\ref{eq:Zeff0}). More recent
calculations \cite{FG06}\footnote{This work also corrected a normalization
error that overestimated $\Z$ values for diatomics reported earlier
\cite{GM00}.} employed an enhancement factor in the integrand
of Eq.~(\ref{eq:Zeff}). It depends on the electron density and describes
its local increase at the positron \cite{Arp78,BN86}.

The main result of these fixed-nuclei calculations is that they fail to
reproduce large experimental  $\Z$ values ($>10^2$) for molecules such as
C$_2$H$_2$, C$_2$H$_4$, C$_2$H$_6$, and C$_6$H$_6$ \cite{OG03}.
This is an indirect confirmation of the role of resonant annihilation
involving nuclear vibrations in these molecules
(see Secs. \ref{subsec:forbid}, \ref{subsec:size} and \ref{subsec:other}).
For smaller polyatomics, such as H$_2$O and CH$_4$, the computed
room-temperature $\Z$ values, 167 and 65, respectively, are a factor of two
or three smaller than the experimental values \cite{GMO01}. Here an adjustment
in the CPP and/or the use of the enhancement factor in the calculation of $\Z$
could bring theory and experiment into agreement. Indeed, according to the
energy-resolved $\Z$ measurements, neither of these molecules shows clearly
discernible contributions of resonant annihilation (Sec.~\ref{subsec:nonres}).
Similarly, for diatomics, such as H$_2$, O$_2$, N$_2$, NO, and CO, the
calculation with enhancement factors gives $\Z$ values between 10 and
40~\cite{FG06}, which are within a factor of two of experimental values for
room-temperature positrons.

There are few annihilation calculations that include the dynamic interaction
between the positron and molecular vibrations. The VCC calculations of
\textcite{GM99} for CO$_2$, and \textcite{GM00} for O$_2$, N$_2$, NO, and CO,
showed that vibrational coupling has a relatively small effect on the
annihilation rates. This is to be expected, since small molecules such as
these most likely do not bind the positron (cf. Sec.~\ref{sec:bind}), and so the
VFR mechanism is ``switched off'' for them. The true magnitude of the effect
of vibrational coupling for these molecules remains somewhat uncertain,
as the reported theoretical $\Z$ values probably suffer from an uncertainty
related to the normalization error (see above). It is also surprising that the
$\Z$ values obtained in these calculations, using a static potential, change
little upon inclusion of the CPP, since practically all other calculations show
that correlations have a large effect on positron scattering and annihilation.

The zero-range-potential (ZRP) model calculations for Kr$_2$ (a
weakly bound van der Waals dimer) demonstrated a
number of key features of {\em resonant} annihilation \cite{Gri02,GL06b}.
First, they showed that, while
the potential representing each of the Kr atoms has no bound states,
the dimer is capable of binding a positron. Second, they
showed that positron binding has a relatively small effect on the vibrational
frequency of the complex. It changes by only about 10\% compared to the
frequency of Kr$_2$, in spite of the fact that the binding energy is
relatively large (i.e., two times the vibrational quantum). Finally, this
model showed the emergence of VFR in a dynamic
positron-molecule calculation.
The resonant contribution 
leads to a large increase in $\Z$ (e.g., for thermal positrons at 300~K,
from $\Z =250$ for direct annihilation to $\Z=700$--950, depending on the
details of the model). Unfortunately, van der Waals molecules
such as Kr$_2$ are difficult to study experimentally, and the ZRP method is
in general too crude to predict
the binding energies or $\Z$ spectra for molecules for which the annihilation
has been measured. That being said, this approach does provide an easily
solvable and instructive model for resonant positron-molecule
phenomena (see Sec.~\ref{subsubsec:bind}).

Recently a theory has been proposed \cite{SLV09}, that described
vibrationally enhanced annihilation using the Feshbach projection operator
formalism. It assumed that the positron is captured into a resonant electronic
state that determines the subsequent vibrational dynamics. The full
implications of this theory have yet to be elucidated. A similar mechanism
drives many electron-molecule attachment processes. However, in the case of
the positron, there is no experimental evidence to date of the required
resonant electronic states.

\subsubsection{Positron-molecule binding}\label{subsubsec:bind}

Calculation of positron binding has proven to be exceedingly challenging.
The electrostatic interaction between positrons and neutral atoms or
molecules (without large dipole moments) is dominated by the nuclear repulsion.
At large separations, the electric field of the positron gives rise
to the attractive $-\alpha _d/2r^4$ potential. At short range, there is an
additional attraction due virtual Ps formation \cite{DFG95,GL04}. Together
with polarization, these forces can overcome the static repulsion and thus
enable the formation of virtual levels or bound states.

In the case of atoms, reliable calculations for positron bound states have been
done for about ten species with one or two valence electrons (such as Li, Be,
Na, Mg, Cu, Zn, etc.). The work by \textcite{MBR02} is a good review of the
state of the
field a few years ago. The calculations were done using a variety of methods:
many-body theory and its combination with the configuration interaction (CI)
\cite{DFG95,DFG99}; stochastic variational method (SVM) \cite{RM97,RMV98};
and the CI method with core polarization potentials \cite{MR99,BM00}.

\begin{figure}[ht]
\includegraphics*[width=7.5cm]{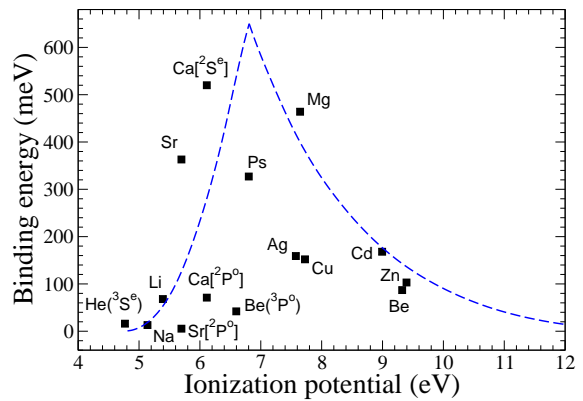}
\caption{Binding energies of positron-atom complexes as a function
of their ionization potential: squares, SVM and CI calculations
\cite{MBR02,BM07,BM10}; dashed curve, model ``alkali atom'' \cite{MBR99}.
For positron binding to metastable states the latter are indicated in
round brackets. The two bound states for Ca and Sr are shown in square
brackets.}
\label{fig:atomsIP}
\end{figure}

These calculations provide useful insights into the physics of
positron binding. One important parameter is the ionization potential of the
atomic system $E_i$ and its relation to the Ps binding energy $E_{\rm Ps}$.
For systems with $E_i>E_{\rm Ps}$, the electrons are relatively tightly
bound in the target. Since the positron is repelled by the atomic core, it
then forms a loosely bound state and stays outside the atom (i.e., represented
asymptotically as $A + e^+$). For $E_i<E_{\rm Ps}$ however, the positron
can attract a valence electron forming a ``Ps cluster'' \cite{RM98}. In this
case, the bound state is asymptotically a Ps atom orbiting the residual
positive ion ($A^+ + \,{\rm Ps}$).
Figure~\ref{fig:atomsIP} shows the calculated binding energies $\eb $ for atoms
as a function of $E_i$. A calculation for a model ``alkali atom'' \cite{MBR99}
shown by a dashed curve, suggests that $\eb $ peaks at $E_i=E_{\rm Ps}$,
and the calculations for real atoms generally support this picture.

In the ``alkali atom'' model, a positron and a single valence electron move
in the field of a fixed atomic core. As the core potential is varied, both
the ionization potential and the dipole polarizability change in such a
way that $\alpha _d\propto E_i^{-2}$ \cite{MBR99}. In real systems,
and in particular, in molecules, $\alpha _d$  and $E_i$ can be regarded more
akin to
independent parameters. As shown later (see Sec.~\ref{sec:bind}),
both are found experimentally to influence the observed binding energy.
Theoretical studies of atoms and experimental studies of molecules indicate
that their maximum binding energies are comparable, $\eb \lesssim 0.5$~eV,
while the values for the strongly polar alkali hydrides can be as large
as 1~eV (see below).

A major difference between positron binding to atoms and to molecules is that
molecules can have permanent dipole moments. Theoretically, a static molecule
with dipole moment $\mu > \mu _{\rm cr}=1.625~\mbox{D}~=0.639$~a.u., possesses
an infinite number of positronic, as well as electronic bound states
\cite{Cra67}.
For a molecule that is free to rotate, the critical dipole moment for binding
is greater than $\mu _{\rm cr}$. This critical value increases as
the molecular moment of inertia decreases or the angular momentum of the
molecule increases \cite{Gar71}.

At present, positron-molecule binding has been predicted theoretically for
a few strongly polar molecules
\cite{Str96,Str01,Str04,Str99,DT88,CS06,BLM05,BLT06,GFB06,TBK03,PTB05,%
SW76,SM01,MBM01,BMM98,TSB01,KJ81,ADB08}. These calculations employed
a variety of methods such as Hartree-Fock (HF), CI and diffusion Monte Carlo
(DMC), as well as explicitly correlated Gaussians (ECG).
Positron bound states have also been found using the CPP method for the
nonpolar cage-like molecule C$_{20}$, which has a large $\alpha _d=25.4$~\AA$^3$
\cite{CLG08}.
A representative selection of recent results is given in
Table~\ref{tab:calc_BE}. For simple molecules, such as LiH,
the results obtained by different methods are generally in good agreement.
The binding energies obtained at the static HF level increase considerably
when correlations are included [e.g., see $\eb $ values for HCN, urea
(NH$_2$)$_2$CO, and acetone (CH$_3$)$_2$CO]. In fact, the current CI values
for the larger polyatomics may still considerably underestimate the true
$\eb $ values due to incomplete CI expansions.

\begingroup
\begin{table}[ht]
\squeezetable
\caption{Calculated positron-molecule binding energies.}
\label{tab:calc_BE}
\begin{ruledtabular}
\begin{tabular}{lcccl}
Molecule & $\mu $\footnotemark[1] & $\eb $ & Method & Reference \\
& (D) & (meV) & & \\
\hline
LiH & 5.88 & 1000 & DMC & \textcite{MMB00}\\
 &  & 909 & FCSVM\footnotemark[2] & MR\footnotemark[3] \\
 &  & 1005\footnotemark[4] & ECG & \textcite{Str01} \\
 &  & 1043\footnotemark[5] & ECG & BA\footnotemark[3] \\
 &  & 1626 & NEO\footnotemark[2] & \textcite{ADB08} \\
LiF & 6.33 & 599 & DMC & \textcite{MBM01} \\
BeO & 6.26 & 680 & DMC & \textcite{MBM01} \\
 &  & 375 & MRD-CI\footnotemark[2] & \textcite{BLP07} \\
NaH & 6.40 & 1031 & MRD-CI & \textcite{GFB06} \\
RbH & 9.03 & 1639 & MRD-CI & \textcite{GFB06} \\
MgO & 6.42 & 472 & MRD-CI & BL\footnotemark[3] \\
LiO & 6.84 & 304 & MRD-CI & BL\footnotemark[3] \\
CH$_2$O 
 & 2.33 & 19 & CI & \textcite{Str04} \\ 
HCN & 2.98 & 2 & HF & CS\footnotemark[3] \\
 &  & 35 & CI & CS\footnotemark[3] \\
 &  & 38 & DMC & \textcite{KMT09} \\
Urea & 3.99 & 6 & HF & \textcite{TBK03} \\
& & 13 & CI & \textcite{TBK03} \\
Acetone & 2.88 & 1 & HF & \textcite{TBK03}\\
& & 4 & CI & \textcite{TBK03}\\
C$_{20}$ & $-$ & 780\footnotemark[6] & CPP & \textcite{CLG08} \\
&  & 230--250\footnotemark[6] & CPP & \textcite{CLG08}
\end{tabular}
\footnotetext[1]{Dipole moments from \textcite{CRC,BLM05,GNB97} or as
cited.}
\footnotetext[2]{FCSVM, fixed-core SVM; NEO, nuclear-electronic orbital
method; MRD-CI, multireference single- and double-excitation CI.}
\footnotetext[3]{\textcite{MR00} (MR); \textcite{BA04} (BA); \textcite{BL08}
(BL); \textcite{CS06} (CS).}
\footnotetext[4]{Adiabatic positron affinity.}
\footnotetext[5]{Non-Born-Oppenheimer variational calculation.}
\footnotetext[6]{The $C_i$ isomer of C$_{20}$ is predicted to have a
deeply bound $s$-type state and three weakly bound $p$-type states.}
\end{ruledtabular}
\end{table}
\endgroup

\begin{figure}[ht]
\includegraphics*[width=7cm]{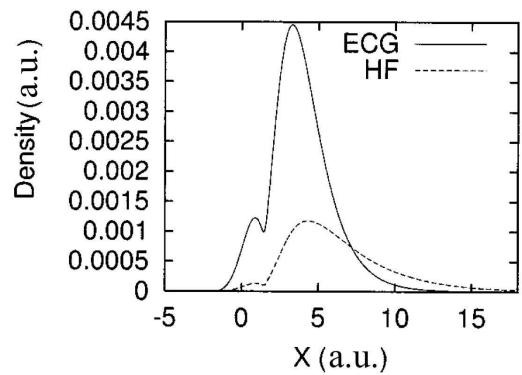}
\caption{Positron density in $e^+$LiH along the
molecular axis: dashed curve, HF ($\eb =0.16$~eV); solid curve,
ECG ($\eb =0.94$~eV); Li atom is at $X=-1.51$~a.u., and H at $X=1.51$~a.u.
From \textcite{Str99}.}
\label{fig:LiH}
\end{figure}

The positron density in bound states with polar molecules is asymmetric,
with a strong pile-up outside the negatively-charged end of the
molecule \cite{Str99,Str01,BLP07}. This is shown by
Fig.~\ref{fig:LiH} for LiH. This figure also shows that correlations
(e.g., included through ECG) produce a large increase in the positron
density at the molecule, compared to the static HF calculation.
This increase in the density follows the increase
in $\eb$, as described by Eq.~(\ref{eq:rhoep}) for
bound states with nonpolar species.

Both the electron and positron densities change little with vibrational
excitation of the molecule \cite{GFB06}. In the alkali hydrides, positron
binding noticeably increases the
bond lengths and softens the vibrational modes. For example, the energies of
the first vibrational excitations in LiH and $e^+$LiH are 168 and
109~meV, respectively \cite{MMB00,GFB06}. This is related to the fact that
the structure of this molecule is closer to Li$^+$HPs than to a loosely bound
positron orbiting the neutral molecule. In contrast, in BeO and MgO,
positron binding changes the bond lengths and the vibrational frequencies
by less than 1\% \cite{BL08,BLP07}.

At present there are no {\em ab initio}
calculations of positron binding to alkanes or other nonpolar or weakly
polar molecules (except for C$_{20}$).
Positron binding in such systems is exclusively due to electron-positron
correlation effects. To obtain a bound state, the calculation must include
them accurately, since binding does not exist at the static (e.g., HF) level.
As an alternative, and given the availability of experimental data
(Secs. \ref{sec:large} and \ref{sec:bind}), \textcite{GL06b,GL09}
explored positron-molecule binding to alkanes using ZRP. The ZRP is the
simplest form of a model potential, suited to studying low-energy
processes \cite{DO88}. The idea of the model potential approach is to fit the
potential to experimental data (e.g., the binding energy for a given
molecule) and then use it to study binding for a range of similar molecules.

In the ZRP method, the bound-state wave function of the positron in the
field of $N$ centers placed at ${\bf R}_i$ has the form \cite{DO88},
\begin{equation}\label{eq:bound}
\Psi ({\bf r})= \sum_{i=1}^N A_i \frac{e^{-\kappa |{\bf r}-{\bf R}_i|}}
{|{\bf r}-{\bf R}_i|},
\end{equation}
where $\kappa >0$ is related to the bound-state energy by
$\eps _0=-\kappa ^2/2$. The interaction with each center is parameterized by
$\kappa _{0i}$ through the boundary condition,
\begin{equation}\label{BCmole}
\Psi\mid _{{\bf r}{\rightarrow}{\bf R}_i} \simeq \, {\rm const}
\times (|{\bf r}-{\bf R}_i|^{-1}- \kappa_{0i}).
\end{equation}
Subjecting $\Psi $ from Eq.~(\ref{eq:bound}) to $N$ conditions (\ref{BCmole})
yields a set of linear homogeneous equations for $A_i$, whose solvability
determines the allowed values of $\kappa $.

\begin{figure}[ht]
\includegraphics*[width=7cm]{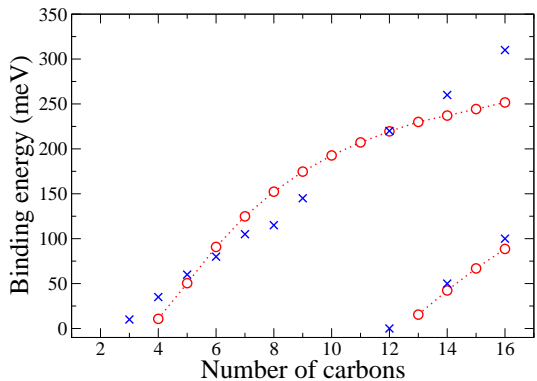}
\caption{Positron binding energies for alkanes from experiment
(crosses) (Table \ref{tab:Zeff_N}) and ZRP model calculation (circles). The
ZRP model is fit to $\eps _b=220$~meV for dodecane.}
\label{fig:bind}
\end{figure}

The alkanes, C$_n$H$_{2n+2}$, were modeled by a planar zig-zag chain of $n$
ZRPs, each representing the CH$_3$ or CH$_2$ group. The distance between the
neighboring ZRPs is given by the length of the C--C bond 2.91~a.u.,
and the angle between adjacent bonds is equal to 113$^\circ $. The parameter
$\kappa _{0i}=-0.69$~a.u. was chosen to reproduce
the binding energy for dodecane ($n=12$, $\eps _0=-220$~meV) \cite{GL09}.
Figure~\ref{fig:bind} compares the results of this calculation with the
measured binding energies for alkanes up to $n=16$ (cf. Table \ref{tab:Zeff_N}).

Figure \ref{fig:bind} shows that the model gives a good overall description
of positron binding to alkanes, including the prediction of the second bound
state. However, it fails to capture in a quantitative way some of the details.
The model predicts binding for
$n\geq 4$, whereas ethane ($n=2$) is observed to bind positrons. The
model also predicts that a second bound state emerges for $n=13$, while
experimentally this state is observed at $n=12$ \cite{BYS06,YS08a}
(cf. Fig.~\ref{fig:alk_14_16}).

\begin{figure}[ht]
\includegraphics*[width=4.2cm]{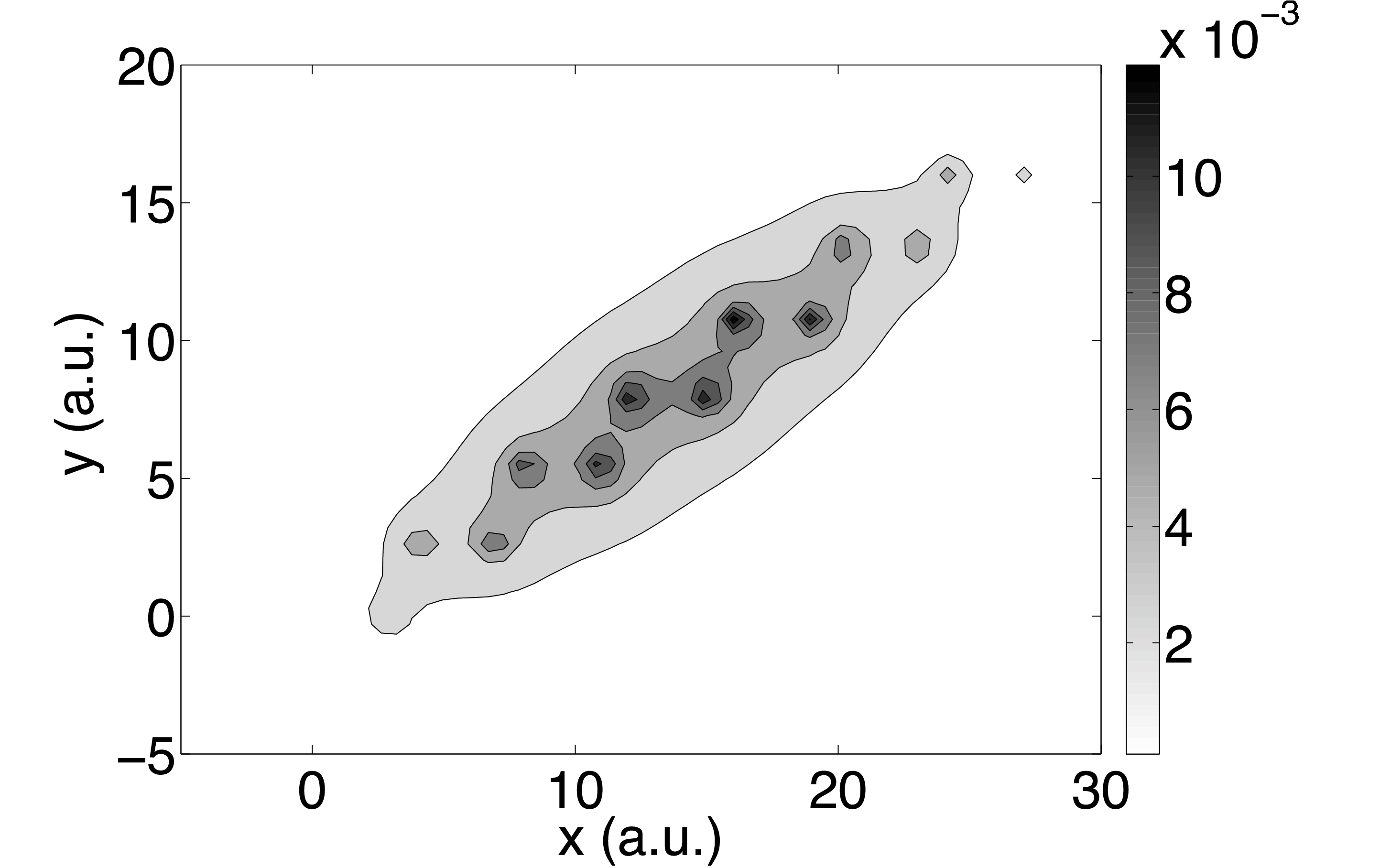}
\includegraphics*[width=4.2cm,]{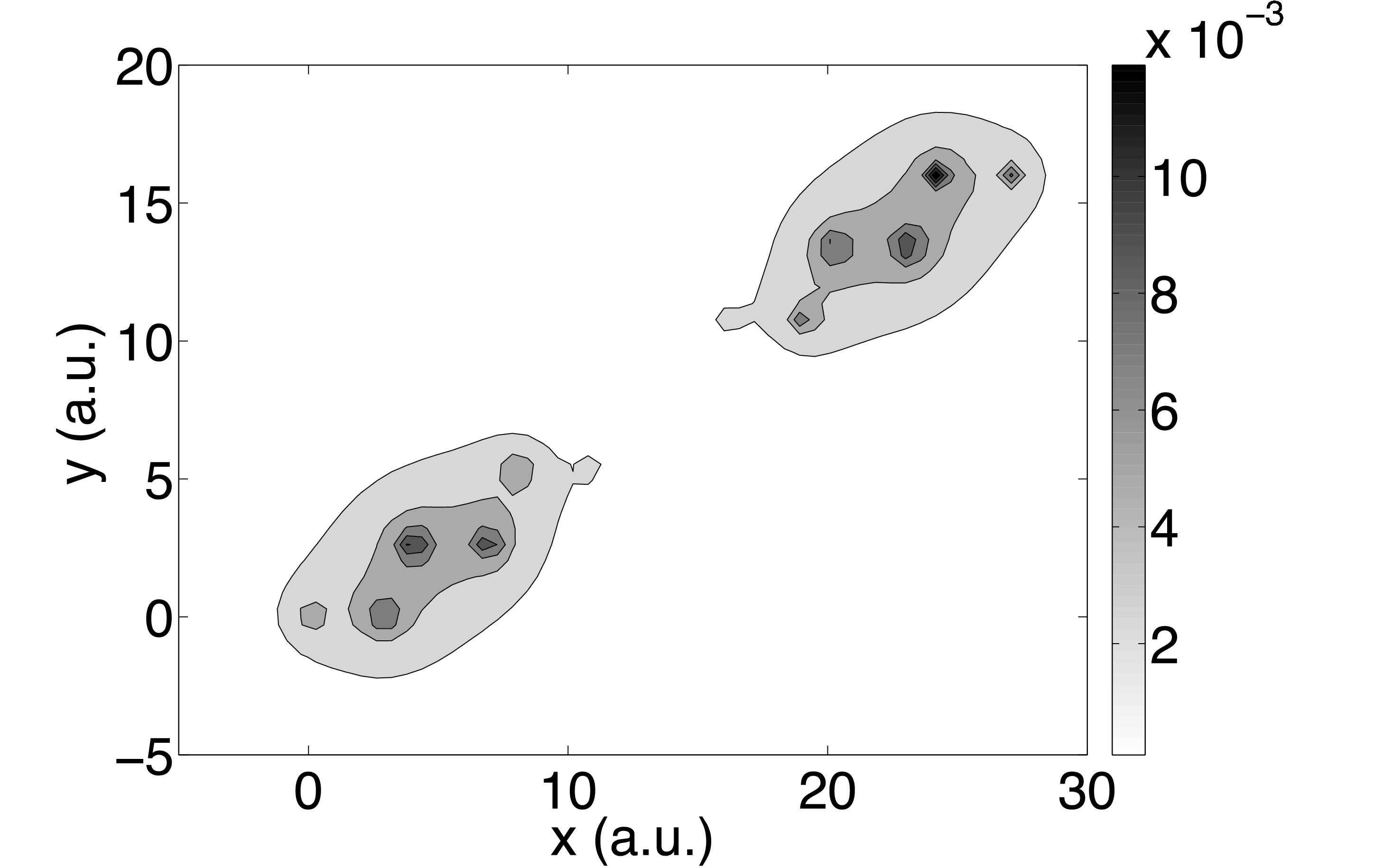}
\caption{Two-dimensional density of the positron wave functions in
the ZRP model for the 1st (left) and 2nd (right) bound states
in tetradecane ($n=14$). In these plots, the first carbon atom is at the origin,
and the C--C bonds are alternately parallel and at 67$^\circ $ degrees to
the $x$ axis.}
\label{fig:c14}
\end{figure}

To visualize the bound states, the two-dimensional density
\begin{equation}\label{2dDensity}
\rho(x,y)=\int_{-\infty}^{+\infty} |\Psi(x,y,z)|^2dz,
\end{equation}
where $x$ and $y$ are in the plane of the carbon chain, is shown in
Fig.~\ref{fig:c14} for the first and second bound states of
tetradecane ($n=14$). Both states are quite diffuse, with the positron
spread over the whole molecule. The wave function of the second bound state
must be orthogonal to the ground state, and so it changes sign on a nodal
surface close to the center of the molecule. On the density plot
(Fig.~\ref{fig:c14}, right), this corresponds to an area of low density near
the midpoint.
The actual positron wave functions are expected to differ from that
given by Eq.~(\ref{eq:bound}) in that the latter does not exclude the
positron from the regions inside the atomic cores (which are of ``zero range''
in the model). However, the atomic cores are relatively small compared
to the extent of the positron wave function, in keeping with the main
assumption of the ZRP model. Thus the model captures the main
features of positron-molecule bound states.

\section{Experimental tools and procedures}\label{sec:exp}

In this section, experimental techniques are described that have been 
used to investigate low-energy positron-molecule annihilation and 
related effects. The methodology has varied considerably 
over the past half century of these studies, and no attempt is made to be 
complete. Emphasis is placed upon typical and/or best practices.
The reader is referred to the original papers for further details. 

\subsection{Annihilation-rate measurements with thermalized positrons in 
atmospheric pressure gases}\label{subsec:exp_therm}

Positrons from conventional sources such as radioisotopes or electron 
accelerators typically have energies ranging from $\sim 1$~keV to 
$\sim 0.5$~MeV. Thus, to study positron interactions at low energies, 
some method must be used to slow the positrons. Early annihilation rate
measurements were done
using the test species themselves as ``moderators'' and 
measuring the spectra of time delays between the positron production
and annihilation signals \cite{SD49,Deu51a,Deu51b,Osm65b,HCG82,WCC83,GH78}.
In a typical 
experiment, the 1.28~MeV gamma ray that accompanies  positron emission
from a $^{22}$Na positron source provides a start signal, and the detection
of a single, 511~keV gamma ray from a two-gamma annihilation event is used
as a stop signal. Achieving an adequate signal-to-noise ratio requires
working at gas densities $\geq 0.1$ amagat. 

Annihilation rates are obtained from the dependence of the exponential time
decay of the 511 keV annihilation signal (i.e., after the positrons come to
thermal equilibrium) on the test-gas density. Using fast 
electronics, it was also possible to measure the slowing down of the 
fast positrons \cite{SM78}. Where necessary, the rate of thermalization was 
increased by adding a light species with a small $\Z$ value
such as molecular hydrogen \cite{WCG85}, and the long-time 
ortho-positronium component was quenched using a small 
admixture of a gas with unpaired electron spins such as NO or O$_2$
\cite{Deu51a,Deu51b}.

\subsection{Buffer-gas positron traps as tailored sources of positrons}
\label{subsec:traps}

Buffer-gas (BG) traps \cite{SPL88,MS92,SGG99} proved to be a useful tool
to tailor positron gases, plasmas and beams for positron annihilation
studies \cite{SG04}.
Positrons from a sealed $^{22}$Na radioactive source are slowed to
electron-volt energies using a solid-neon reflection moderator
[efficiency 1--2\% \cite{GS96}]. A 50 mCi $^{22}$Na source and neon moderator
produces $\sim 5\mathrm{-}10\times 10^6$ slow positrons per second.
The slow positrons are then guided magnetically into a buffer-gas,
Penning-Malmberg trap \cite{MS92}. It consists of a uniform
magnetic field ($B \sim 0.1$~T) coaxial with a set of cylindrical electrodes
biased to form a stepped, three-stage, potential 
well. The stages contain a nitrogen buffer gas with successively lower 
pressures. Positrons become trapped by losing energy through electronic
excitations of the N$_2$ molecules. They then cool to the ambient (i.e., room)
temperature in the third stage by additional collisions with the N$_2$. More
recently, a small amount of CF$_4$ was added to increase the cooling rate
\cite{GS00,SGM02a}.
The result is a magnetized, thermal gas (or plasma) of as many as $10^8$
positrons at a temperature of 300 K. The trapped positrons are in a background
gas pressure $\leq 10^{-6}$ torr, which can then be pumped out, depending upon
the experiment.

\subsection{Annihilation-rate measurements in positron traps}
\label{subsec:anntrap}

Positron traps have enabled an 
improved method to study the interaction of thermal positrons with 
a large variety of test species \cite{IGM95,SPL88,MS91,IGG00}. In this case,
the positrons are trapped, the buffer gas is pumped out, and a test gas
species is introduced into the trap at a low pressure. Low pressures
ensure that the annihilation events are due to two-body positron-molecule
interactions and thus that three-body processes are negligible. This technique
also permitted study of low vapor pressure targets. Furthermore, the
positron temperature (e.g., 300~K) could be measured to verify that they are
in thermal equilibrium with the test species. This was done using standard
plasma techniques, dumping the positrons and measuring their energy
distribution using a retarding potential analyzer \cite{MS92}.

\begin{figure}[ht]
\includegraphics*[width=8cm]{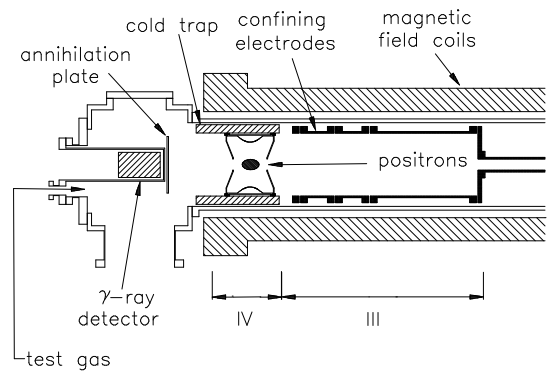}
\caption{Schematic diagram of the apparatus used to study the 
interaction of trapped, thermal positrons with low-pressure gases. The 
positrons are contained in a fourth trapping stage surrounded by a cold 
surface to minimize the effects of impurities present in the vacuum 
system. From \textcite{IGM95}.}
\label{fig:old_trap}
\end{figure}

The apparatus for these thermal annihilation-rate measurements
is shown in Fig. \ref{fig:old_trap} \cite{IGM95}. In this case, the
trapped positrons were shuttled to a separate
confinement stage surrounded by a cryogenically cooled surface to 
reduce the level of impurities in the vacuum system. It operated with
liquid nitrogen (77 K) or an ethanol-water mixture ($-7\,^\circ $C)
depending upon the test species.
The annihilation was monitored by holding the positrons for a given time,
then dumping those that had not annihilated
onto a collector plate and measuring the gamma ray
signal using a NaI(Tl) scintillator and a photomultiplier. 
Experiments were also conducted in which the thermal positrons were heated by
short bursts of radio-frequency noise to provide a measure of the annihilation
rate as a function of positron temperature \cite{IGG00}.

For test species that are gases at 300~K, the pressure is controlled to
micro-torr precision
by feedback using a capacitance manometer and a piezoelectric valve.
A liquid delivery system, while offering less control, permitted studies 
of a broader range of test species. It can be used with species that have 
relatively low boiling points (i.e., close to room temperature) and 
pulverized solid samples, such as naphthalene \cite{You07}. 
For liquid samples, a freeze-pump-thaw procedure is used to eliminate
volatile contaminants. The sample is 
placed in a temperature-regulated bath, and a needle valve is used 
to leak vapor into the annihilation cell. While there was no 
feedback control of the pressure, it was still reasonably 
stable \cite{IGM95}. 

Some vapors must be run at pressures below the sensitivity of the 
capacitance manometer to avoid detector saturation. 
In this case, the pressure of the test species is measured with an ion gauge
calibrated against the manometer at higher pressures. A few species studied
(e.g., naphthalene) have such low vapor pressures that accurate pressure
calibration was not possible \cite{You07,YS08a}.

The major uncertainty in these $\Z$ measurements is estimated to be a
$\pm 20$\% systematic error in the measurement of test-gas pressures.

\subsection{Trap-based cold positron beams}\label{subsec:beam}

To obtain detailed information about positron interactions with matter, 
the tool of choice is a beam of positrons with a well defined and 
tunable energy. In early experiments, the beam energy resolution was
limited by the energy spread of positrons emerging from a 
moderator (e.g., a fraction of an electron volt).
While the use of a cold primary moderator \cite{BCM86}
or remoderator \cite{GM87} could reduce the energy spread to 30--40~meV,
these techniques were rarely, if ever used in atomic physics experiments.
The advent of the BG positron accumulator enabled an efficient, pulsed
and tunable, low-energy positron beam with a comparable, and potentially
smaller energy spread \cite{GKG97,Gil00,KGG98}.

In this technique, $\sim \!10^4$ positrons are
accumulated in a time $\sim 0.1$~s and cooled for a similar
time \cite{GBS02,SBM04}. Then the exit-gate electrode is lowered to a
potential $V_E$, which sets the beam energy. The bottom of
confining potential well is then raised
(i.e., to $\sim 0.25$~V above $V_E$) in a time $\sim 5~\mu $s to produce
a positron pulse with a similar time duration. The beam-transport energy
(typically $\sim 2$ eV) is set below the threshold for positronium formation
to avoid positron loss and extraneous gamma-ray signals. All this is done
in the BG-trap magnetic field (i.e., $B\sim 0.15$~T), producing
pulses of $1\mathrm{-}3\times 10^4$ positrons at a few Hz rate.
The pulses are magnetically guided to the annihilation cell in
fields ranging from 0.03 to 0.1~T.

The beam energy distribution is measured using 
the cylindrical electrode of the gas cell as a retarding potential analyzer 
(RPA) \cite{You07}. The mean
energy of the beam in the annihilation cell can be verified independently 
by measuring the time-of-flight delay of the positrons passing through the 
cell as a function of increasing cell potential $V_C$ \cite{Mar05,SGM02a}. 
Typical uncertainties in the mean beam energy are $\sim $10~meV.

The parallel energy
distribution can be modeled by a single Gaussian \cite{You07}. At a more
accurate level, there is typically a high-energy tail containing $\sim 10$\%
of the beam particles, depending upon the beam-formation 
protocol and the relative magnitudes of the magnetic field in the buffer-gas
trap and the measurement region.\footnote{The ratio $\eps _\perp /B$ and the
total positron energy $\eps = \eps _\parallel + \eps _\perp $ are both
conserved in this magnetic beam-transport system, where $\eps _\parallel $ is
the parallel beam energy, and $\eps _\perp $ is the transverse
energy of the particles' gyromotion in the plane perpendicular to the
magnetic field $B$ \cite{BGS03,SGM02a}.} Parallel energy spreads as small as
18 meV (FWHM) can be achieved using this technique \cite{GKG97}, when the
beam is kept in the same strength magnetic field as the buffer-gas positron
accumulator. Typical values of parallel energy spread in the annihilation
experiments described here are somewhat larger, $25\mathrm{-}30$~meV,
due to the fact that the annihilation cell was at a lower magnetic field
than that of the trap \cite{BGS03}.
The spread in transverse energies in the BG trap is set by the ambient
temperature $T_\perp = 25$~meV (i.e., corresponding to an electrode temperature
of $\sim 300$ K).

The positron energy distribution in the beam is modelled as
\begin{equation}\label{eq:f}
f(\eps _\parallel , \eps _\perp )=\frac{1}{k_BT_\perp \sqrt{2\pi \sigma^2}}
\exp \left[ -\frac{\eps _\perp }{k_BT_\perp}
-\frac{(\eps _\parallel -\bar \eps )^2}{2\sigma^2}\right],
\end{equation}
where  $\eps _\parallel $ and $\eps _\perp $ and the parallel and transverse
positron energies, $\sigma $ is the root-mean-squared width of the parallel
energy distribution (i.e., corresponding to a FWHM of $\sigma\sqrt{8\ln 2}$),
and $\bar \eps $ is the mean parallel energy of the positron beam.

This distribution has been
verified using the resonant annihilation peak for the C-H stretch modes in
propane (cf. Sec. \ref{sec:large}). The line width of the VFR peak is assumed
to be negligible (i.e., $\lesssim 1$ meV; cf. Sec. \ref{sec:theory}),
as is the spread of the propane C-H stretch frequencies. 
The distribution in Eq.~(\ref{eq:f}) is convolved with a delta
function, yielding the distribution of (total) positron energies of the
beam \cite{GL06a},
\begin{eqnarray}
f_{\rm b}(\eps -\bar \eps )&=&\int f(\eps _\parallel ,\eps _\perp )
\delta (\eps _\parallel +\eps _\perp -\eps )d\eps _\parallel d\eps _\perp 
\label{eq:posen1} \\
&=&\frac{1}{2k_BT_\perp }
\exp \left[ \frac{\sigma ^2}{2(k_BT_\perp )^2}\right]
\exp \left(-\frac{\eps -\bar \eps }{k_BT_\perp }\right)\nonumber \\
&\times & \left\{ 1+\Phi \left[ \frac{1}{\sqrt{2}}
\left(\frac{\eps -\bar \eps }{\sigma }-\frac{\sigma}{k_BT_\perp }\right)
\right]\right\},\label{eq:posen}
\end{eqnarray}
where $\Phi (x)$ is the error function. Note that this
energy distribution depends on the difference between the total energy
$\eps $ and mean parallel energy $\bar \eps $.

\begin{figure}[ht]
\includegraphics*[width=7cm]{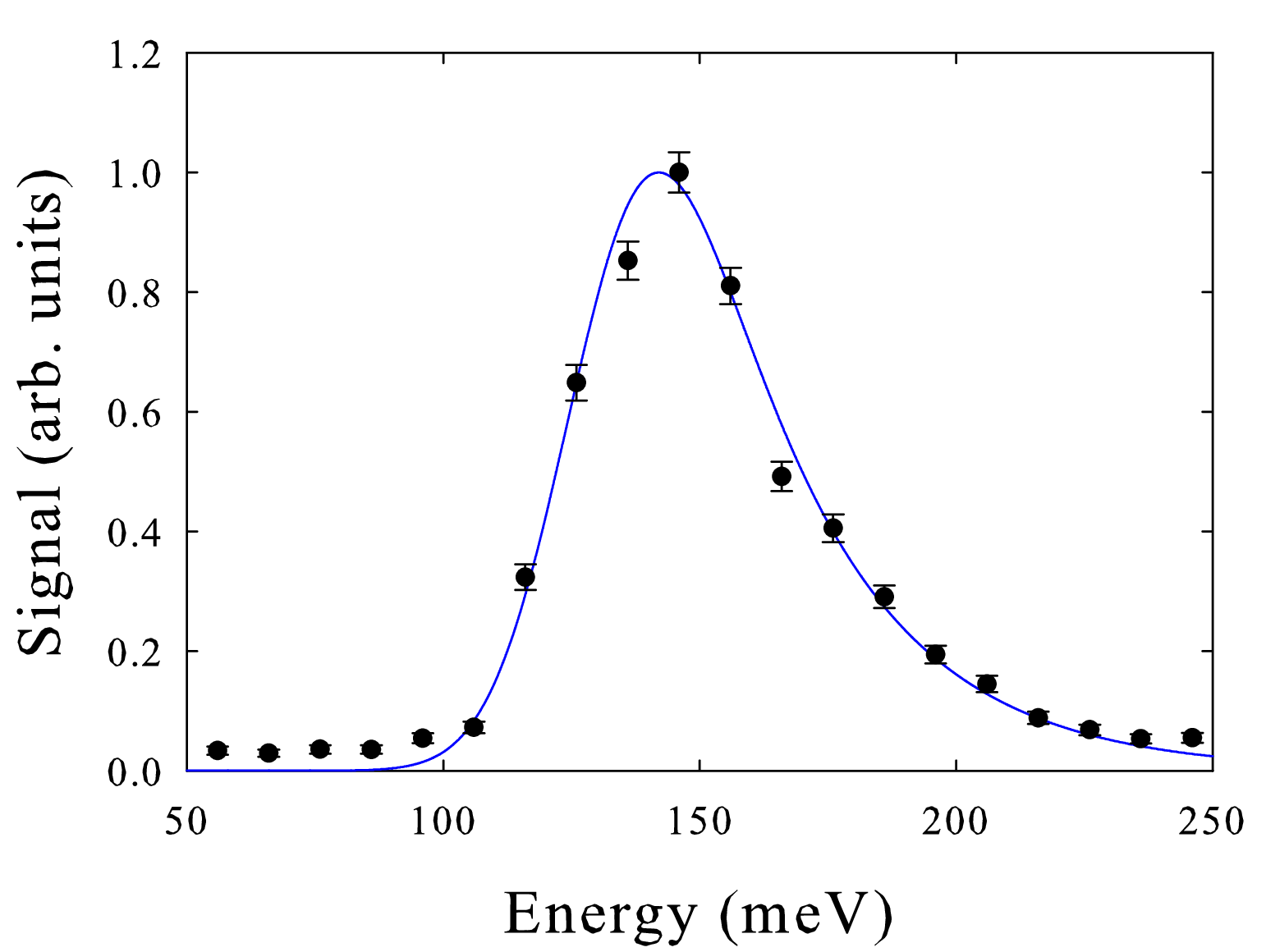}
\caption{Measurement of the distribution of total positron energies in the
trap-based beam using a
vibrational-Feshbach annihilation resonance: solid curve, prediction
of Eq.~(\ref{eq:posen}), normalized arbitrarily, shifted in
energy, and fitted to the energy-reversed, normalized, C-H stretch peak in
$\Z$ for propane ($\bullet$), with $T_\perp =26$~meV and a parallel energy 
spread (FWHM) of $27$~meV. From \textcite{You07}.}
\label{fig:energy_dist}
\end{figure}

The resulting fit, shown in Fig.~\ref{fig:energy_dist}
is in excellent agreement with the data. Due to the spread in perpendicular
energies, the position of the peak observed as a function of the
mean parallel energy (which is set and measured by varying $V_C$)
is about 12~meV below the true energy of the peak (i.e., as a 
function of the {\em total} positron energy). The annihilation spectra
in this review are presented as a function of the total positron energy,
which is taken to be 12~meV higher than the mean parallel energy set in the
experiment.

\subsection{Energy-resolved annihilation measurements}\label{subsec:en_res_ann}

\begin{figure}[ht]
\includegraphics*[width=8.5cm]{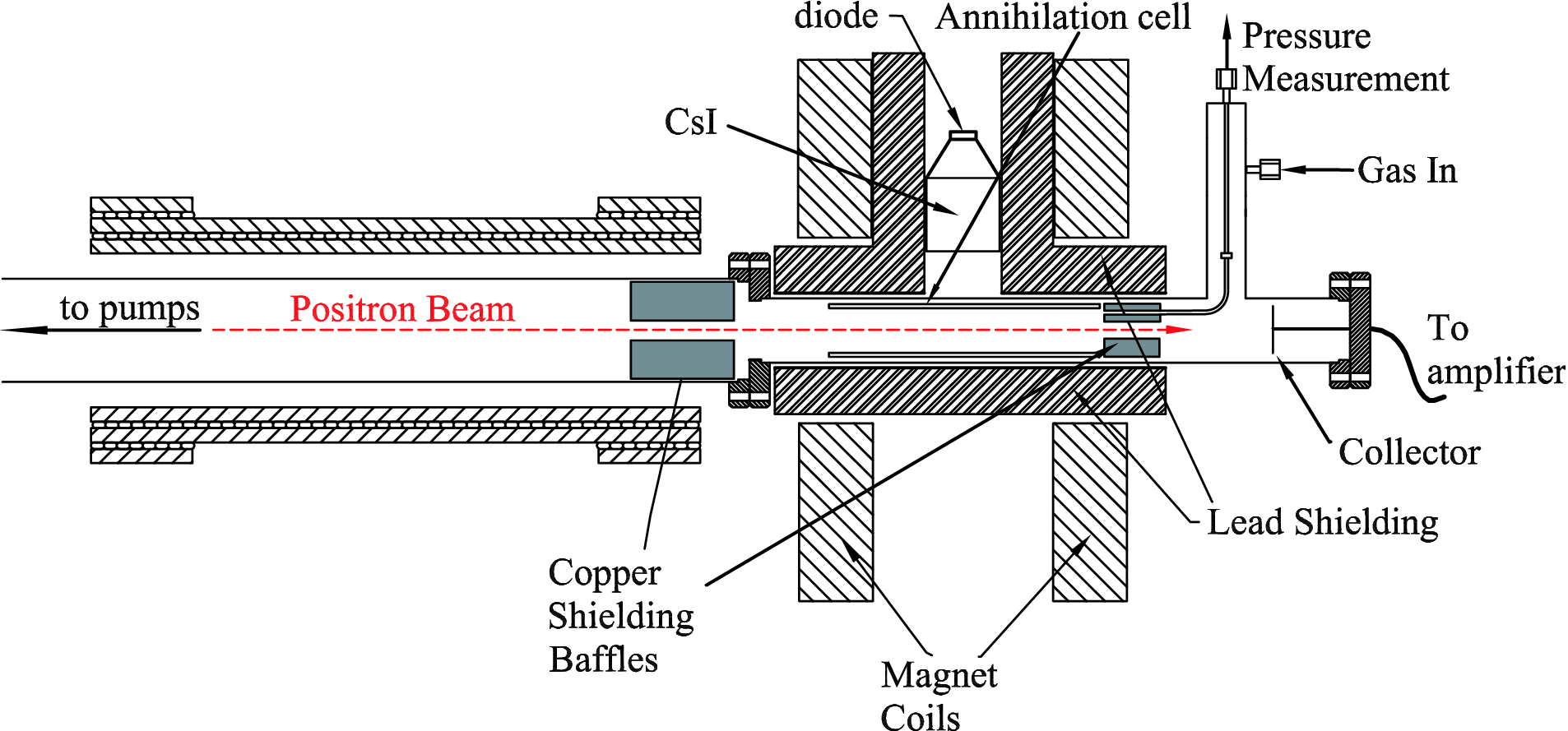}
\caption{Schematic diagram of the gas cell, shielding, and detection 
apparatus used for energy-resolved positron annihilation 
measurements (not to scale). From \textcite{You07}.}
\label{fig:ann_cell}
\end{figure}

Central to this review are measurements of positron annihilation on
molecules, resolved as a function of incident
positron energy \cite{BYS06,GBS02,BGS03,YS08a}. The annihilation cell
is shown schematically in Fig.~\ref{fig:ann_cell}. It consists of
a cylindrical, gold-plated electrode 4.4~cm in diameter and 17~cm long.
The gamma-ray detector and associated shielding restrict
the detector field-of-view to a region $\leq 15$~cm in length along
the axis of the cell. Magnet coils outside the cell impose a field of
$\sim 0.075$--0.095~T, with the lowest value in region viewed by the
detector. Metal baffles shield
the gas cell and the detector from spurious gamma-ray decays. The system for
handling gases and vapors is described in Sec.~\ref{subsec:anntrap} above.

Single gamma rays are detected 
using a CsI crystal and a photodiode, followed by a single-channel 
analyzer centered on the 511 keV annihilation gamma-ray line. The 
absolute detector efficiency and the sensitivity along the path of the 
positron beam are measured using a calibrated gamma-ray test 
source \cite{GBS02}.

Pulses of positrons pass through the gas cell 
several times while annihilation events are recorded, with the total 
scattering kept below 15\%. A typical time window for this 
measurement is $\sim 15~\mu\mathrm{s}$ \cite{YS08a}.
Positrons are kept in flight while the annihilation events are recorded to
avoid spurious gamma-ray 
background signals. To avoid detector saturation, the average signal level
is adjusted to $\sim 1$ count per 10 positron pulses. Typical 
test-gas pressures range from 0.1 to 100 $\mu $torr, due to the large
variations in annihilation rate for different chemical species. 
Background signals can be as low as one count 
per $10^9$ positrons cycled through the annihilation cell.

A typical spectrum consists of $\sim 10$--25 pulse-train measurements at 
each energy, taken at 10--15~meV intervals over the relevant range of 
positron energies (e.g., $\leq 500$~meV). This measurement is then 
repeated a few hundred times. 
Complications due to scattering restrict measurements to energies
$\geq 50$~meV. Absolute 
values of $\Z$ are obtained from measurement of the detector 
efficiency, the pulse strength, the detector-sensitivity-averaged path 
length, and the test-gas pressure.

Uncertainties in these parameters are 
estimated to result in a 20\% overall systematic uncertainty in the absolute 
magnitude of $\Z$. The error bars shown in this review indicate
the statistical uncertainty due to the finite number of counts at a given
energy. In many cases, these statistical errors are smaller than the size
of the data points. The linearity of the signal with test-gas pressure is 
checked to ensure that scattering and three-body effects are negligible.

A separate annihilation cell was used for studies of annihilation as a
function of the temperature of the test species \cite{YS08c}. It is capable
of cooling target gases down to 100~K. To ensure that the 
test species are actually thermalized, the annihilation cell was operated
at temperatures below the sticking temperature of the
test species on the cell walls.

\subsection{Gamma-ray spectral measurements}\label{subsec:gam_meas}

The center-of-mass momentum of the annihilating electron-positron pair
contains important information about the momentum distribution of the
bound electrons in an atomic or molecular target. It produces both Doppler
shifts of the photon energies and changes in their directions (see
Sec. \ref{subsec:anspec}). Using such measurements, one can 
distinguish, for example, annihilation on different atoms, or 
annihilation on valence and inner-shell electrons in a 
given species \cite{IGG97}.

In a two-gamma event, the photon momentum distribution can be determined 
by measuring either the angular deviation of the gamma rays
[the so-called angular correlation of annihilation radiation 
(ACAR) technique \cite{CRJ94}], or the Doppler broadening of the 
annihilation gamma-ray line. For the tenuous samples studied here,
Doppler broadening is the method of choice. 

The apparatus for these measurements is shown in Fig.~\ref{fig:old_trap} 
\cite{IGS97}. The gamma rays are
detected using a high-purity Ge detector with an energy resolution of 
1.16 keV (FWHM). Positrons are trapped and cooled, then the N$_2$ trapping
gas is pumped out, and the test gas is introduced. Gamma-ray data are 
recorded, and then this cycle is repeated. The errors in the measurement of
gamma-ray energies are estimated to be $\pm 0.02$~keV. The errors in
spectral intensity are predominantly statistical due to the finite number
of counts \cite{IGS97}.

\subsection{Annihilation-induced fragmentation}\label{subsec:ann_frag}

Positron annihilation on atomic and molecular species produces a 
spectrum of positive ions that can be measured using 
time-of-flight techniques. The first experiment of this kind arranged for 
positrons in a Penning trap to interact with molecular species for a 
short time. Then the resulting ions were dumped and the mass spectrum measured
using a time-of-flight technique \cite{PSL89,GGM94}.
Subsequently, more detailed studies were conducted using 
improved techniques to measure the ion mass spectra
\cite{MSL00,HXM96a,XHL95,XHL94,XHL93,HDX93,DHE90}. In this case, 
positrons from an electron LINAC were moderated 
and then accumulated in a Penning trap, where they were allowed 
to interact with the test species. The product ions were 
accelerated and detected using a microchannel plate. Significantly 
higher mass resolution was achieved using a spatially varying (quadratic)
potential to arrange the same arrival time for ions of the same mass 
starting at different initial positions in the trap.

\section{Annihilation on small molecules}\label{sec:small}

Research on positron annihilation on atomic and molecular targets has
typically been focused in two areas: atomic and small molecular species
where modest annihilation rates are observed, and large hydrocarbons
that are characterized by very large annihilation rates.
Some small polyatomic molecules 
occupy a middle ground \cite{GL06a}. Here annihilation proceeds in a
different manner than in other small targets due to the fact that these
molecules can temporarily bind a positron via VFRs.
Further, in contrast to larger hydrocarbons, the theoretical description
of annihilation in these targets is relatively simple. The focus of
this section is VFR-mediated annihilation in small molecules and the
contrasting case of annihilation in small molecules that do not bind positrons.

Energy-resolved $\Z$ spectra, such as those shown in
Fig.~\ref{fig:butane}, reveal peaks corresponding to the resonant transfer
of energy from the positron to specific molecular vibrations. In each of these
resonances the positron becomes temporarily attached to the molecule,
resulting in a greatly enhanced annihilation rate.
The energies of the resonances are given by Eq.~(\ref{eq:enVFR}), namely,
$\eps _\nu =E_\nu -\eb$, where $E_{\nu}$ is the energy of the excited vibration
$\nu$ and $\eb$ is the positron-molecule binding energy.
Fundamental vibrations for which $E_\nu =\omega _\nu $, can produce strong
annihilation. As a result, the molecular annihilation spectra are
somewhat similar to infrared-absorption spectra. However,
the magnitudes of the annihilation resonances are not
proportional to the IR absorption strengths. They follow a different scaling
with molecular size. While IR-active modes dominate the spectra of many
molecules, there are cases in which nominally IR-inactive modes, as well as
combinations and overtones, also appear to produce annihilation resonances.

\subsection{Halomethanes as a benchmark example of VFR}\label{subsec:halom}

\begin{figure}[ht]
\includegraphics*[width=8cm]{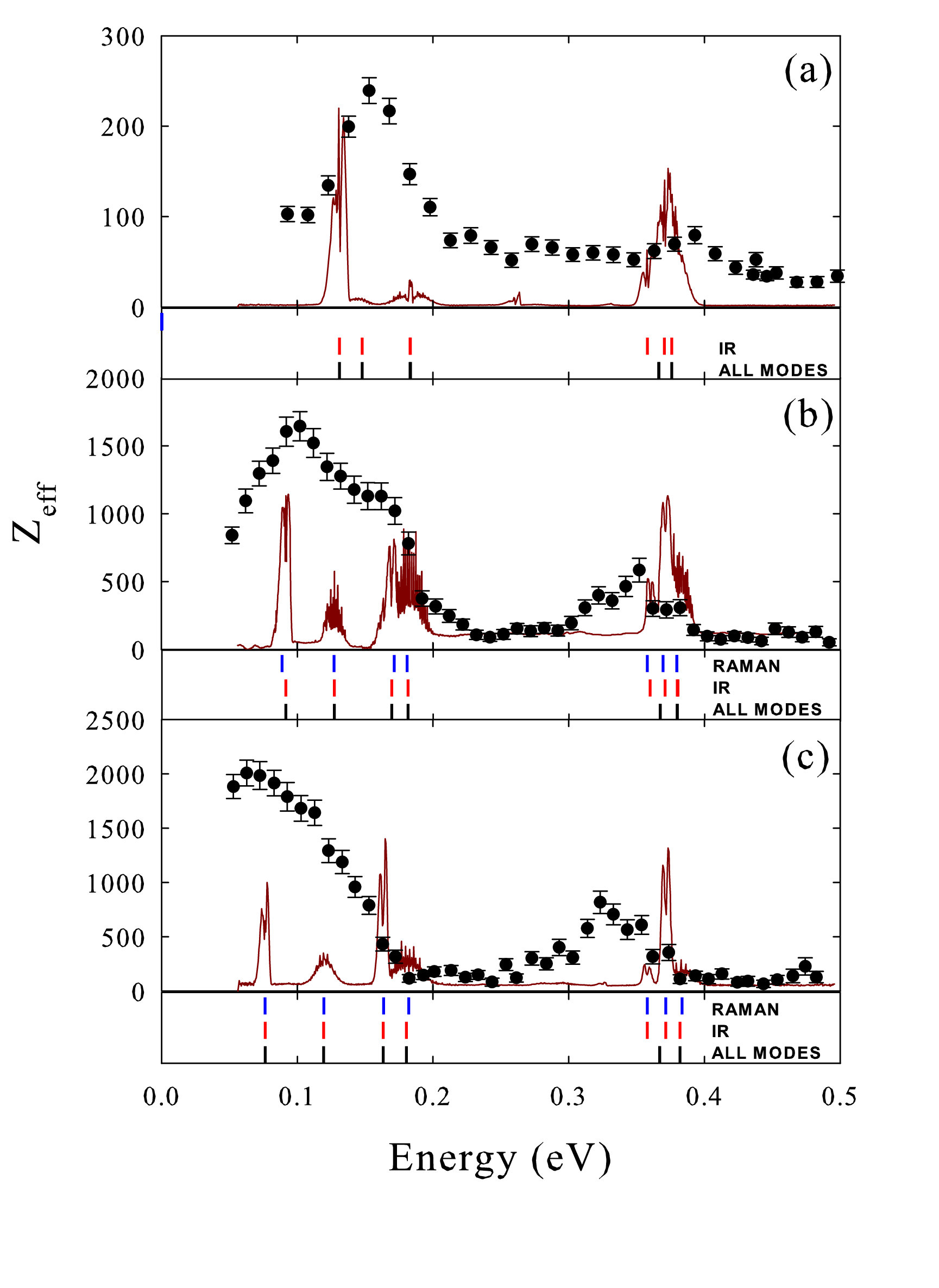}
\caption{Annihilation rates $\Z$ ($\bullet $) and IR absorption (solid curves)
for methyl halides: (a) CH$_3$F, (b) CH$_3$Cl, and (c) CH$_3$Br \cite{BYS06}.
Vertical bars below each plot indicate the vibrational energies from IR and
Raman measurements, and selected mode frequencies (``all modes'')
from \cite{NIST}.}
\label{fig:halom}
\end{figure}

The singly halogenated methanes are a near-perfect set 
of molecules to test theories of positron VFRs. Each molecule has 
only six vibrational degrees of freedom, all of which are dipole-active.
There are three energy-separated pairs of fundamental vibrations: the C-H
stretch modes, the C-H bend modes, and the C-$X$ modes where $X$ is the
halogen. The energy-resolved annihilation 
spectra and the infrared absorption spectra for these molecules are shown in 
Fig. \ref{fig:halom}. In CH$_3$Cl and CH$_3$Br, one can discern VFR from
all of the infrared-active modes. The high-energy peak is due to the C-H
stretch mode, and the broad low-energy feature is due to the C-H bend and
C-$X$ modes. There is no evidence of multimode VFRs.

As shown in Fig. \ref{fig:halom}, the position of the C-H stretch resonance
shifts downward in energy as the size of the halogen atom is increased,
reflecting an increase in the positron binding energy. The binding energies
range from near zero in CH$_3$F to about 40~meV in CH$_3$Br.
Since vibrational Feshbach resonances cannot occur unless the positron is
bound to the molecule, the small positive energy shift of the C-H
stretch peak in CH$_3$F is likely the result of a very small binding
energy \emph{and} a small positive shift in the mode energy.

\begin{figure}[ht]
\includegraphics*[width=8cm]{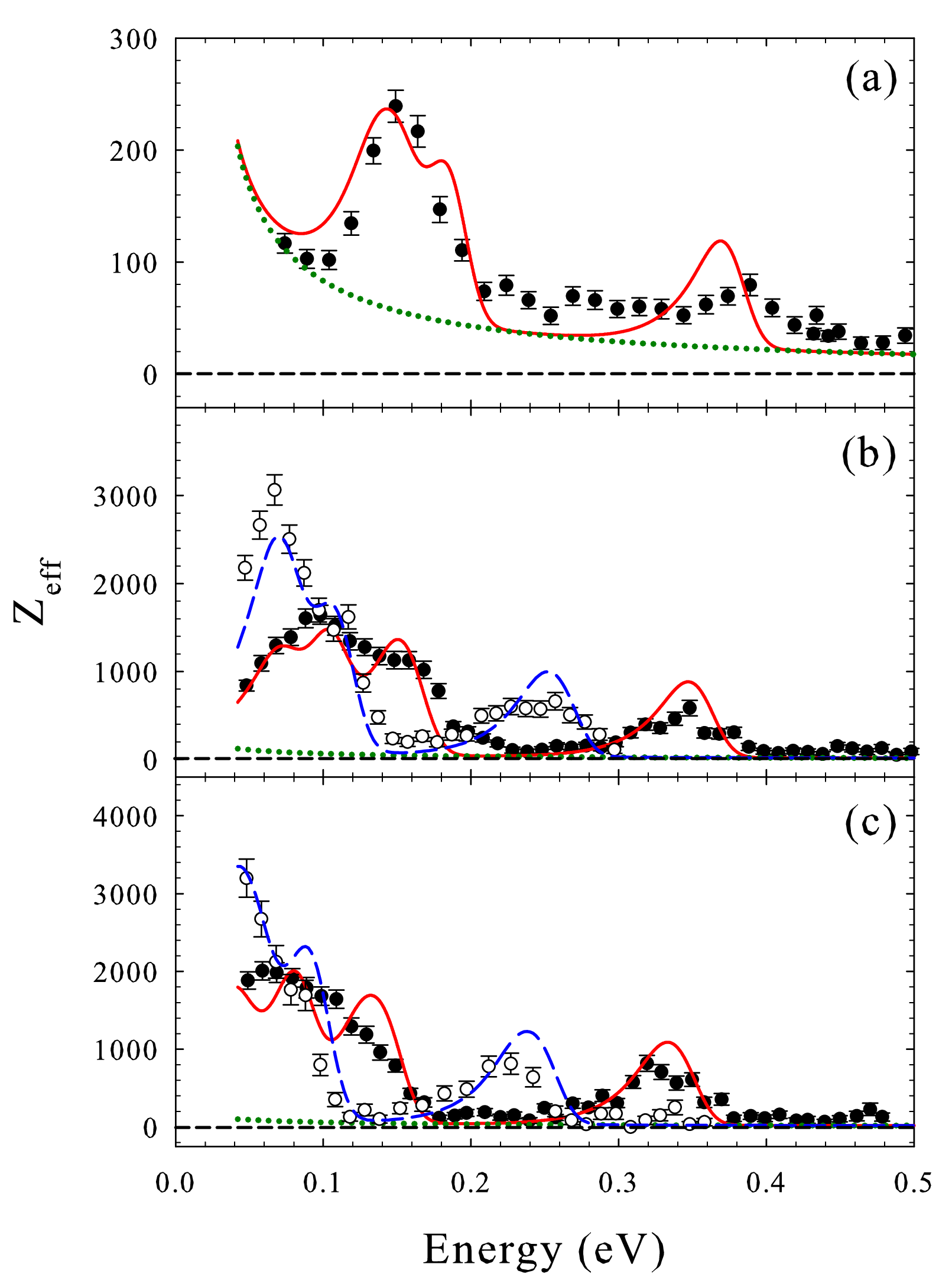}
\caption{Comparsion of experimental and theoretical $\Z$ for methyl halides
CH$_3X$ ($\bullet $ and solid curves) and CD$_3X$ ($\circ $ and dashed curves)
for (a) $X$$=$F, $\eb=0.3$~meV; (b) $X$$=$Cl, $\eb=25$~meV; and (c) $X$$=$Br,
$\eb=40$~meV \cite{BYS06,GL06a,YGL08}. Dotted
curves show the contributions of direct annihilation.}
\label{fig:halom1}
\end{figure}

According to
Eq.~(\ref{eq:Zeff_fin}), the annihilation rate in small molecules can be
described  by a sum of Breit-Wigner resonances for each mode, $\nu$,
convolved with the instrumental positron energy resolution function
$f_{\rm b}(\eps )$.
All of the modes are dipole active in the halomethanes. Thus the elastic
capture rates, which are roughly proportional to the IR strengths, are
expected to be much larger than the annihilation rates
(see Sec.~\ref{subsec:IR}).
As a result, $\Gamma \simeq \Gamma ^e$, and Eq. (\ref{eq:Zeff_fin}) is
greatly simplified. Using Eq.~(\ref{eq:rhoep}), one obtains
\begin{equation}\label{eq:Zeff_fin1}
\bar Z _{\rm eff}^{(\rm res)}(\bar \eps )=\pi F
\sum_\nu g_\nu b_\nu \Delta (\bar \eps -\eps _\nu),
\end{equation}
where $g_{\nu} = \sqrt{\eb/\eps _\nu}$. Since $F$ is assumed to be
constant (see Sec.~\ref{subsec:dir}), the relative magnitudes
of the dipole-active resonances in a given small molecule are determined by
the $g_\nu $ factors. The only adjustable parameter is the
binding energy, which can be determined by comparison with experiment.
The application of
Eq.~(\ref{eq:Zeff_fin1}) to the halomethanes and their deuterated counterparts
is shown in Fig.~\ref{fig:halom1}. The small contribution from nonresonant
direct annihilation, described by Eq.~(\ref{eq:Z_virt}), is included.

For the two larger halogens, the binding energy was obtained from the 
position of the C-H stretch peaks, while for CH$_3$F, the (small) binding 
energy was determined by fitting to the magnitude of $\Z$ \cite{GL06a}.
The resulting binding energies for CH$_3$F, CH$_3$Cl, and CH$_3$Br are 
0.3, 25, and 40~meV, respectively, and the agreement between the
theoretical and experimental  $\Z$ 
spectra is remarkably good. As the binding increases, each 
annihilation peak shifts downward in energy and increases in magnitude in 
accord with the predicted $g$-factor scaling. The theory also predicts
successfully the \emph{absolute magnitudes} of each of these features.

For a wide variety of hydrocarbons, experiments have established that the 
positron binding to the molecule is unchanged when the hydrogen atoms are 
replaced with deuterium atoms. This is plausible, since the binding energy is 
expected to be primarily a function of the electronic rather than the 
vibrational, degrees of freedom \cite{BGS03}. The only 
difference is that the deuterated species 
have lower vibrational energies due to the larger reduced masses
associated with the C-D modes. The binding 
energies for the hydrogenated species were used to predict the $\Z$ 
spectra for the deutrated species, thus providing a stringent test of 
the theory. The results of this 
comparison are shown in Fig.~\ref{fig:halom1}. Note that the $g$-factors and
hence the magnitudes of the peaks, are larger for the deuterated species,
because the resonances occur at smaller positron impact energies. The theory,
now with no free parameters, works
well for the deuterated halomethanes just as it did for the hydrogenated 
species. To our knowledge, no other theory has demonstrated such close
agreement, in both magnitude and shape, with the observed positron-molecule
annihilation resonances.

\subsection{Methanol: a case of multimode VFR}\label{subsec:meth}

\begin{figure}[ht]
\includegraphics*[width=8cm]{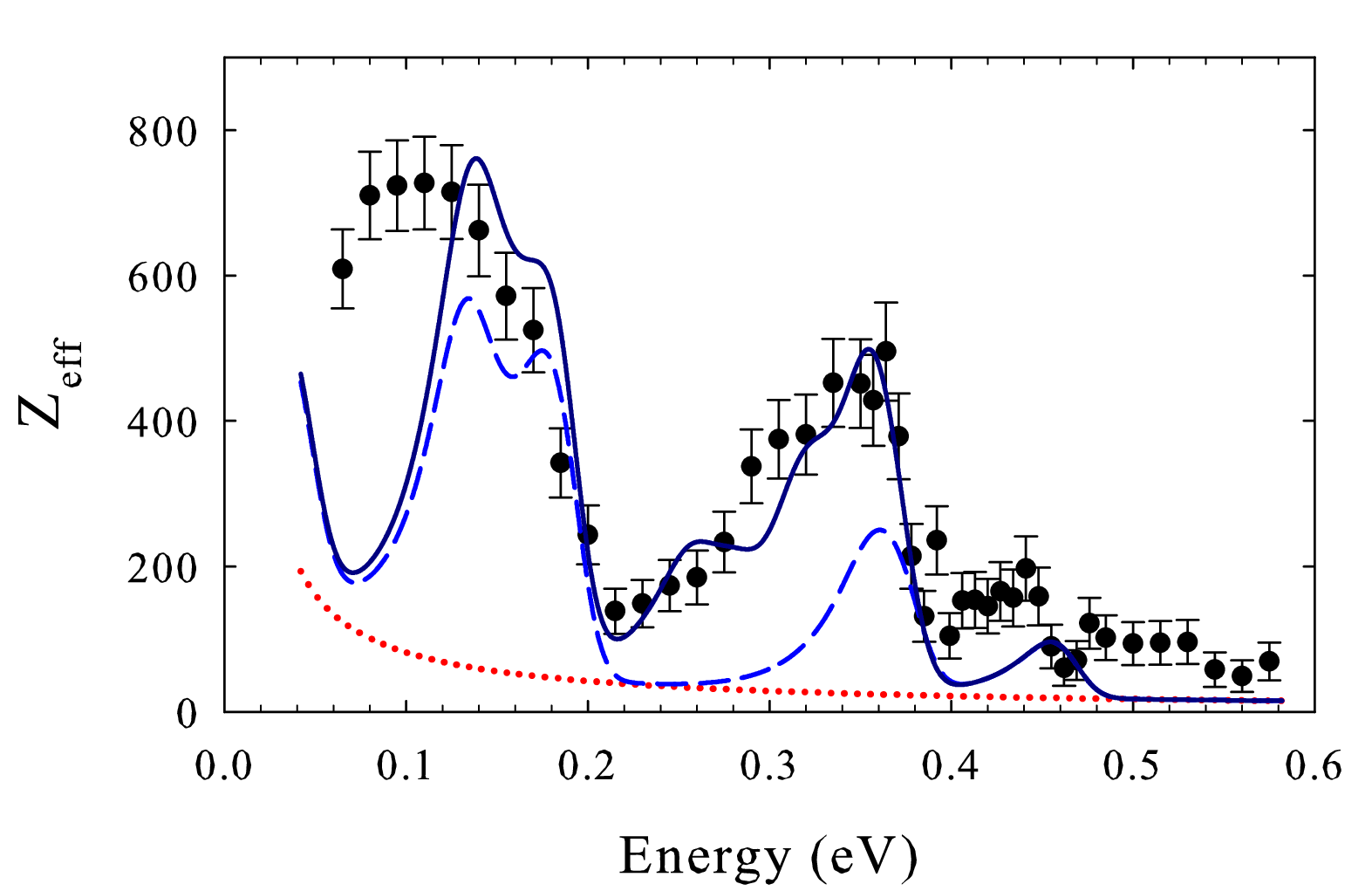}
\caption{Comparsion of experimental $\Z$ ($\bullet $) for methanol (CH$_3$OH)
with theory: dotted curve, contribution of direct annihilation
$\Z^{\rm (dir)}$; dashed curve, total $\Z$ due to VFR of vibrational
fundamentals for $\eb =2$~meV; solid curve, total $\Z$ due to resonant
annihilation involving the 12 modes and 9 overtones and combinations.
See \textcite{YGL08} for details.}
\label{fig:methanol}
\end{figure}

In methyl halides, each 
vibrational mode produces a measurable VFR with a relative magnitude given
by the $g$ factor, and there is no evidence of multimode
excitations. Experiments show that this  ``selection rule'' must 
be relaxed for a variety of other molecular species. 
As in electron-molecule collisions, positrons can be expected to excite 
vibrations that are nominally dipole forbidden. The 
nature of annihilation VFRs, as described by the theory, will be to even out
fairly large variations in the capture rate as long as 
$\Gamma_{\nu}^e \gg \Gamma ^a$. As shown in Fig.~\ref{fig:methanol},
there is evidence of such higher-order
vibrations in the $\Z$ spectrum of methanol CH$_3$OH. This molecule is
isoelectronic with CH$_3$F. Its vibrational spectrum is also similar except
for the additional O-H stretch vibration. However, the $\Z$ spectrum of
methanol is quite different than that of CH$_3$F and the other methyl halides.
There is a significant increase in magnitude of the high- and low-energy peaks
in methanol relative to those of CH$_3$F.

The positions of the C-H stretch peak and the peaks at lower energies
indicate that the binding energy in methanol is small. The fitted binding
energy of CH$_3$F is 0.3~meV; in methanol it could be
an order of magnitude larger but still remain within the experimental energy
uncertainty. There is evidence of an additional peak above the C-H stretch
modes, presumably due to the O-H stretch mode which has the energy
$\omega _{\rm OH}=456$~meV. If this interpretation is correct, this VFR 
is downshifted relative to $\omega _{\rm OH}$ by an amount
somewhat greater than the positron binding energy.

While one can
vary the binding energy to describe better some of the enhancement in $\Z$
seen in Fig.~\ref{fig:methanol}, there are features in the spectrum that
cannot be explained by mode-based VFRs.
In particular, the higher-energy peak in methanol is much broader than the 
corresponding C-H stretch peaks in the halomethanes and the prediction 
based on the IR-active modes. It is also much closer in 
magnitude to the low-energy peak. This discrepancy can only be resolved by
considering additional resonances.

Figure~\ref{fig:methanol} shows the results of two calculations using
a binding energy of 2~meV. The dashed curve is the calculation which
includes only the fundamental vibrations, all of which are dipolemactive.
It falls significantly short of explaining the spectrum. The IR 
absorption measurements in methanol reveal a number of relatively weak  
overtones and combination vibrations (likely some of which are
Fermi resonances) \cite{BZ97}. Using these data allows one to estimate the
elastic rates $\Gamma_{\nu}^e$. They are generally smaller
than those of the fundamentals, but still satisfy the relation
$\Gamma_{\nu}^e \gg \Gamma ^a$ [see Table~I by \cite{YGL08}]. The result of
adding these nine two-quantum overtones and combinations to the twelve
fundamentals is shown by the solid curve in Fig.~\ref{fig:methanol}. This
prediction is clearly in better agreement with experiment.
This comparison provides strong evidence that
multimode vibrations can make significant contributions to the $\Z$ spectra 
of small molecules. There are two remaining discrepancies. One is the
significantly downshifted O-H stretch peak. The second is the higher
experimental $\Z$ values below 100~meV. In methanol there is a torsion
mode at $\sim $40~meV. Its overtones and combinations could provide
the missing spectral weight in this region, but there is at present no
estimate of this effect.

\subsection{VFR from dipole-forbidden vibrations}\label{subsec:forbid}

\begin{figure}[ht]
\includegraphics*[width=7cm]{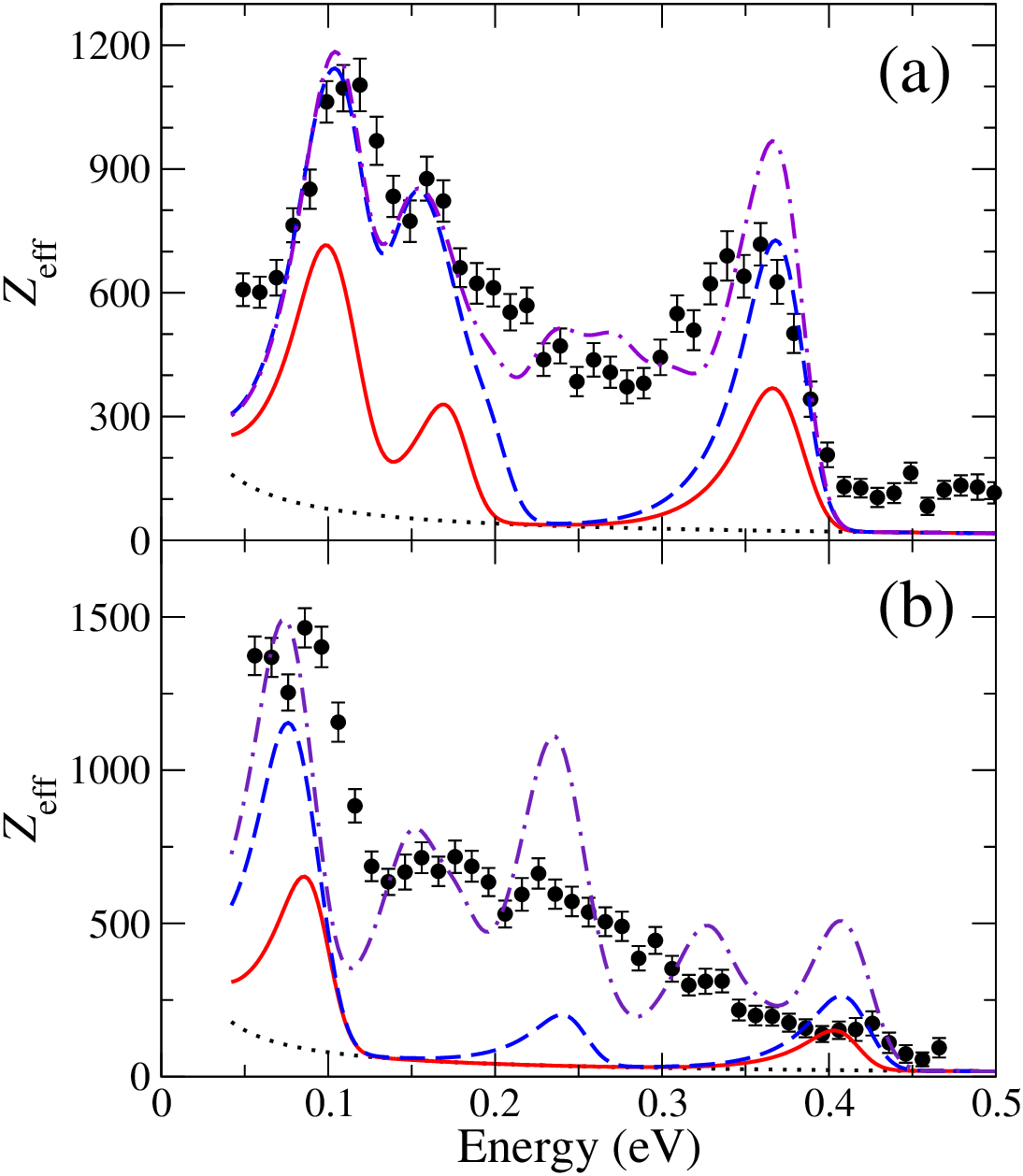}
\caption{Comparsion of the experimental $\Z$ ($\bullet $) with theory for
(a) ethylene (C$_2$H$_4$, $\eb =10$~meV), and (b) acetylene (C$_2$H$_2$,
$\eb =5$~meV). Dotted curves, direct annihilation $\Z^{\rm (dir)}$;
solid curves, total $\Z$ due to all IR active modes. In (a)
dashed curve, total $\Z$ due to $A_g$, $B_g$, and $B_u$
modes that capture $s$-, $p$-, and $d$-wave positrons; chain curve, same
with the addition of 14 IR-active overtones and combinations listed
in \cite{GBH99}. In (b) long-dashed curve, all modes;
chain curve, all modes with $\Sigma _{g,u}$,
$\Pi _{g,u}$, and $\Delta _{g}$ symmetries. The contributions of all overtones
and combinations are {\em ad hoc} weighted by a factor $1/n$,
where $n$ is the number of vibrational quanta.}
\label{fig:c2h4c2h2}
\end{figure}

The theoretical analysis of methanol indicates that multimode vibrations can 
contribute significant spectral weight to an annihilation spectrum. However,
there is also strong evidence that modes with very weak (or nominally zero)
dipole coupling can also produce VFRs. They could, for example,
arise from higher-order nondipole coupling (e.g., electric-quadrupole-active
modes), but there is no simple way to assess their possible contributions.
These nondipole features are exemplified by the experimental and
theoretical $\Z$ spectra of ethylene shown in Fig.~\ref{fig:c2h4c2h2} (a). This
molecule has five IR-active modes and six IR-inactive modes. The shift of the 
C-H stretch peak indicates a binding energy of about 10~meV.

Including only IR-active ($B_u$ symmetry) modes
underestimates $\Z$ by a factor of 2 at the peaks located at 100 and
350~meV. This calculation, which uses the IR strengths from \textcite{BC82},
completely misses contributions in
the interval of energies between these modes. However, the form of
Eq.~(\ref{eq:Zeff_fin}) does not depend on the nature of positron-vibrational
coupling, and all vibrations with
$\Gamma _\nu \approx \Gamma ^e_\nu \gg \Gamma ^a$
contribute equally to $\Z $ (i.e., to within the $g$ factor), as per
Eq.~(\ref{eq:Zeff_fin1}). As shown in Fig.~\ref{fig:c2h4c2h2}, including the
remaining modes (i.e., those with $A_g$ and $B_g$ symmetry, corresponding to
the capture of $s$-, $p$-, and $d$-wave positrons) greatly improves the
agreement. The further inclusion of the fourteen IR-active combination
vibrations (not shown) results in $\Z$ values exceeding those observed
experimentally. However, as shown in Fig.~\ref{fig:c2h4c2h2},
if one weighs their contributions by an empirical factor
$\Gamma_{\nu}^e/\Gamma_{\nu} =1/n$, where $n$ is the number of vibrational
quanta involved, the prediction agrees well with the
experimentally measured spectrum. This analysis shows that nearly all of the
spectral weight between the two largest peaks in ethylene can, at least in
principle,
be attributed to multimode VFR. However, there is presently only an {\em ad hoc}
approach available to decide whether ``borderline'' vibrational capture
channels sit above or below the cutoff value of the coupling strength
set by $\Gamma ^a$.

Shown in Fig.~\ref{fig:c2h4c2h2} (b) is the spectrum of acetylene (C$_2$H$_2$).
Its binding energy is too small to measure directly, and hence it is
obtained by fitting to the spectrum. As in the case of ethylene, the
theoretical $\Z$ which includes only IR-active fundamentals, does not agree
with the measured spectrum. Adding all of the other modes provides little
improvement. It is only after adding the overtones and combinations that
the calculated $\Z$ spectrum matches the magnitude of that observed.
Still, the peaks in the theoretical $\Z$ spectrum are much more
prominent. A similar analysis has been performed
for ammonia \cite{Gri10}. It also highlights the role of overtones
and combinations, and suggests that rotational broadening of VFR
and rotational Feshbach resonances may be required to explain the $\Z$
spectrum in this small molecule.

The modeling discussed here indicates that, to explain the $\Z$ spectra of
molecules such as C$_2$H$_4$ and C$_2$H$_2$, one needs to include the 
VFR of IR-inactive modes, overtones, and combinations. However,
determining $\Gamma _\nu ^e$ and $\Gamma _\nu$ for these resonances
requires calculations of the full vibrational dynamics, and so at present
the theory is incomplete.

\subsection{Effect of molecular size on the magnitudes of VFR}
\label{subsec:smallsize}

\begin{figure}[ht]
\includegraphics*[width=7cm]{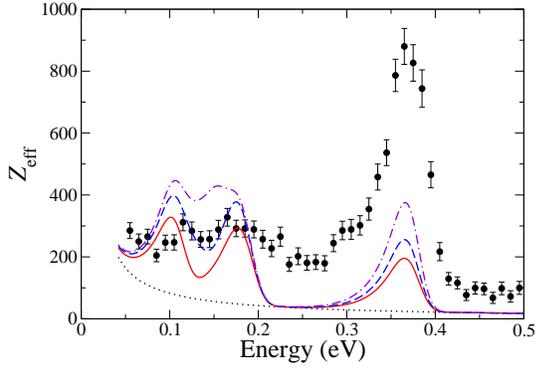}
\caption{Experimental $\Z$ spectrum ($\bullet $) and theory
for ethane (C$_2$H$_6$). The calculations use $\eb=1$~meV and include:
dotted curve, direct $\Z$; solid curve, same with VFR of IR-active
modes; dashed curve, same with the addition of $A_{1g}$ modes; chain curve,
same with the addition of $E_g$ modes.}
\label{fig:ethane}
\end{figure}

As discussed in the next section, larger molecules exhibit
annihilation VFRs that cannot be explained by positron coupling to
fundamentals or combination and overtone vibrations. In these
molecules, an additional enhancement mechanism appears to be operative that
causes the magnitudes of the fundamental resonances to grow rapidly with
molecule size. The smallest molecule to show evidence of such
enhanced VFR 
is ethane. As shown in Fig.~\ref{fig:ethane}, the $\Z$ spectrum for this
molecule has a distinctly different spectral shape than that of
the other molecules discussed in this chapter. The high-energy C-H stretch
peak is three times larger than the low-energy C-H bend peaks, which
is inconsistent with the simple $g$ scaling of VFR magnitudes.
The calculated $\Z$ for ethane, shown in Fig.~\ref{fig:ethane}, indicates
that the VFRs of IR-active and dipole-forbidden modes, populated
by the positron $s$-, $p$-, and $d$-wave capture, can explain the spectral
weight at lower energies. However, they do not account for the magnitude of
the C-H stretch peak, nor for the magnitude of $\Z$ between the low- and
high-energy peaks, which may be due to overtones and combinations.

\begin{figure}[ht]
\includegraphics*[width=7cm]{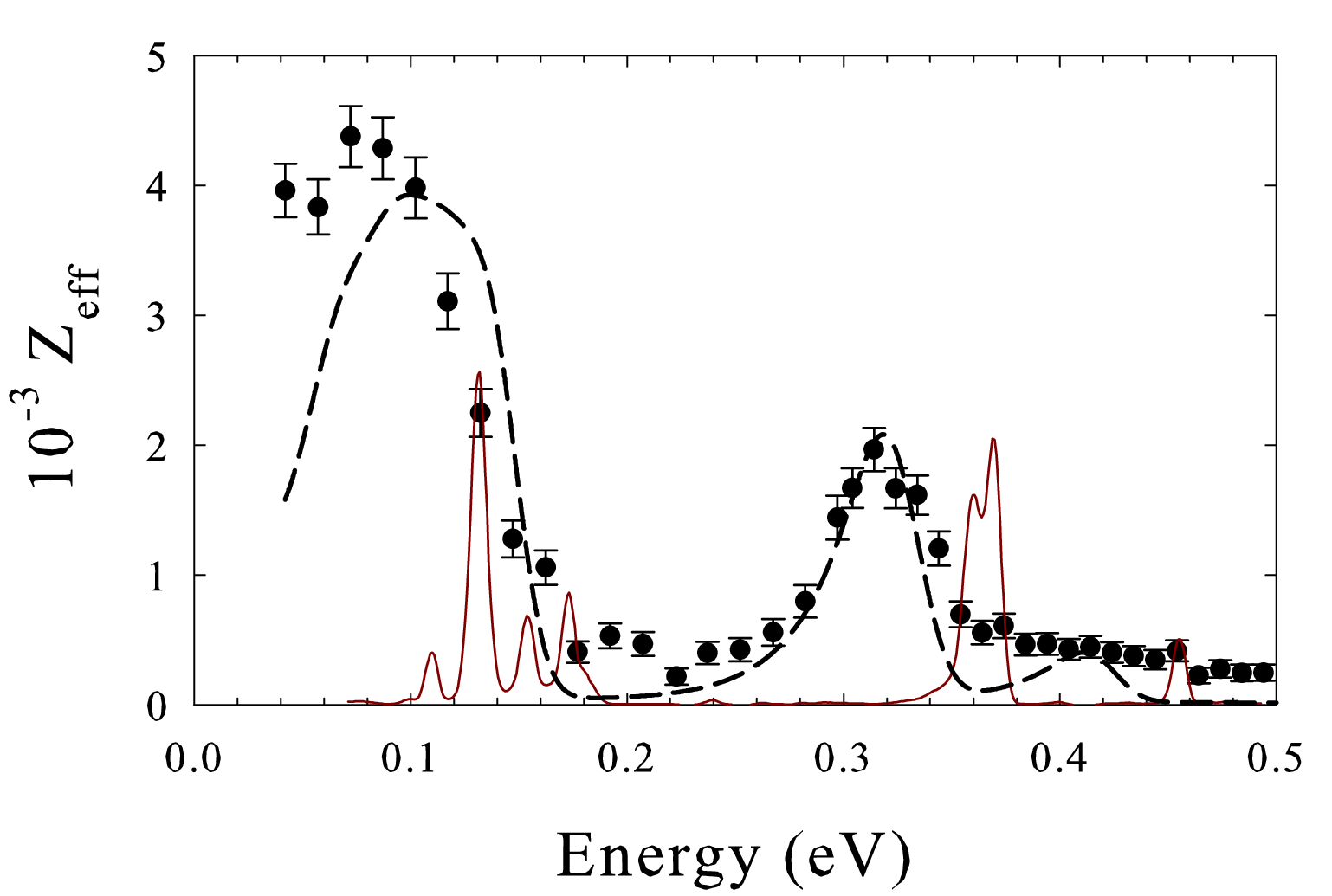}
\caption{Comparsion of experimental $\Z$ ($\bullet $), IR absorption
spectrum (solid curve), and theoretical $\Z$ (dashed curve) for ethanol
(C$_2$H$_5$OH).
The calculation [cf. Eq. (\ref{eq:Zeff_fin})] uses $\eb=45$~meV and
the mode frequencies and transition dipole amplitudes from \textcite{SWD90}.}
\label{fig:ethanol}
\end{figure}

Similar analyses for propane and cyclopropane (see Sec.~\ref{subsec:other})
show that these molecules exhibit even stronger
enhancements of the high- and low-energy peaks, while
the ``background'' of multimode excitations between these resonances
does not appear to be similarly enhanced. However, molecular size alone does
not appear to be an accurate predictor of the scaling of $\Z$ magnitudes.
As shown in Fig.~\ref{fig:ethanol}, the analysis for ethanol (C$_2$H$_5$OH;
one more atom than ethane) indicates that there is no evidence of
such an enhancement for this molecule. Here the agreement with
experimental $\Z$ of the simple IR-active-mode VFR theory without overtones
or combinations is remarkably good.

In summary, the theoretical framework of \textcite{GL06a} allows us to make a
clearer distinction between those molecules that exhibit
``small molecule'' behavior and those that exhibit ``large molecule'' behavior
(i.e., where $\Z$ values can be orders of magnitude larger than those
predicted by the mode VFR theory). However, the physics that is responsible
for this threshold is still poorly understood.

\subsection{Nonresonant annihilation in small molecules}\label{subsec:nonres}

Many molecules do not exhibit resonant annihilation but display instead
relatively smooth featureless $\Z$ spectra.
To better understand these molecules, recall that the minimum requirements
to observe annihilation-mediated VFR are the existence of a positron-molecule
bound state and a vibrational mode that couples to the positron. 

The existence of weak binding with $\eb =\kappa ^2/2$ is linked to a large
positive value of the positron-molecule scattering length $\kappa ^{-1}$
(Sec.~\ref{subsec:dir}). The magnitudes of the resonances are then determined
by $g=\kappa/k$. If $\kappa$ is negative, the bound state is replaced by a
virtual state in the continuum, and VFR are absent. In both cases,
however, one expects a nonresonant background due to direct 
annihilation, proportional to $(k^2+\kappa^2)^{-1}$ [cf. Eq. (\ref{eq:Z_virt})].

Small nonpolar or weakly polar molecules are far less likely to bind
positrons, and so it is not surprising that many of these
molecules lack VFRs. One such molecule is CO$_2$. It has a relatively flat
spectrum with $\Z\approx 35$ above 150~meV \cite{YS08b}, and the
thermal $\Z$ of 54.7 \cite{WCG85}. Since this molecule
has 22 electrons, these values are not far from the uncorrelated electron gas
prediction. Using a 
vibrational close coupling formalism, \textcite{GM99} predicted a 
resonance-free spectrum with a nearly constant $\Z \approx 50$,
which is in reasonably good agreement with the measurements.

\begin{figure}[ht]
\includegraphics*[width=7cm]{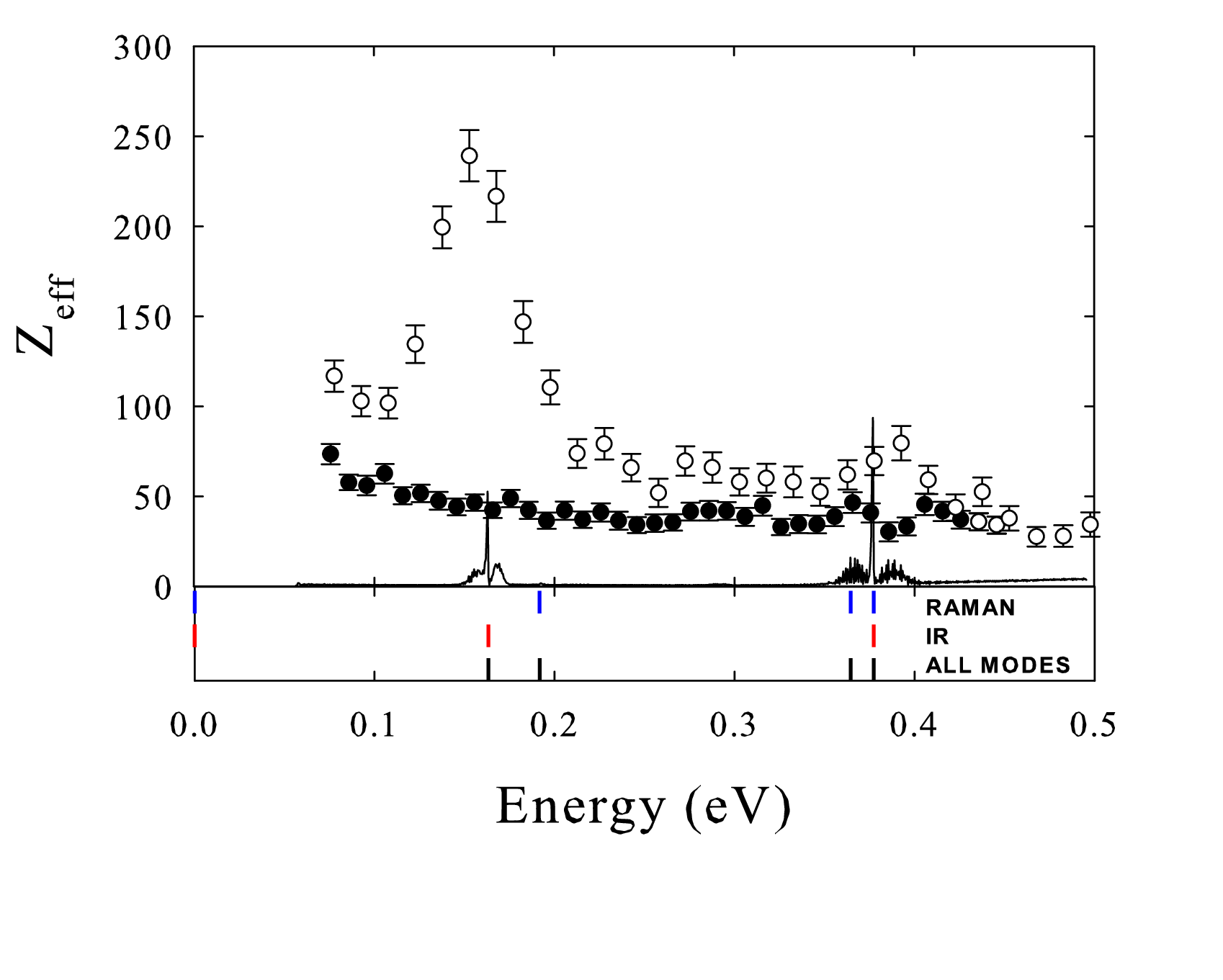}
\caption{Measured $\Z$ spectra for CH$_4$ ($\bullet $) and CH$_3$F
($\circ $), and IR absorption spectrum of methane (solid curve).}
\label{fig:CH4}
\end{figure}

The spectrum of methane CH$_4$, shown in Fig.~\ref{fig:CH4}, is also
relatively featureless, and similar in both magnitude and energy
dependence to the $\Z$ for CF$_4$ (not shown) \cite{BGS03}. This
is consistent with these molecules not supporting the positron bound states.
It refutes an earlier conjecture \cite{Gri00,IGG00}, which was based on the
analysis of room-temperature $\Z$ for methane and its fluoro-substitutes.
However, the difference between molecules close to the border between
VFR-active and VFR-inactive species can be 
stark. Figure~\ref{fig:CH4} shows a comparison of CH$_4$ and CH$_3$F.
The analysis presented in Fig.~\ref{fig:halom1} indicates that CH$_3$F has a
very small binding energy ($\sim 0.3$~meV). Yet this produces a distinct
resonance at 150~meV, and the spectral weight in the vicinity of the C-H
stretch mode is nearly doubled.

\begin{figure}
\includegraphics*[width=7cm]{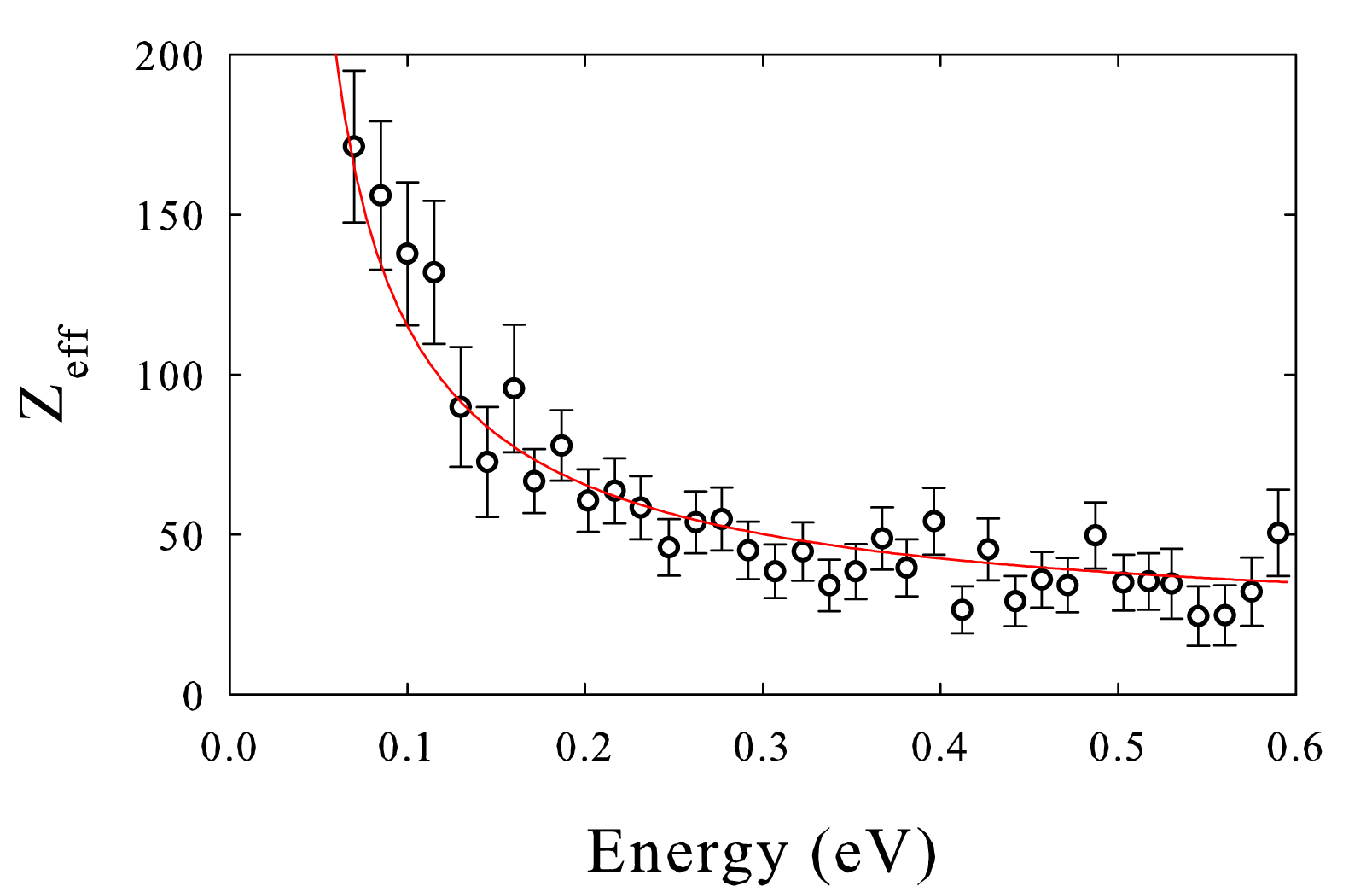}
\caption{Experimental $\Z$ spectrum ($\circ $) for water, and a fit based
on the direct annihilation model: solid curve, $\Z$ from Eq.~(\ref{eq:Z_virt})
with $\kappa ^2/2=0.3$~meV plus a constant offset of 20.}
\label{fig:H2O}
\end{figure}

As shown in Fig.~\ref{fig:H2O}, the water molecule also lacks distinct
VFR \cite{YS08b}. Its $\Z$ rises smoothly with decreasing
energy, reaching a value of 319 for thermal positrons at 300~K \cite{IGM95}.
The spectrum is well represented by the expression for
direct annihilation [Eq. (\ref{eq:Z_virt})].\footnote{In principle,
Eq.~(\ref{eq:Z_virt}) should be modified to account for the permanent
dipole moment of this molecule.} The fitting parameter $\kappa$ is
consistent with a virtual state at $\sim 1$~meV. A constant
$\Z$ offset of 20 was also included in this fit [cf. the first term
in Eq.~(\ref{eq:Zdir})].

The $\Z$ data for these small polyatomics can be compared with the
direct annihilation calculations of \textcite{GMO01}. In that work, the
positron-molecule interaction is described by
a local correlation-polarization potential, and $\Z$ includes the
electron-positron contact density enhancement factor \cite{BN86}.
The calculations predict a steady rise in $\Z$ with decreasing positron energy
below 0.5--1~eV. The calculated annihilation rates for CH$_4$, NH$_3$,
and H$_2$O (including those at thermal positron energies) are about a factor
of 2 lower than the experimental values, while for CF$_4$, the calculated
$\Z$ is about two times larger. With the exception of NH$_3$ \cite{Gri10},
the calculations confirm that VFR are not necessary to explain the
$\Z$ spectra in these molecules.

Some molecules exhibit other spectral features. One example is the
sawtoothlike oscillation centered at $\sim 380$~meV in CH$_4$
(Fig.~\ref{fig:CH4}).
While the IR spectrum indicates strong absorption at this energy,
the magnitude and shape of this feature in $\Z$ are not consistent with
the VFR observed in other molecules.
Such sawtooth features are also observed in the $\Z$ spectra of CO$_2$
and H$_2$O \cite{YS08b}.

The origin of these features is at present unclear. \textcite{NG05a}
predicted that water should have fairly strong vibrational excitation        
cross sections, with sharp onsets. Channel coupling could result in
additional structure in the $\Z$ spectrum near the vibrational excitation
thresholds \cite{YS08b}. Alternatively, \textcite{YS08b} suggested that
these features could represent interference between direct and resonant
annihilation. However, the latter is not compatible with the evidence that
CO$_2$, H$_2$O, and CH$_4$ do not to bind positrons.

Other non-VFR-type features are either predicted
or allowed by theory but have yet to be observed. One example is
a shape resonance which could occur if the positron became temporarily trapped
inside a positive-energy potential barrier.
However,
low-energy positron scattering and annihilation is usually dominated by the
$s$-wave component of the incident positron wave function,
which has no centrifugal barrier. Further, the atomic cores
are repulsive, so a shape resonance is unlikely. Exceptions in other systems
include the prediction of a $p$-wave positron shape resonance in the Mg
atom \cite{MB07,MZB08} and cage-state shape resonances in cubane (C$_8$H$_8$),
C$_{20}$ and C$_{60}$ \cite{GL99,CLG08,GNG05}. Finally, \textcite{NG03} 
hypothesized that the presence of a virtual state, by itself, can lead to a
long-lived intermediate state (e.g., following a vibrational de-excitation
collision), and this is expected to produce a broad spectral feature.
Experimental investigation of these predictions is warranted.

\subsection{Small molecule summary}

The theory of \textcite{GL06a} (Sec.~\ref{subsec:IR}) provides a remarkably
useful framework for understanding resonant annihilation in small molecules.
The relative contribution of each VFR is a competition between resonant
elastic scattering and annihilation. As long as the coupling to the positron 
continuum is strong enough, the magnitude of each resonance is proportional
to $g=\sqrt{\eb/\eps }$, and the annihilation spectrum is a sum of distinct
resonances.
	
In the case of the methyl halides and ethanol, the conditions for the simple
$g$ scaling were confirmed by direct calculation of the capture rates using the
Born-dipole-type approximation (Sec. \ref{subsec:IR}). For methyl halides,
the binding energies were extracted from the measurements on the protonated
species. They were then used to make a prediction of $\Z$ for the deuterated
species, thus providing a stringent test of the theory. Agreement between
experiment and theory is excellent.

In other cases, however, the VFR of the IR-active modes are
insufficient to explain the observed annihilation spectra. In methanol,
it is necessary to include dipole-active multimode vibrations;
and in ethylene and acetylene, one must also include IR-inactive
modes and $n$-quantum overtones and combinations to match the experimentally
measured $\Z$. In these cases, the VFR of overtone and combination
vibrations had to be adjusted arbitrarily (i.e., by factors $1/n$) to match
the experimental $\Z$.
Thus, while the theory of Sec.~\ref{subsec:IR} provides useful
insights into the vibrations that are likely involved,
it lacks quantitative predictive accuracy for many molecules.

This theory has also helped to elucidate the boundary
separating species that have enhanced $\Z$ (i.e., beyond that predicted
by single resonances) from those that do not. The factors responsible for
this transition appear to be molecular size and the density of vibrational
states near a resonance (Sec.~\ref{subsec:large}). This enhancement
is likely due to IVR. There is some evidence that it is already operative
in molecules as small as ethane and is dominant in propane.

\section{IVR-enhanced resonant annihilation in larger
molecules}\label{sec:large}

\subsection{Overview}\label{subsec:over}

As indicated in Table ~\ref{tab:Zeff}, annihilation rates for large molecules
grow rapidly with molecular size. Similar to the case of small molecules,
these large rates are understood to be the result of positron attachment
via VFR. As shown in Fig.~\ref{fig:butane} for butane, the $\Z$
spectra have distinct peaks, the positions of which are 
strongly correlated with those of the IR peaks but are shifted downward 
by the binding energy. As was the typical case for small molecules, the VFR in
larger molecules also occur predominantly at energies corresponding to those
of the fundamental
vibrations. The low-energy plateau in the alkane spectra (e.g.,
below 140 meV in Fig. \ref{fig:butane}) is due to C-H bend modes and C-C modes,
while the high-energy peaks are due to C-H stretch modes. In the case of large
molecules, however, the $\Z$ values in these peaks are enhanced by orders
of magnitude above those predicted for VFR in small molecules.

\begin{figure}
\includegraphics*[width=8cm]{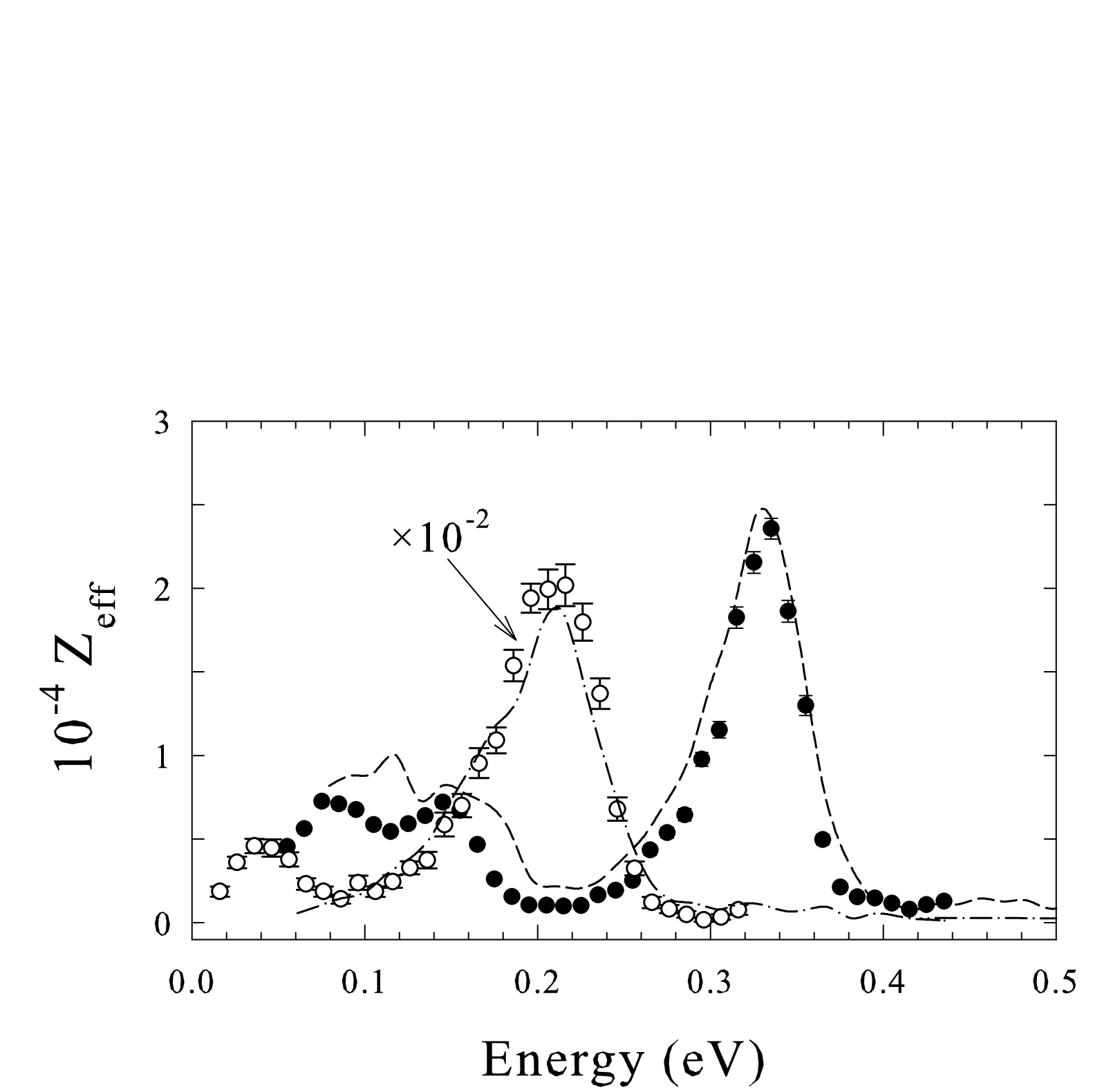}
\caption{$\Z$ spectra of hydrogenated butane (solid circles) \cite{BGS03} and
nonane (open circles) \cite{YS08a} compared with (the appropriately scaled)
fully deuterated analogs (dashed and chain curves) \cite{BGS03}. Mode
energies for the deuterated species have been scaled by the appropriate
reduced mass factor to match
those in the hydrogenated compounds, while the positron binding energy was
assumed to be independent of deuteration. The magnitudes of $\Z$ have been
scaled by the appropriate $g$ factor at each incident positron energy.}
\label{fig:but_non}
\end{figure}

To validate the identification of the resonant peaks with 
the vibrational modes, shown in Fig. \ref{fig:but_non} is the spectrum of
butane and nonane and the corresponding fully deuterated
compounds \cite{BGS03,YS08a}. When the spectra of the deuterated compounds are 
corrected for the change in vibrational mode frequencies,
the deuterated and hydrogenated data are in good agreement.

As discussed in Sec. \ref{sec:theory}, a plausible explanation for the large
magnitudes of $\Z$ in large molecules is that, by the excitation of a
vibrational 
fundamental, a large number of otherwise inaccessible multimode VFR 
(so called dark states) contribute to the annihilation, mediated by the 
process of intramolecular vibrational energy redistribution (IVR) 
\cite{Gri00,GG04}. In this paradigm, the incident positron is 
first captured  into a mode-based doorway state (e.g., 
involving infrared active modes) that then couples to a bath of 
quasidegenerate multimode dark states. This increases 
the multiplicity of the final capture states and causes an approximately 
proportionate enhancement in the resonant annihilation rate. As discussed
below,
it appears that the fundamental vibrations act as doorways in large molecules.
This in turn leads to the excitation of some, but not all of the nearby
multimode states in the IVR process, which results in enhanced annihilation.
A similar coupling to dark states occurs in molecular photoabsorption
\cite{SM83}.

This doorway-state model provides a common thread in the discussion of 
annihilation for large molecules. A particularly important link in this 
model is the connection between the observed rapid rise in the resonant 
annihilation rate with molecular size and IVR. A number of experiments 
and analyses clarify and, in some cases quantify, this physical picture. 

\subsection{The alkane molecule paradigm}\label{subsec:alk}

\begin{figure}
\includegraphics*[width=8.5cm]{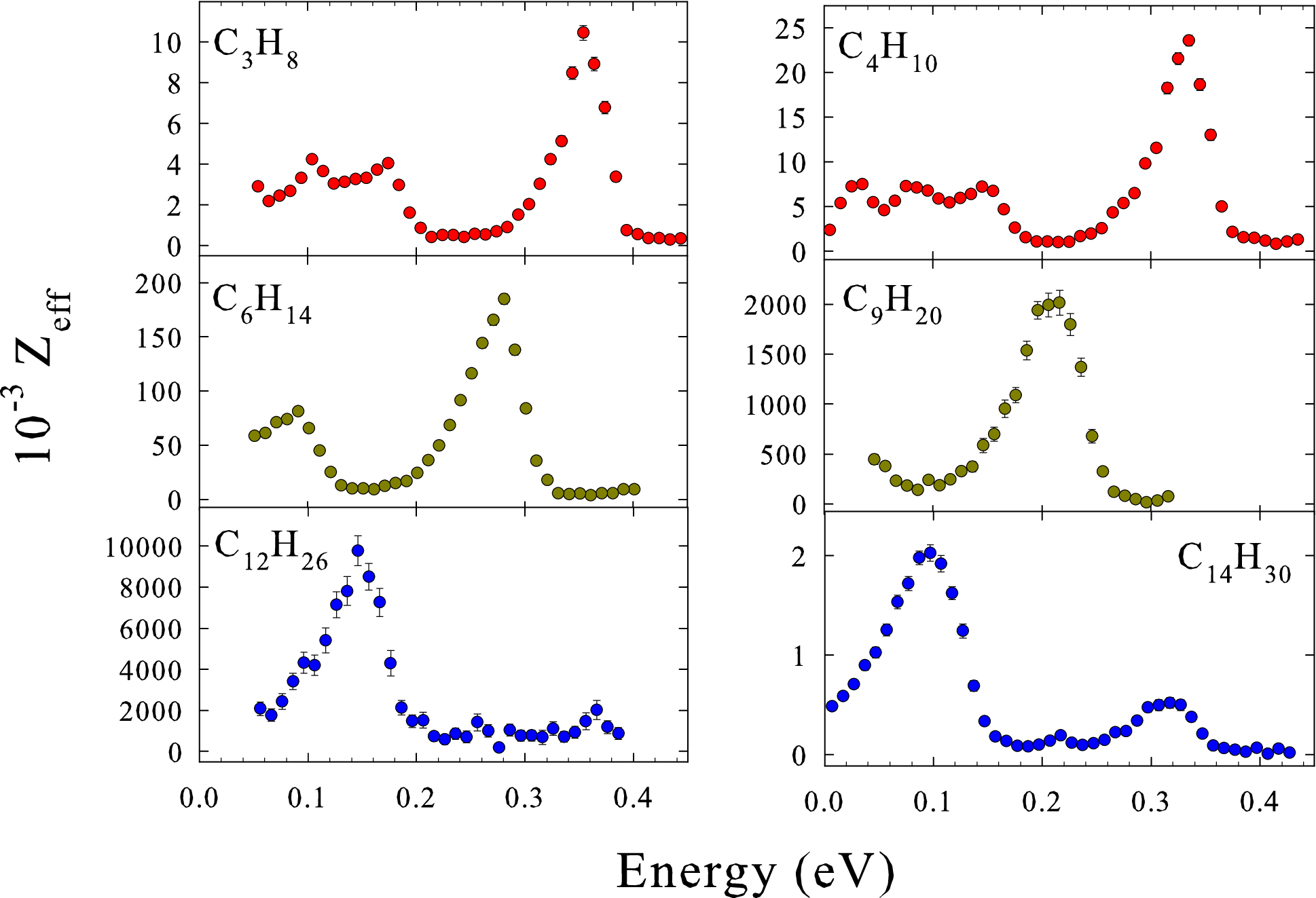}
\caption{$\Z$ spectra as a function of incident positron energy for a
variety of alkane molecules, showing the systematic shift of the spectra
to lower positron energy as molecular size is increased. Values of $\Z$ are
absolute except for C$_{14}$H$_{30}$, which is in arbitrary units due to
difficulties measuring its vapor pressure.}
\label{fig:var_alk}
\end{figure}

Alkane molecules have been studied more extensively than any
other molecular species (cf. Figs. \ref{fig:butane} and \ref{fig:but_non}).
Figure~\ref{fig:var_alk} shows six examples. These spectra resemble closely the 
spectra of the fundamental vibrations, albeit downshifted
by the positron-molecule binding energy. The binding energies increase by 
20--25~meV for each carbon-based monomer added to the alkane chain. 
Thus, the binding energy for propane is 10~meV, while for nonane it is 
145~meV. The magnitudes of the resonant annihilation peaks grow 
rapidly with molecular size, indicating that IVR, albeit likely 
incomplete, is operative.

The theory of resonant annihilation for small molecules (Sec.~\ref{subsec:IR})
prescribes a restricted role for binding energy in determining $\Z$
spectra, namely, 
\begin{equation}
\Z \propto\ b_\nu g(\eps ) = b_\nu \sqrt{\eb /\varepsilon}
\end{equation}
where $b_\nu$ is the multiplicity of the excited resonant capture
states. The vibrational coupling appears only via the
positron capture rate $\Gamma^e$, which usually cancels the total rate
$\Gamma $ for all but very weak resonances. 

\begin{figure}
\includegraphics*[width=8cm]{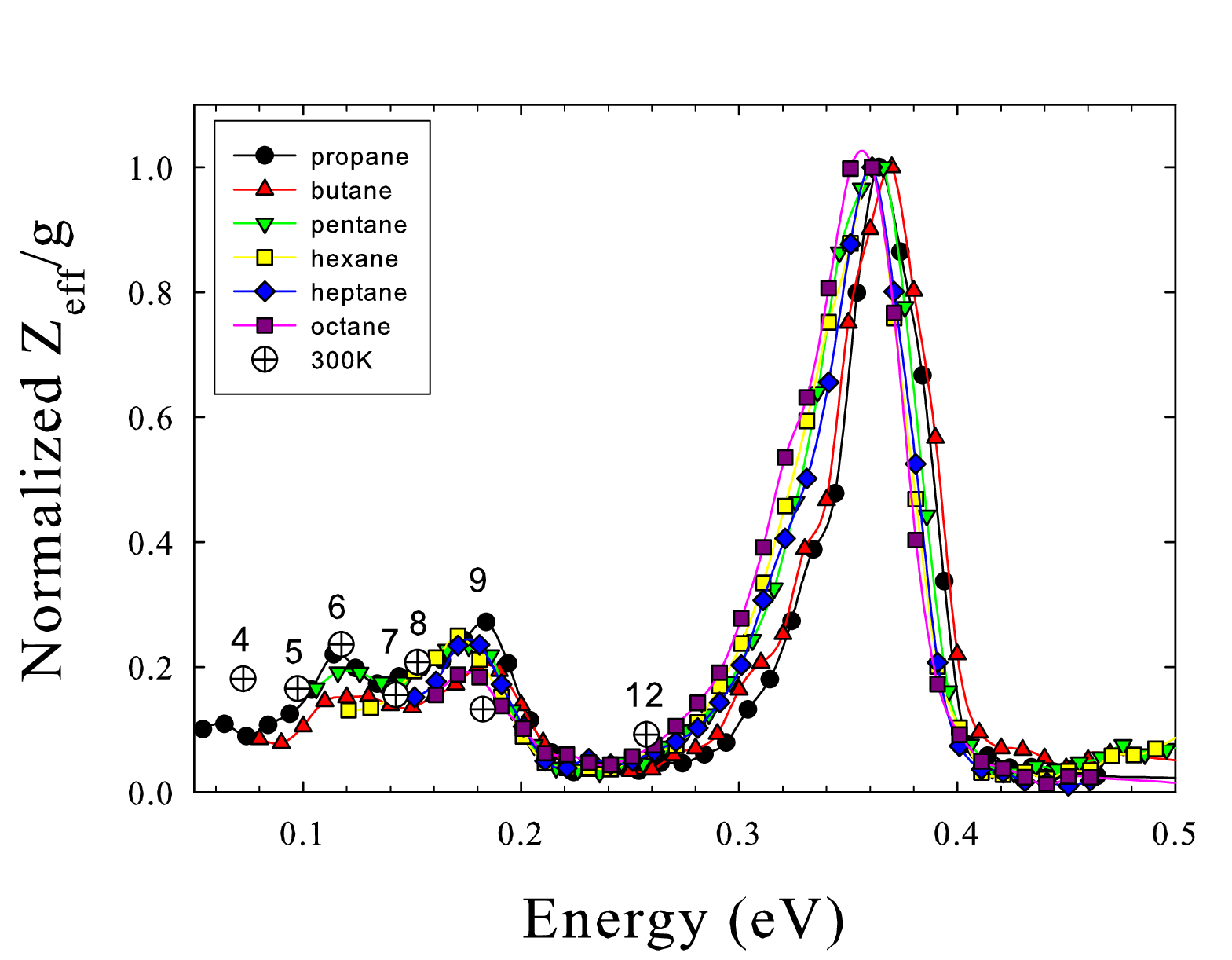}
\caption{Normalized and energy-shifted $\Z /g$ spectra for alkanes with
$n=3$--8 carbons. For comparison, room-temperature $\Z /g$ data for alkanes
are also shown (circles with plus sign) at energies $\eb +\varepsilon_T$, where
$\varepsilon_T=\frac{3}{2} k_BT = 37.5$~meV. Each room temperature datum is
labeled by the number of carbons $n$ in the molecule. See text for details.}
\label{fig:alk_norm}
\end{figure}

As shown in Fig.~\ref{fig:alk_norm}, the $\Z$ spectra for alkanes larger than
ethane are self-similar when they are 
first scaled by the factor $g(\eps )$, then the resulting magnitudes of
$\Z/g(\eps )$ are normalized at the C-H stretch peaks, and, finally, the
spectra are shifted upward by their binding energies \cite{YS08a}.
A surprising result from this analysis, not presently understood,
is that the relative magnitudes of the high-energy and lower-energy 
peaks in the alkanes remain the same over a factor of 3 in molecular
size and a factor of $10^3$ in the magnitude of $\Z$. The resulting
self-similar spectra shift downward with increasing molecular size as the
positron binding energy increases. As discussed
in Sec. \ref{sec:therm}, this has the consequence that the $\Z $ 
spectrum, measured with thermal positrons, can be assigned to
the vibrational modes populated through the corresponding VFRs 
in the thermal energy range. This connects in a quantitative way the
energy-resolved measurements and thermal measurements of $\Z$.

\begin{figure}
\includegraphics*[width=8cm]{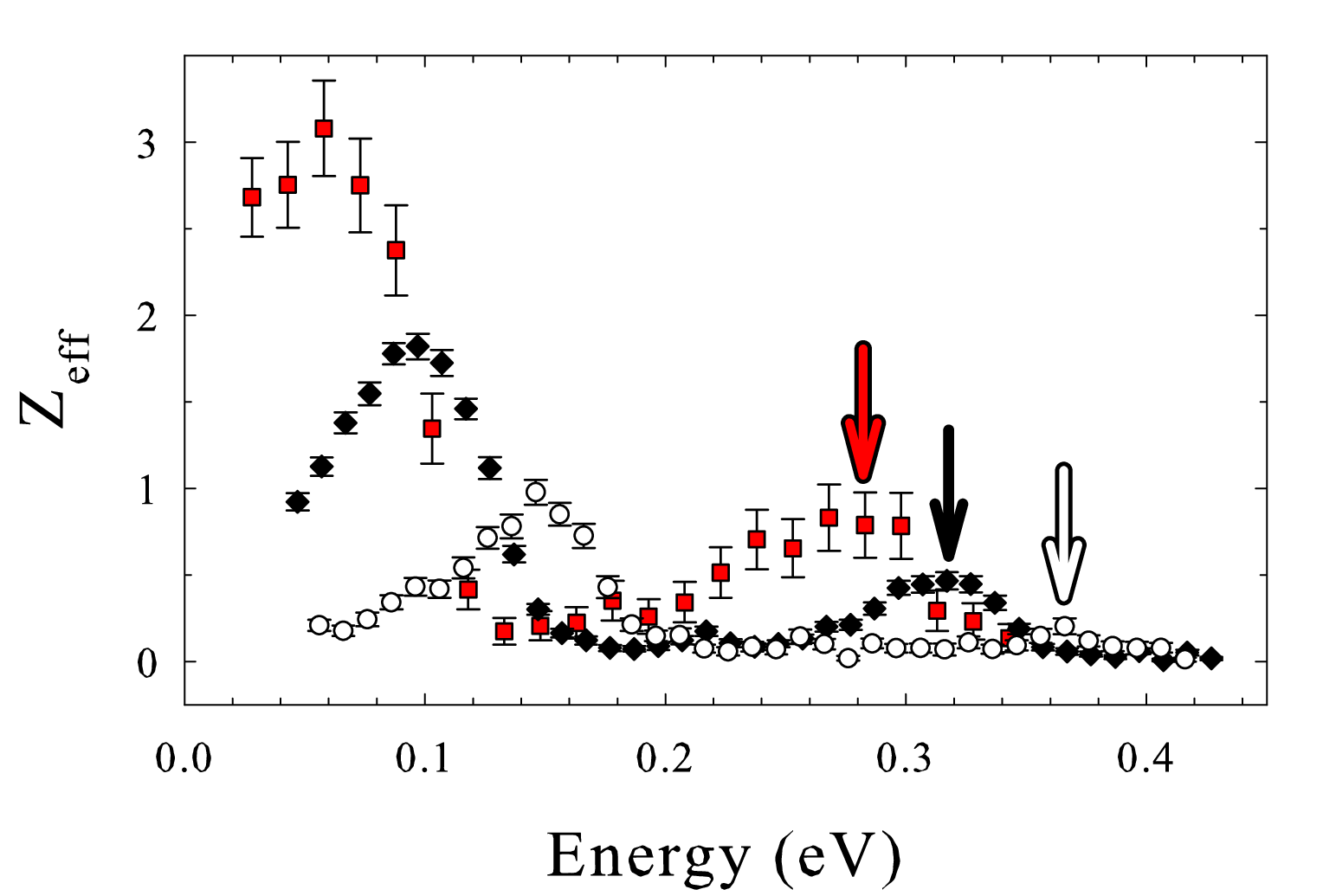}
\caption{$\Z$ spectra for ($\circ $) dodecane, (diamonds) tetradecane,
and (squares) hexadecane. The vertical arrows indicate the positions of the
C-H stretch mode VFR peaks for the second bound state (i.e., the first
positronically
excited state) in each molecule. The spectra for tetradecane and hexadecane
have been normalized arbitrarily since their vapor pressures were too low to
measure reliably. The large peaks at lower energy are the C-H stretch mode
VFR for the first bound states (the positronic ground states).}
\label{fig:alk_14_16}
\end{figure}

As the size of the alkane is increased beyond twelve carbons (dodecane),
a new feature appears in the $\Z$ spectrum at an energy close to that of
the C-H stretch peak. As 
shown in Fig. \ref{fig:alk_14_16}, the magnitude of this peak grows as the
alkane size increases, and it shifts downward in energy,
just as the original C-H stretch peak. In hexadecane (the 16-carbon alkane),
the C-H peak occurs at a mere 55~meV incident positron energy, corresponding
to a binding energy of 310~meV.

This new resonance is attributed to a second positron bound state
(i.e., the first positronically excited state) on 
the molecule. It is populated by a C-H stretch peak VFR in a manner similar 
to the larger ground-state peak that occurs at smaller incident positron
energy \cite{BYS06}.
The small peak at 365~meV in dodecane is identified as the C-H stretch
peak of the first, positronically excited bound state with a binding energy
of a few millielectron volts, while the larger peak at 150~meV is due to the
positron in its ground state. The positions of these peaks
are in good agreement with a model calculation described in
Sec.~\ref{subsubsec:bind} (cf. Fig.~\ref{fig:bind}).

An important feature of these positronically excited resonances is their
magnitude relative to the corresponding ground-state resonance in the same
molecule. For small molecules, it is expected that the contribution of a 
resonance will be proportional to $g=\sqrt{\eb /\varepsilon}$. For 
both tetradecane and hexadecane, the ratios of the magnitudes of the first 
and second bound-state C-H stretch resonances are equal to the ratios of the 
$g$ factors for these resonances. In particular, since the positron
overlap density is proportional to $\sqrt{\eb}$ (Sec.~\ref{subsubsec:bind}),
it is expected to be smaller for the second bound state, due to the
smaller binding energy.
This scaling of $\Z$ for the positronically excited states and the scaling
with $g$ of the spectra shown in Fig. \ref{fig:alk_norm} both demonstrate the
important role of the factor $g$ in determining the magnitudes of
annihilation peaks in large molecules.

\subsection{Dependence of $\Z$ on molecular size}\label{subsec:size}

\begin{figure}[ht]
\includegraphics*[width=8cm]{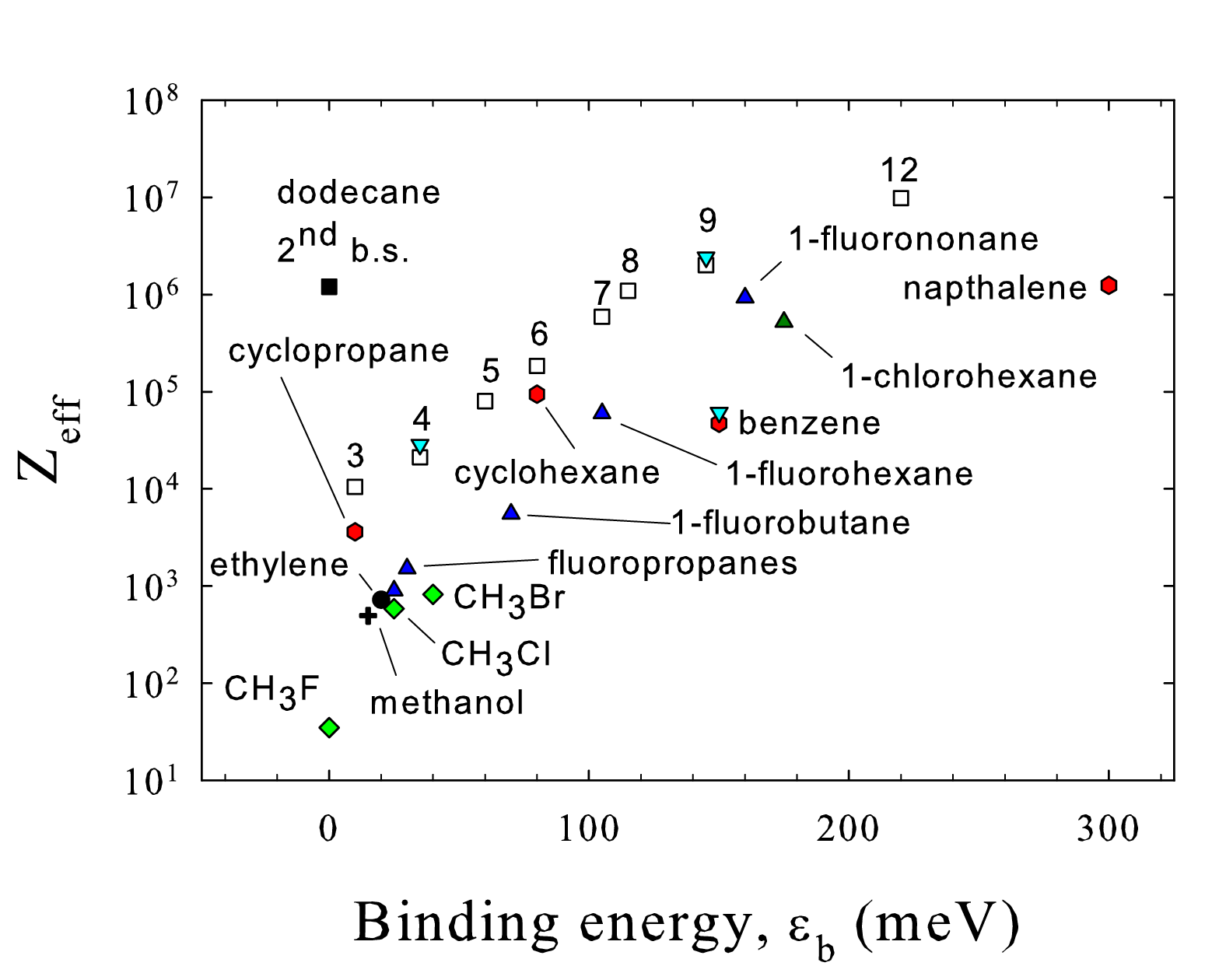}
\caption{$\Z$ at the C-H stretch peak \emph{vs} binding energy $\eb$
for alkanes, C$_{n}$H$_{2n+2}$ ($\square $), with the number of carbons $n$
indicated; rings (hexagons); halomethanes ($\diamond$); ethylene
($\bullet$); methanol (+); 1-chlorohexane ($\bigtriangleup$);
fluoroalkanes ($\bigtriangleup$);
deuterated species ($\bigtriangledown$).}
\label{fig:Zeff_bind}
\end{figure}

As shown in Fig.~\ref{fig:Zeff_bind}, positron-molecule
annihilation has been studied for a variety of chemical species.
While the vibrational modes and energy levels 
in these molecules differ in various ways, all of the hydrocarbons studied
contain strong C-H stretch vibrational modes that result in prominent
annihilation resonances. Consequently, they provide a convenient benchmark
to determine positron-molecule binding energies and the relative magnitudes
of the annihilation rates.

Relatively small changes in molecular structure can have significant 
effects on both the positron binding energies and the overall annihilation 
rates. For example, the C-H stretch peak in 1-chlorohexane is shifted 
downward relative to that of hexane by nearly 100~meV, while the magnitude
of the C-H stretch resonance is increased by nearly a factor 
of 3 \cite{YS08a}.

Based upon the alkane data, one might conclude that binding 
energy and the magnitudes of the annihilation resonances are strongly 
correlated \cite{YS08a}. However, as shown in Fig.~\ref{fig:Zeff_bind}, this 
hypothesis is not borne out by the data. With the exception of the alkanes, 
the binding energies and the magnitudes of the C-H stretch resonances 
appear to be only weakly correlated. For example, the $\Z$ 
values for 1-chlorohexane, benzene, and naphthalene are an order of 
magnitude or more smaller than those for alkanes with similar binding 
energies.

\begin{figure}
\includegraphics*[width=8cm]{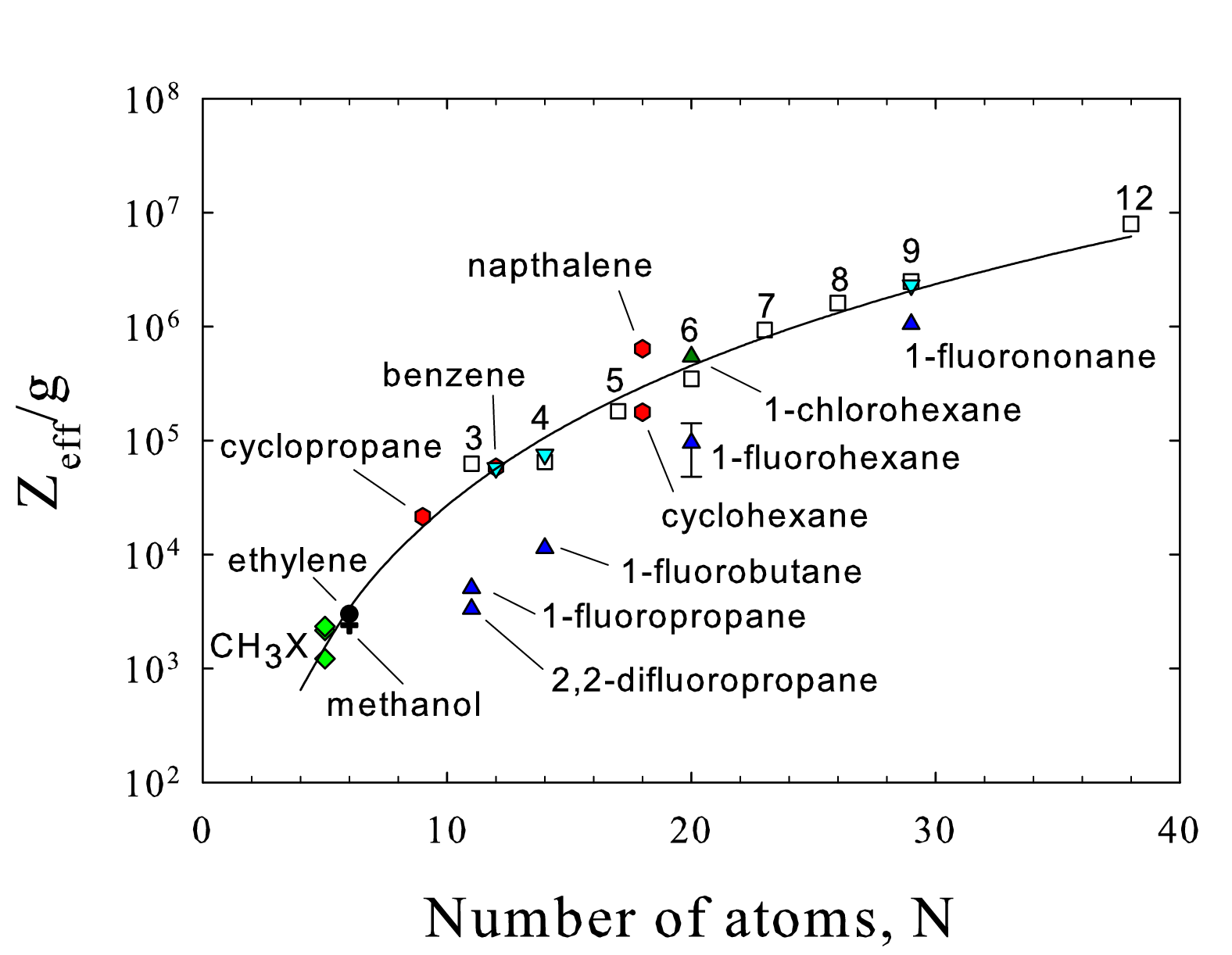}
\caption{$\Z$ at the C-H stretch peak, normalized by the factor
$g=\sqrt{\eb/\eps }$, \emph{vs} the number of atoms in the
molecule. The symbols are as in Fig.~\ref{fig:Zeff_bind}.
The solid line is the fit given by Eq.~(\ref{eq:Zeff_Nat}). The error
bars on 1-fluorohexane indicate the unusually large uncertainty
for this datum.}
\label{fig:Zeff_N}
\end{figure}

This lack of correlation of $\Z$  with binding energy motivated further
analysis of the $\Z$ magnitudes. Shown in Fig. \ref{fig:Zeff_N} and listed in
Table \ref{tab:Zeff_N} are data for $\Z$ the 
C-H stretch peak, normalized by $g$, as a function of the number of 
atoms $N$. With the exception of the partially fluorinated compounds,
which will be discussed below, the magnitudes of the annihilation rates
for the relatively wide variety of molecules studied
lie close to a universal curve. 1-chlorohexane and benzene are no longer
outliers to the extent they were in Fig.~\ref{fig:Zeff_bind}. The empirical
scaling of $\Z$ with $N$ is found to be
\cite{YS07,YS08a},
\begin{equation}\label{eq:Zeff_Nat}
\Z/g = 2.3N^q ,
\end{equation}
with $q =4.1\pm 0.1$. This scaling likely reflects the dependence 
of the total number of accessible positron-molecule vibrational states
(i.e., including dark states) on the number of vibrational degrees of freedom. 

\begingroup
\renewcommand{\arraystretch}{0.8}
\squeezetable
\begin{table}
\caption{Parameters for annihilation on various large molecules: the number
of atoms $N$; positron binding energies; $\Z$  at the C-H stretch mode and
for thermal distributions of positrons at 300~K; and the normalized rate,
$\Z^{\rm (CH)}/g$. The $\Z$ values have an overall uncertainty of
20\% (cf. Sec.~\ref{subsec:en_res_ann}).}
\label{tab:Zeff_N}
\begin{ruledtabular}
\begin{tabular}{lrrrrr}
Species  & $N$ & $\eb$\footnotemark[1]
& $\Z^{\rm (CH)}$\footnotemark[2] & $\Zth$\footnotemark[3]
& $\Z^{\rm (CH)}/g$ \\
 & & (meV) & & & \\
\hline
\multicolumn{6}{c}{Alkanes}\\
Methane  & 5&  $<0$  & -- & 142 & -- \\
Ethane & 8& $>0$& 900 & 660 & -- \\
Propane  & 11  & 10 & 10\,500 &  3\,500 & 63\,000 \\
Butane& 14  & 35  & 21\,000 & 11\,300 & 65\,000 \\
Pentane  & 17  & 60 & 80\,000 & 37\,800 &180\,000 \\
Hexane& 20  & 80  &184\,000 &120\,000 &350\,000 \\
Heptane  & 23  & 105  &590\,000 &242\,000 &930\,000 \\
Octane& 26  & 115 & 1\,090\,000 &585\,000 & 1\,610\,000 \\
Nonane  & 29  & 145& 2\,000\,000\footnotemark[4] & 643\,000 & 2\,500\,000 \\
Dodecane & 38  & 220& 9\,800\,000 & 1\,780\,000 & 8\,000\,000 \\
\hspace{15pt} 2nd B.S. & 38  & $>0$ & 1\,200\,000 &  -- &-- \\
Tetradecane & 44  & 260  & 11$x$\footnotemark[5]& -- & 6.8$x$\footnotemark[5]\\
\hspace{15pt} 2nd B.S. & 44  & 50 & 2.8$x$\footnotemark[5]
& -- & 7.0$x$\footnotemark[5]\\
Hexadecane  & 50  & 310 & 15$y$\footnotemark[5] & 2\,230\,000
& 6.4$y$\footnotemark[5]\\
\hspace{15pt} 2nd B.S. & 50  & 100 & 4.0$y$\footnotemark[5]
& -- & 6.5$y$\footnotemark[5]\\
\multicolumn{6}{c}{Alkane isomers}\\
Isopentane  & 17  & 60 & 80\,000 & 50\,500 &180\,000 \\
\multicolumn{6}{c}{Rings}\\
Cyclopropane& 9& 10  &  3\,600 & -- & 21\,500 \\
Cyclohexane & 18  & 80& 94\,000 & 20\,000 &180\,000 \\
Benzene  & 12  & 150& 47\,000 & 15\,000 & 58\,000 \\
Naphthalene & 18  & 300  & 1\,240\,000 &494\,000
&640\,000 \\
\multicolumn{6}{c}{Haloalkanes}\\
1-fluoropropane& 11  & 30  &  1\,520 & -- &  5\,100 \\
2,2-difluoropropane  & 11 & 25  & 900 &  8\,130 &  3\,300 \\
1-fluorobutane & 14  & 70&  5\,600 & -- & 11\,500 \\
1-fluorohexane & 20  & 105 & $60\pm 30\!\times \!10^3$ &269\,000
& $94\pm 47\!\times \!10^3$ \\
1-chlorohexane & 20  & 175  &520\,000 & -- &540\,000 \\
1-fluorononane & 29  & 160 &930\,000 & -- & 1\,050\,000 \\
\multicolumn{6}{c}{Deuterated}\\
d-benzene& 12  & 150 & 61\,000 & 36\,900 & 57\,500 \\
d-butane & 14  & 35  & 28\,500 & -- & 75\,000 \\
d-nonane & 29  & 145 & 2\,400\,000 &641\,000 &2\,300\,000\\
d-naphthalene  & 18  & $\sim 300$& -- & -- & -- \\
\end{tabular}
\end{ruledtabular}
\footnotetext[1]{Determined from the
energy shift of the C-H peak in $\Z$.}
\footnotetext[2]{At the C-H stretch peak
\cite{GBS02,BGS03,BYS06,YS07,YS08a}.}
\footnotetext[3]{For thermal positrons at 300~K
\cite{HCG82,IGM95,IGS94,IGS97,WCC83,Iwa97}.}
\footnotetext[4]{$\Z^{\rm (CH)}$ for nonane is from \textcite{YS08a} as
opposed to \textcite{BGS03}, as the latter had a narrower positron energy
distribution for this molecule.}
\footnotetext[5]{Absolute $\Z$ could not be determined,
so the values are multiplied by arbitrary factors $x$ and $y$.}
\end{table}
\endgroup

It is not surprising that $\Z$ depends linearly on
$g=\kappa /k$, so long as the weak binding picture (i.e., $\kappa \ll 1$ a.u.)
is valid. Furthermore, the $1/k$ factor arises from the normalization of
the incident positron wave function.
The surprising aspect of this $g$ scaling is that it appears to incorporate
the only dependence of $\Z$ on $\eb $. While the \emph{overall}
vibrational density of states is correlated with the number of atoms $N$,
it can certainly change without $N$ changing (e.g., by chemical substitution).
Further, small-molecule theory assumes that the lifetime of a 
vibrationally excited positron-molecule complex is short compared to the 
annihilation time. However, if the states accessible by IVR are 
especially long lived, the 
magnitudes of the resonances might then depend upon the capture rate 
rather than $g$. The annihilation rate would then be 
expected to saturate, growing only linearly with molecular 
size. The fact that the scaling of Eq.~(\ref{eq:Zeff_Nat}) remains valid for
the largest molecules studied to date is evidence that this
saturation limit has not yet been reached.

\subsection{Toward a model of annihilation in large molecules}

Our present theoretical understanding of the annihilation spectra of large 
molecules is incomplete. As discussed in Sec.~\ref{sec:theory},
it is not possible to explain the large values of $\Z$ that are observed
using only VFR involving the vibrational fundamentals. This difficulty could
possibly be overcome by considering mode-based resonances as doorways for
positron capture in complex multimode VFRs. However, estimates assuming
complete IVR show that $\Z$ would increase much faster with molecular size
than is observed (Sec. \ref{subsubsec:dens}). Further, when all combination
and overtone vibrations are assumed to couple to the positron continuum, the
resulting spectrum is predicted to be featureless, bearing little resemblance
to the experimentally measured spectra. Nevertheless, the presently available 
theoretical models do provide a useful framework with which to interpret some
of the experimental results. In the following, additional
experiments and analysis are discussed that elucidate details of 
VFR-mediated annihilation in large molecules and place constraints on 
viable theoretical models of this process.

\subsection{Inelastic autodetachment}\label{subsec:inel}

Inelastic escape channels are a potentially important mechanism that 
can limit the magnitudes of resonant annihilation peaks
(Sec.~\ref{subsec:large}). They can occur when a resonantly captured positron
is released 
from the molecule by the de-excitation of a vibration other than that produced
during the initial capture. Such vibrations can either be excited through IVR 
following resonant capture or by thermal excitation. This process is expected
to lead to a reduction in the annihilation rate, since the positron will
spend less time on the molecule.

An important consideration in such an inelastic process is how the 
resonant $\Z$ is affected by the binding energy. If the positron is in a 
weakly bound state, many vibrational modes will have sufficient 
energy to eject it; while for deeply bound states, fewer modes are 
able to do this. Thus, if inelastic escape channels were present, one 
would expect an additional dependence of the $\Z$ on $\eb$, beyond the scaling
with $g$. However, this contradicts a number of experimental 
results. For example, the C-H stretch-peak magnitudes for the first and 
second bound states of tetradecane and hexadecane strictly follow the
$g$ scaling with no additional dependence on $\eb $.
Furthermore, the $\Z$ at the C-H stretch peaks for nearly all molecules studied
adhere strictly the $\Z/g \propto N^{4.1}$ scaling, with relatively little
deviation due to their disparate binding energies.
This appears to rule out inelastic escape channels as being generally
important in determining $\Z$ values.

\subsubsection{Fluorine-substituted alkanes}\label{subsubsec:fluo}

There is, however, one notable exception. As shown in
Fig.~\ref{fig:Zeff_N}, 
partially fluorinated alkanes, such as the 1-fluoroalkanes, deviate 
\emph{significantly} from the scaling of Eq. (\ref{eq:Zeff_Nat}). The data
indicate that the substitution of a single fluorine atom for a terminal
hydrogen in a given alkane reduces the height of the C-H stretch peak in
$\Z$ by as much as an order of magnitude. The C-F stretch mode 
might be expected to play a significant role in this apparently inelastic 
process since the cross section for positron-impact excitation of the C-F
stretch vibration in CF$_4$ is unusually large \cite{MGS06,SGM02b}. 

Such an inelastic channel can significantly reduce the magnitude of $\Z$
by increasing the total post-capture escape rate. In this case
[cf. Eq. (\ref{eq:Zeff_fin})],
\begin{equation}\label{eq:Zeff_nu}
\Z (\nu) \propto \frac{\Gamma^e_\nu }
{\Gamma^a_\nu +\Gamma^e_\nu +\Gamma^i_\nu },
\end{equation}
where $\Gamma^i_\nu $ is the inelastic escape rate (e.g., via de-excitation
of mode $n$ with energy $\omega _n$), so that
increasing $\Gamma^i_\nu $ reduces $\Z$. This process requires that the
multimode state $\nu $ contains quasidegenerate components in which mode $n$
is excited. This restriction imposes a threshold at $\eps =\omega_n -\eb$,
above which $\Z$ is reduced.

\begin{figure}
\includegraphics*[width=8cm]{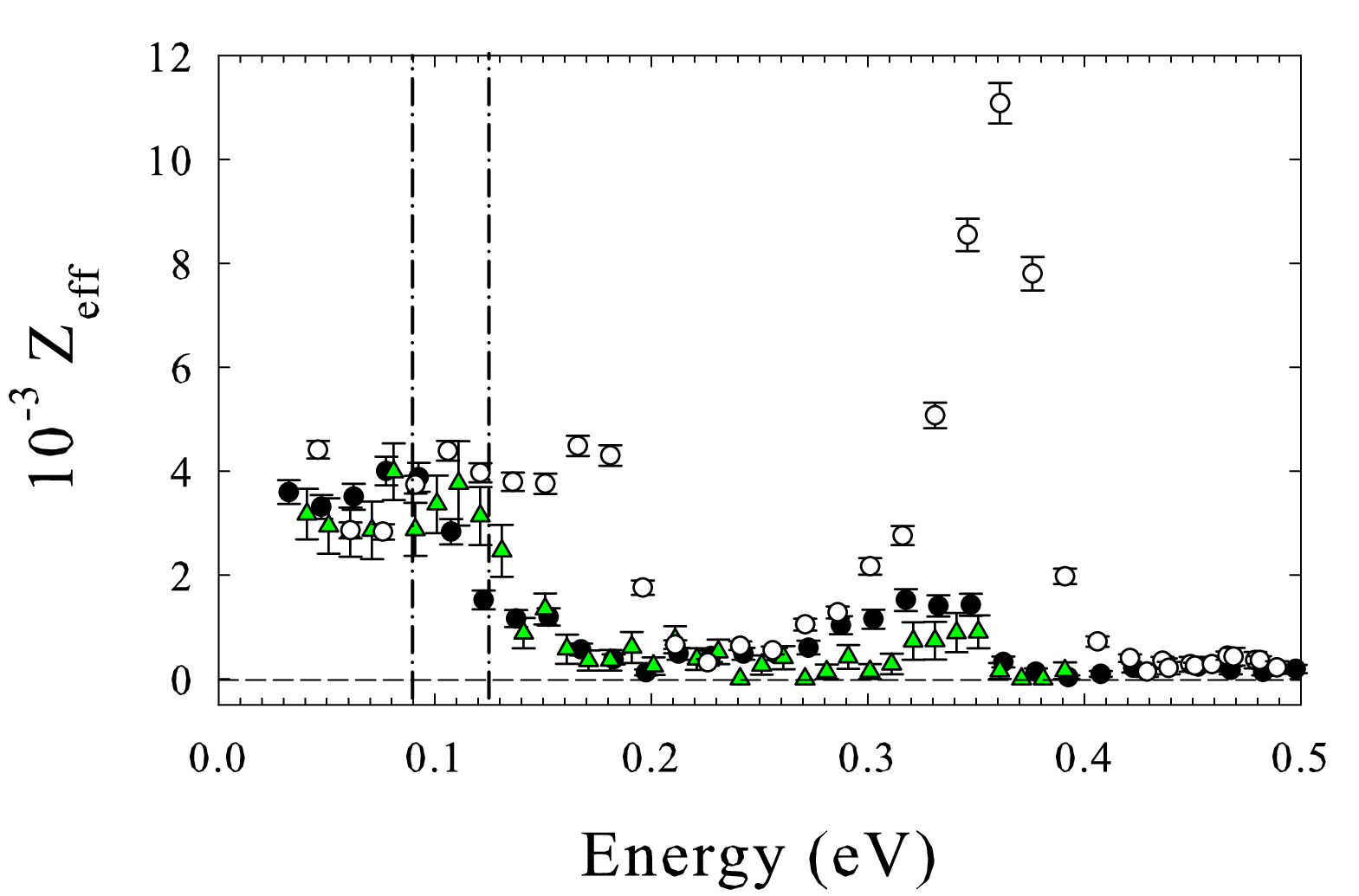}
\caption[$\Z$ for propane, 1-fluoropropane, and 2,2-difluoropropane]
{The $\Z$ spectra for ($\circ $) propane and ($\bullet$)
1-fluoropropane \cite{YS08a}, and ($\bigtriangleup $) 
2,2-difluoropropane \cite{GBS02}. The vertical dot-dashed
lines indicate the expected energies of C-F stretch resonances and escape
thresholds in each molecule, based on C-F stretch mode energies
from \textcite{GZD99,DGS81}. See text for details.}
\label{fig:fluoroprop}
\end{figure}

As shown in Fig.~\ref{fig:fluoroprop}, the $\Z$ spectra for
1- and 2,2-fluoropropane display just such a suppression at larger incident
positron energies as compared with the analogous hydrogenated compounds.
In 1-fluoropropane, the C-F stretch 
annihilation resonance is expected to occur at $\sim 90$~meV,
assuming $\eb = 30$~meV and $\omega_{\rm CF} \approx 120$~meV
\cite{GZD99}. In 2,2-difluoropropane, the C-F stretch peak is expected to
occur at $\sim 125$~meV, assuming $\eb = 25$~meV and
$\omega_{\rm CF} \approx 150$~meV \cite{DGS81}. In both cases, the suppression
of $\Z$ occurs near these threshold energies
(shown by dot-dashed lines in Fig. \ref{fig:fluoroprop}).

As discussed by \textcite{YS08a}, there are similar decreases in the
magnitudes of the $\Z$ at higher incident positron energies
in the larger partially fluorinated alkanes, 1-fluorobutane,
1-fluorohexane, and 1-fluorononane, relative to the analogous hydrogenated
molecules. However, there is 
also an {\em increase} in the magnitudes of the $\Z$ spectra at 
smaller positron energies, making the thresholds for escape-channel behavior
less distinct. The behavior of 1-fluorononane is
more difficult to explain. The positron binding energy of 160 meV in this
molecule would appear to preclude the C-F stretch mode
[$\omega_n\lesssim 127$~meV \cite{CL83}]
from acting as an inelastic escape channel. However, a decrease in
annihilation is observed at higher energies nonetheless \cite{YS08a}.
This is currently not understood.

\subsubsection{Effects of molecular temperature on $\Z$}\label{subsubsec:temp}

For molecules at finite temperatures, the energy for positron detachment
which suppresses $\Z$ can be supplied by pre-existing thermally excited 
modes rather than those excited in the attachment process. Thus $\Z$ might be 
expected to increase significantly with decreasing molecular 
temperature. At one point, it was thought this could explain the
empirical observation that $\Z \propto (2n + 2) \exp (\eb /k_BT)$
in alkane molecules \cite{Bar04}.
A qualitatively different effect of varying molecular temperature was
suggested by \textcite{NG05}. They proposed that vibrational excitation
and the associated change in molecular geometry could be required to induce
or increase positron binding to molecules, and this, in turn, could affect the
activation of annihilation resonances. In this model, one would expect that
increasing the molecular temperature should result in an increase in
$\Z$.

\begin{figure}
\includegraphics*[width=7.5cm]{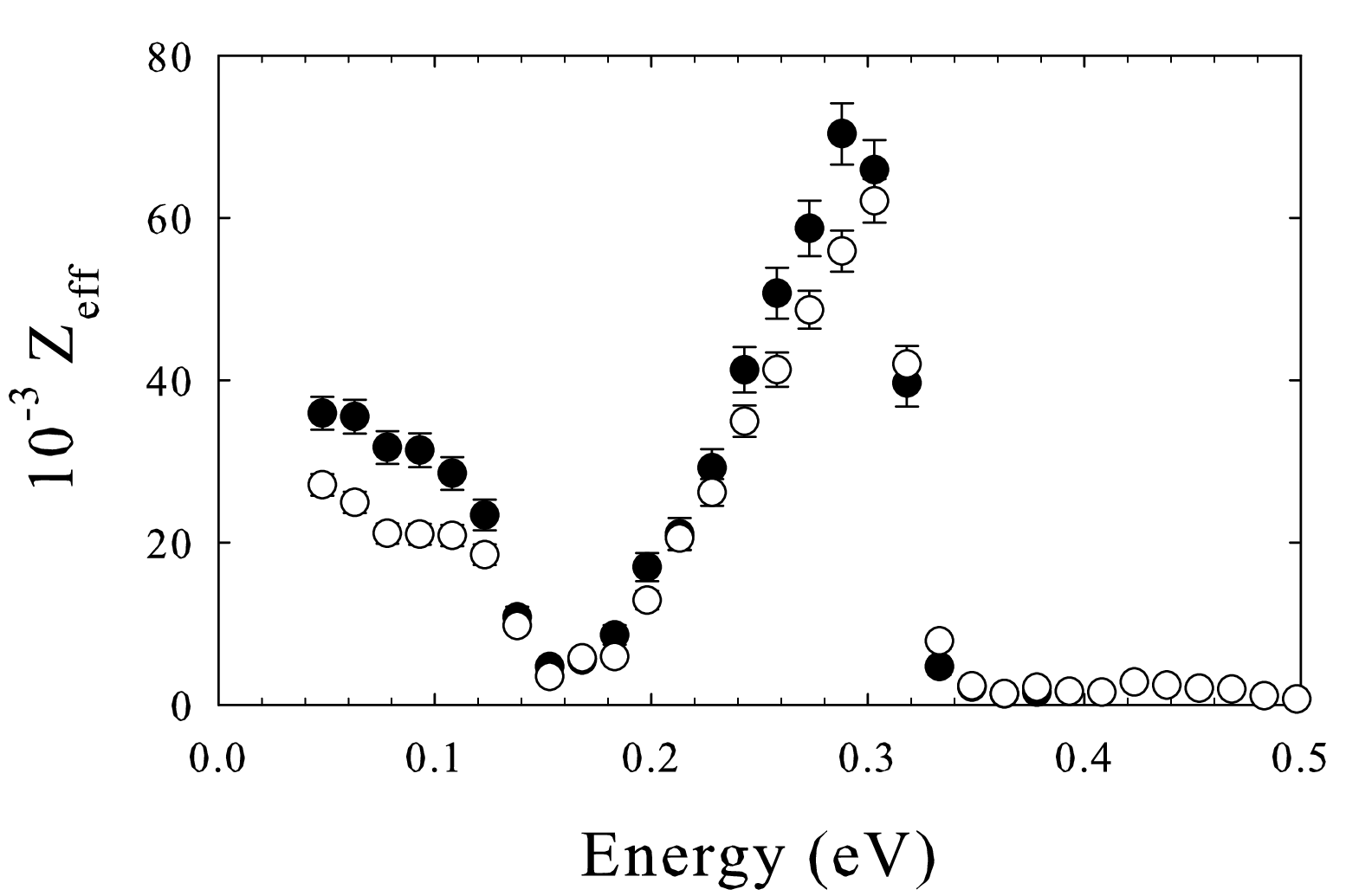}
\caption[Energy-resolved $\Z$ for cold pentane]
{Energy-resolved measurements of $\Z$ for pentane (C$_{5}$H$_{12}$) at
300 K ($\circ$) and 153 K ($\bullet$) using a cold cell \cite{YS08c}.}
\label{fig:Zeff_cold}
\end{figure}

Experiments were done to test these ideas using a specially constructed 
cold cell, so that $\Z$ spectra could be measured at different molecular
temperatures \cite{YS08c}. The apparatus is described in
Sec.~\ref{subsec:en_res_ann}. Care was taken to 
ensure that the test gas in this flowing-gas system actually cooled to the 
ambient temperature of the cold cell. In addition, the test-gas pressure 
was maintained a safe margin below the equilibrium vapor pressure at each 
temperature in order to avoid condensation on surfaces near and inside 
the cell. The spectra for pentane at 153 and 300~K are shown
in Fig.~\ref{fig:Zeff_cold}. There is only a small ($\sim $10\%) increase in
the magnitude of the C-H stretch resonance with the change
in molecular temperature, while at lower incident positron
energies, there is a somewhat larger increase in $\Z$ (i.e., $\sim $30\%).
Similar results were obtained for heptane at 195 and 300~K, but
the increase in the low-energy portion of the spectrum was somewhat larger
($\sim $50\%) \cite{YS08c}.
The spectra of both molecules indicate that their binding energies
do not change with changes in the molecular temperature.

These experimental results indicate that there is clearly no
Boltzmann-factor-like 
dependence of $\Z$, as that considered by \textcite{Bar04}. If there 
were, the pentane C-H stretch peak would have grown by a factor of 10.
These results also tend to rule out the suggestion, made
by \textcite{NG05} for C$_2$H$_2$, C$_2$H$_4$, and C$_2$H$_6$ that
thermal deformation of the molecular
bonds can enhance significantly the binding energy and hence increases the 
rate of VFR-mediated positron attachment and annihilation.
The data show the opposite trend, namely, that increasing the molecular
temperature leads to smaller values of $\Z$.

These findings also confirm other results indicating the absence of thermally
excited escape channels.
For thermally excited modes to provide effective escape, their
energies must exceed the positron binding energy. As the binding
energy increases (e.g., with alkane size), one would expect these channels
to be switched off. Hence thermally activated detachment would produce an
additional dependence of the resonant $\Z$ on $\eb$ beyond the $g$ factor.
Such an effect is not apparent in the alkanes,
where the empirical scaling is described by Eq.~(\ref{eq:Zeff_Nat}) for both
the positron ground and first excited states.

A question remains as to why the temperature effect appears to be stronger
at low impact energies than at high energies, yet the $g$-normalized spectra
for alkanes at 300~K (cf. Fig.~\ref{fig:alk_norm}) remain self-similar.

\subsection{Other IVR-related phenomena}\label{subsec:other}

While the $N^{4.1}$ scaling provides strong evidence of IVR, there are still
significant questions as to how the IVR, induced by positron capture,
proceeds in large molecules. We discuss here a few of the outstanding
issues.

The partially fluorinated alkanes
(Sec. \ref{subsubsec:fluo}) provide a clear example of post-capture
vibrational energy transfer, which suppresses the $\Z$ spectra due to
inelastic detachment. 
The so-called ``intermediate'' multimode state containing the excited C-F 
stretch mode is suggestive of a tiered IVR model, in which vibrational energy 
redistribution occurs incrementally through an ever-growing set of 
dark states \cite{NF96}. This could explain why only a few multimode
excitations involving the C-F stretch (out of all possible 
excitations) have such a disproportionate influence. A similar tiered model
has been used to describe laser-excited vibrational dynamics in
phenols \cite{YKE07} and to calculate IVR rates for acetylenic
stretch modes \cite{SM93}.

In general, the haloalkanes have larger values of thermal $\Z$ at 300~K
than their hydrogenated counterparts. There 
is also evidence that thermal values of $\Z$ for partially deuterated
benzenes and other substituted benzenes are similarly enhanced
\cite{IGM95}. These observed increases in
$\Z$ may be due to the increase in the density of vibrational ``dark'' states.
Similar physics may account for the large thermal $\Z$ values in CCl$_4$ 
and CBr$_4$ (9000 and 40\,000, respectively).

\begin{figure}
\includegraphics*[width=7.5cm]{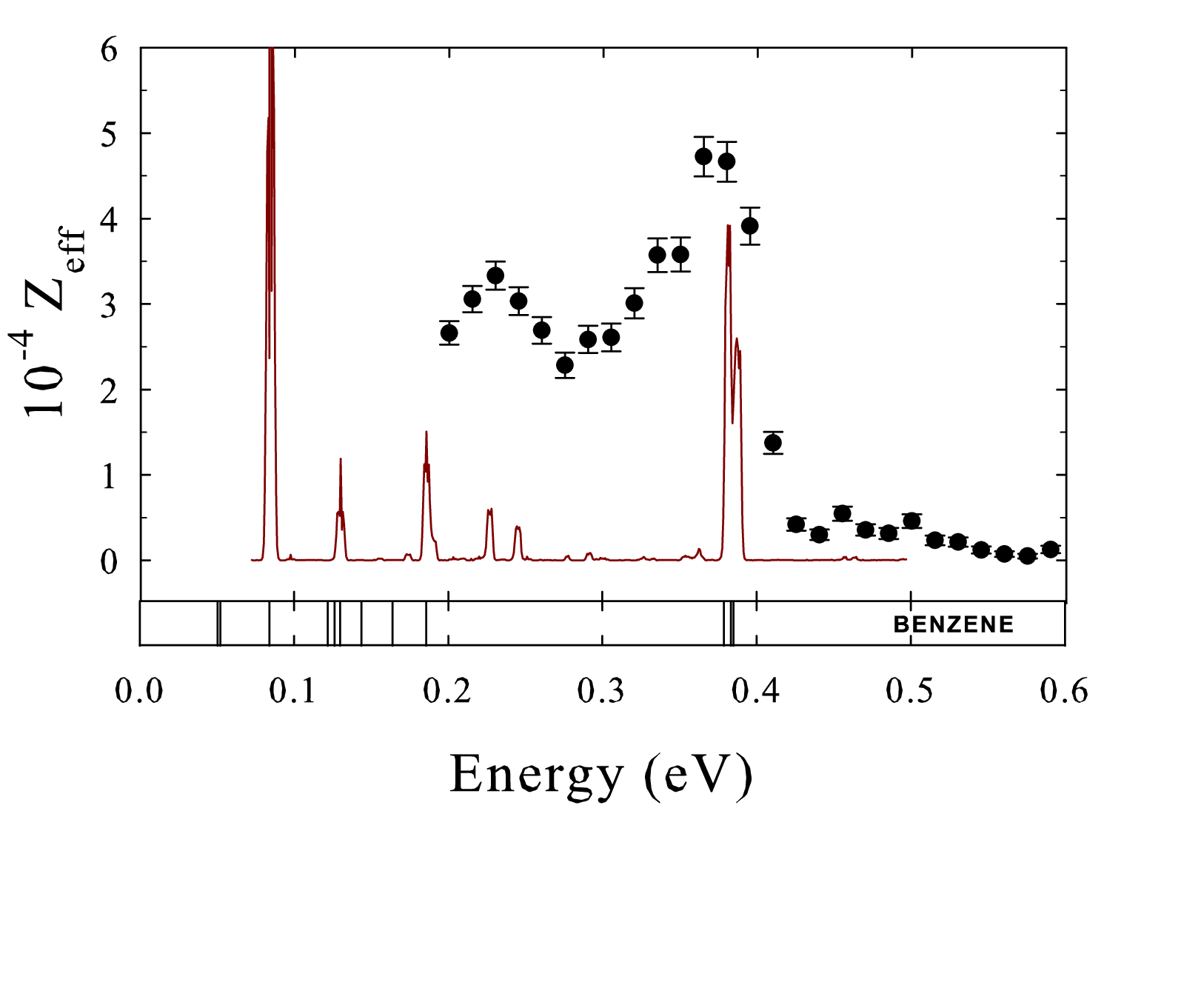}
\caption{Energy-resolved $\Z$ ($\bullet$) and infrared absorption (solid curve)
for benzene. The $\Z$ spectrum has been shifted upward by the binding energy
($\eb =150$~meV) for direct comparison. The normalization of the IR
absorption \cite{NIST} is arbitrary. Vertical lines indicate the positions
of the vibrational modes.}
\label{fig:benzene} 
\end{figure}

It is not clear at present what makes a ``good'' vibrational doorway 
state. In most large hydrocarbons, it is only the fundamental vibrations 
that appear to produce VFR; however, there are exceptions. 
Shown in Fig.~\ref{fig:benzene} is the $\Z$ spectrum of benzene, shifted
upward by its binding energy and compared with the infrared absorption
spectrum. Note the distinct peak at $\sim $235~meV in the shifted $\Z$
spectrum. While there are no nearby fundamental vibrations, there are two
IR-active combination vibrations at 227 and 244~meV in the IR
spectrum. Thus, the additional peak in $\Z$ appears to be evidence of
unusually strong capture into multimode doorways, likely enhanced by IVR.

\begin{figure}
\includegraphics*[width=7.5cm]{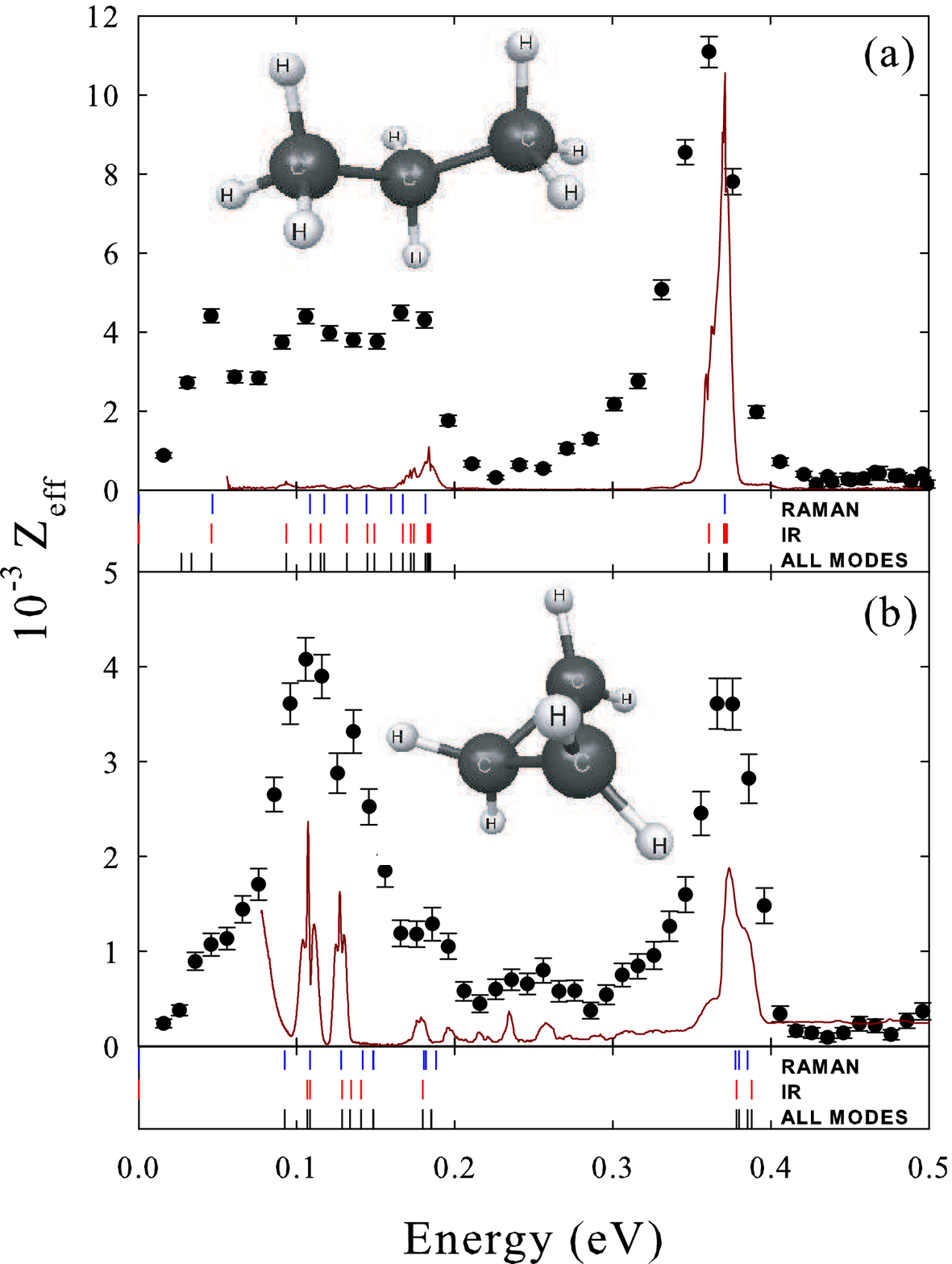}
\caption{Energy-resolved $\Z$ ($\bullet$) for (a) propane and (b) cyclopropane
\cite{BYS06}. The solid curves are the normalized IR
absorption spectra, and the vertical bars below each plot show the
vibrational modes \cite{NIST}. Shown as insets are the molecular
structures.}
\label{fig:cyclop}
\end{figure}

In elucidating the role of IVR it is of interest to compare propane and
cyclopropane. Figure~\ref{fig:cyclop} shows that transforming the propane
molecule into a ring reduces the C-H stretch peak by a factor of 3,
approximately in agreement with the $N^4$ scaling in Eq.~(\ref{eq:Zeff_Nat}).
At the same time the plateau at lower energies
is narrowed so that only one broad peak remains. However,
its magnitude in cyclopropane is practically the same as in propane.
Thus the cyclopropane spectrum does not
obey the self-similar scaling observed in the alkanes. 
It is possible that the low-energy peak contains mode-based
VFRs, with little enhancement due to IVR. The reduction in the magnitude
of the C-H peak may be due to the absence of low-frequency modes in
cyclopropane. This is consistent with
Table~\ref{tab:dens} which shows that cyclopropane has a markedly lower
vibrational density at the C-H stretch energy as compared with propane.

As shown in Fig.~\ref{fig:cyclop}, there is an identifiable feature in
cyclopropane at $\sim 250$~meV
(i.e., in the gap between the C-H stretch and lower-energy modes)
that does not appear in propane or larger hydrocarbons.
It occurs in the energy range where there are peaks in the IR spectrum and it
is likely due to combination and overtone vibrations (similar to that observed
in benzene). This again points to the possibility that appreciable IR
coupling is a predictor of the strength of multimode doorways in large
molecules. In cyclopropane this peak is not particularly enhanced
(e.g., relative to the C-H stretch resonance), and looks closer to the
effects due to combination and overtone VFRs observed in small molecules (see
Sec.~\ref{sec:small}, e.g., ethylene, Fig.~\ref{fig:c2h4c2h2}).
This phenomenon also bears further scrutiny.

\subsection{Large molecule summary}

There are several defining characteristics of the $\Z$ spectra of large 
molecules. They exhibit a series of peaks, the positions of which bear 
close resemblance to those of the fundamental vibrations, downshifted by 
the positron-molecule binding energy. The amplitudes of these 
resonances grow rapidly with molecular size but exhibit a relatively weak 
dependence on binding energy and incident positron energy {\em via} the 
factor $g$. The amplitudes of the C-H stretch resonances in 
hydrocarbons obey the empirical scaling of Eq.~(\ref{eq:Zeff_Nat}). 
This scaling likely reflects the rapid increase in the number of
vibrational degrees of freedom of the molecule with molecular size.
It suggests that the VFR are enhanced by the IVR process. However, the
extent of this IVR appears to be limited and does not reach the
statistically complete final state. 
Inelastic escape channels appear to be relatively unimportant, at least
in the hydrocarbons studied thus far.  

Our theoretical understanding of annihilation processes in large molecules
is not nearly as well developed as it is for small molecules. Qualitatively,
a positron appears to excite a vibrational fundamental (or in some cases,
such as benzene, a 
combination or overtone)  and populates a doorway resonance. The 
vibrational energy imparted to the molecule can then flow to some set of 
quasidegenerate dark states.
However, if this picture is correct, there must be an operative escape
channel. If not, all VFR will give approximately the same contribution
to the annihilation rate. Considering the available experimental evidence,
it seems plausible that the resonant $\Z$ magnitudes are determined by a 
balance between diffusion to some (limited) set of dark states,
and positron ejection through either the doorway state that it entered
or other nearby doorway states in what might be termed a ``quasielastic''
escape process.

Beyond this qualitative picture, there are an assortment of relatively poorly
understood effects. The most prominent of them is arguably the self-similar
spectra observed in alkanes. There is also the occasional occurrence 
of multimode VFR, and the inelastic detachment observed in partially
fluorinated alkanes. These effects provide tantalizing 
hints of the many types of complex internal dynamics that can be 
responsible for VFR-enhanced annihilation depending upon the 
particular situation. At a minimum, further investigations are warranted. 
For example, experiments with higher positron energy resolution would 
be useful to determine additional details of the processes involved. 

\section{Positron-molecule binding energies}\label{sec:bind}

\subsection{Relation to molecular properties}\label{subsec:molprop}

As described in Secs.~\ref{sec:small} and \ref{sec:large}, positron-molecule
binding energies have now been measured or estimated for about 30
molecules \cite{GBS02,BGS03,BYS06, YS08a,YS08b}. 
These data were analyzed recently by \textcite{DYS09} with a view toward
determining which molecular parameters affect $\eb$. Table~\ref{tab:bind}
lists the available binding energies together with molecular dipole
polarizabilies $\alpha _d$, dipole moments $\mu $, and ionization
energies $E_i$.

\begingroup
\renewcommand{\arraystretch}{0.8}
\squeezetable
\begin{table}[ht]
\caption{Positron binding energies, normalized thermal annihilation
rates $\Zth /Z$, and physical parameters for selected molecules.}
\label{tab:bind}
\begin{ruledtabular}
\begin{tabular}{llccccccc}
Name & Formula & $\eb$\footnote{Measured in energy-resolved
annihilation experiments \cite{YS08a,YS08b}.} &
$\eb$\footnote{Predicted by Eq.~(\ref{eq:fit}).} & $Z$ &
$\Zth /Z$\footnote{$\Zth$ is measured at
room temperature with thermalized positrons \cite{IGM95,Iwa97}.}
& $\alpha _d$\footnotemark[4] & $\mu $\footnotemark[4] &
$E_i$\footnotemark[4] \\
 & & (meV) & (meV) & & & (\AA$^3$) & (D) & (eV) \\
\hline
\multicolumn{9}{c}{Alkanes and related molecules} \\ 
Ethane & C$_2$H$_6$ & $>0$ & $-14$	& 18 &  37 & 4.4 & 0 & 11.5 \\
Propane & C$_3$H$_8$ & 10 & 10 & 26 & 135 & 6.3 & 0.1 &	11.1 \\
Butane & C$_4$H$_{10}$ & 35 & 31 & 34 & 330 & 8.1 & 0 & 10.6 \\
Pentane	& C$_5$H$_{12}$	& 60 & 54 & 42 & 900 & 10.0 & 0 & 10.4 \\
Hexane & C$_6$H$_{14}$ & 80 & 77 & 50 & 2400 & 11.8 & 0 & 10.2 \\
Heptane & C$_7$H$_{16}$ & 105 & 100 & 58 & 4200 & 13.7 & 0 & 9.9 \\
Octane & C$_8$H$_{18}$ & 115 & 123 & 66 & 8800 & 15.5 & 0 & 10.0 \\
Nonane & C$_9$H$_{20}$ & 145 & 146 & 74 & 8700 & 17.4 & 0 & 10.0 \\
Dodecane & C$_{12}$H$_{26}$ & 220 & 214 & 98 & 18000 & 22.9 & 0 & 9.9 \\
Tetradecane & C$_{14}$H$_{30}$ & 260 & 261 & 114 & -- & 26.6 & 0 & 9.9 \\
Hexadecane & C$_{16}$H$_{34}$ & 310 & 306 &130 & -- & 30.3 & 0 & 9.9 \\
Butane-d10 & C$_4$D$_{10}$ & 35 & 31 & 34 & -- & 8.1 & 0 & -- \\
Nonane-d20 & C$_9$D$_{20}$ & 145 & 146 & 74 & 8700 & 17.4 & 0 & -- \\
Acetylene & C$_2$H$_2$ & $>0$ & $-28$ & 14 & 230 & 3.3 & 0 & 11.4 \\
Ethylene & C$_2$H$_4$ & 20 & $-17$ & 16 &  75 & 4.2 & 0 & 10.5 \\
Isopentane & C$_5$H$_{12}$ & 60 & 57 & 42 & 1200 & 10.0 & 0.1 & 10.3 \\
Cyclopropane & C$_3$H$_6$ & 10 & 0.7 & 24 & -- & 5.7 & 0 & 9.9 \\
Cyclohexane & C$_6$H$_{12}$ & 80 & 68 & 48 & 420 & 11.1 & 0 & 9.9 \\
\multicolumn{9}{c}{Aromatics} \\
Benzene & C$_6$H$_6$& 150 & 149 & 42 & 360 & 10.4 & 0 & 9.3 \\
Benzene-d6 & C$_6$D$_6$ & 150 & 149 & 42 &  730 & 10.4 & 0 & 9.3 \\
Naphthalene & C$_{10}$H$_8$ & 300 & 296 & 68 & 7300 & 16.6 & 0 & 8.2 \\
\multicolumn{9}{c}{Alcohols}\\
Methanol & CH$_3$OH & 2 & 5 & 18 & 84 & 3.3 & 1.7 & 10.9 \\
Ethanol & C$_2$H$_5$OH & 45 & 27 & 26 & -- & 5.1 & 1.7 & 10.5 \\
\multicolumn{9}{c}{Partially halogenated hydrocarbons}\\
Methyl fluor. & CH$_3$F & $>0$ & $-3$ & 18 & 77 & 2.4 & 1.85 & 12.9 \\
Methyl chlor. & CH$_3$Cl & 25 & 23 & 26 & 580 & 4.4 & 1.9 & 11.2 \\
Methyl brom. & CH$_3$Br & 40 & 35 & 44 & -- & 5.6 & 1.8 & 10.5 \\
1-fl.propane & C$_3$H$_7$F & 30 & 45 & 34 & -- & 6.0 & 2.0 & 11.3 \\
2,2-difl.prop. & C$_3$H$_6$F$_2$ & 25 & 51 & 42 & 190 & 5.9 & 2.4 & 11.4\\
1-fl.butane & C$_4$H$_9$F & 70 & 27\footnotemark[5]
& 42 & -- & 7.8 & -- & -- \\
1-fl.hexane & C$_6$H$_{13}$F & 105 & 73\footnotemark[5]
& 58 & 46000 & 11.5 & -- & -- \\
1-fl.nonane & C$_9$H$_{19}$F & 160 & 141\footnotemark[5]
& 82 & -- & 17.0 & -- & -- \\
1-chl.hexane & C$_6$H$_{13}$Cl & 175 & 138 & 66 & -- & 13.6 & 2.0 & 10.3
\end{tabular}
\end{ruledtabular}
\footnotetext[4]{Polarizabilities, dipole moments and ionization energies
from \textcite{CRC,Mil90,McC63}.}
\footnotetext[5]{Lacking values of $\mu$
for these species, the predictions of Eq.~(\ref{eq:fit}) are lower bounds on
$\eb$.}
\end{table}
\endgroup

\begin{figure}[ht]
\includegraphics*[width=8.5cm]{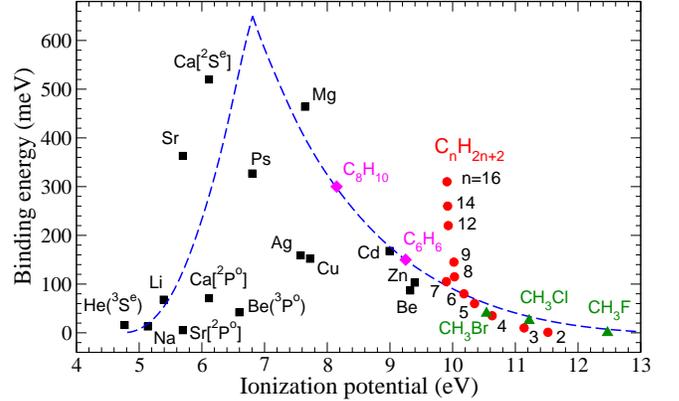}
\caption{Binding energies of positron-atom and  positron-molecule
complexes as a function of their ionization potential. Calculations: squares,
various atoms \cite{MBR02,BM10}; dashed curve, model alkali atom \cite{MBR99}.
Experiment: circles, alkanes with $n$ carbons; diamonds, the aromatic
molecules, benzene and naphthalene; triangles, methyl halides
(see Secs.~\ref{sec:small}, \ref{sec:large} and \ref{sec:bind}).}
\label{fig:atmolIP}
\end{figure}

Figure~\ref{fig:atmolIP} shows the experimental binding energies for
alkanes, aromatic molecules, and methyl halides and calculated
binding energies for several atoms, as a function of their ionization
potential. 
The magnitudes of $\eb $ calculated for atoms and measured for molecules
are quite similar (i.e., $\eb \le 0.5$~eV).
The binding energies for smaller alkanes, aromatics, and methyl halides
follow a ``model atom'' curve \cite{MBR99}. However, the ionization energies
for alkanes with more than $n=7$ carbons remain practically constant, while
the positron binding energies continue to grow with $n$.

\begin{figure}[ht]
\includegraphics*[width=7cm]{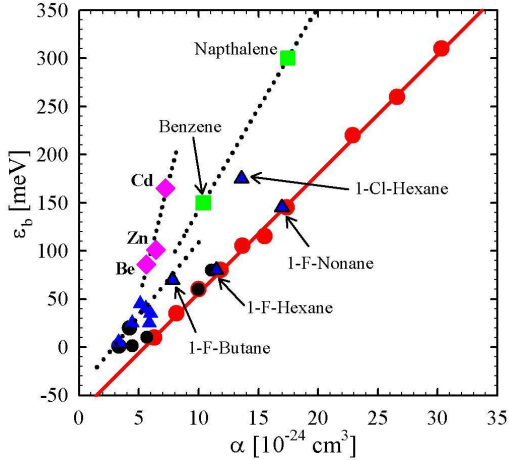}
\caption{Measured positron binding energies $\eb$ as a
function of the dipole polarizability $\alpha _d$:
large circles, alkane molecules used in the linear fit shown by the solid line;
smaller circles, alkane-related
molecules; triangles, molecules with permanent dipole moments; squares,
aromatics with $\pi $ bonds. Dotted lines, guides to show the linearity of
$\eb$ for the different series; diamonds, calculated $\eb $ for the
atoms Be, Zn and Cd \cite{MBR02}, shown for comparison.}
\label{fig:bind_alpha}
\end{figure}

The dipole polarizability $\alpha _d$ characterizes the strength of
the long-range positron-molecule attraction.
Figure~\ref{fig:bind_alpha} shows positron binding energies for molecules
as a function of $\alpha _d$. Also shown are the theoretical
values for three atoms, Be, Zn, and Cd \cite{MBR02}. This figure
suggests an approximately linear relationship between $\eb$ and $\alpha _d$
for groups of homologous species. Based on this, the analysis of
\textcite{DYS09} began with a linear fit of $\eb$ to $\alpha _d$ for
alkanes.
However, as seen in Fig.~\ref{fig:bind_alpha}, this fit generally
underestimates the binding energies for other classes of molecules,
most notably those with permanent dipole moments and aromatic molecules.
This motivated the inclusion of two additional parameters in the analysis,
namely the molecular dipole moment $\mu $ and the number
of $\pi $ bonds $N_\pi$ for aromatic molecules. Such use of the bonds
is similar to the approach used by \textcite{Mil90} to
parametrize the molecular polarizability.

\begin{figure}[ht]
\includegraphics*[width=7cm]{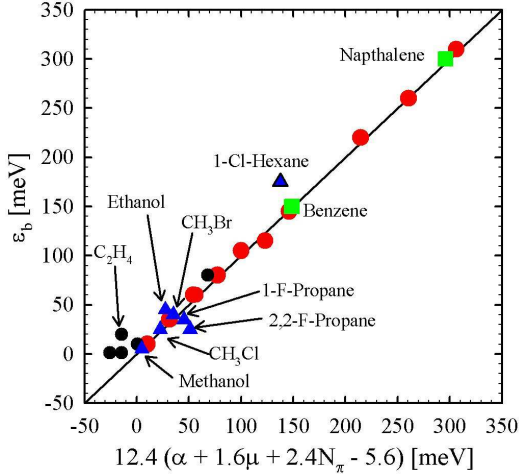}
\caption{Binding energy from the fit Eq.~(\ref{eq:fit})
(solid line) using the polarizability $\alpha _d$,
dipole moment $\mu $, and the number of $\pi $ bonds $N_\pi$ for
aromatic molecules. Symbols as in Fig.~\ref{fig:bind_alpha}.}
\label{fig:fit}
\end{figure}

The resulting fitting function, using the numerical values of the molecular 
parameters as listed in Table~\ref{tab:bind}, is \cite{DYS09},
\begin{equation}\label{eq:fit}
\eb = 12.4 ( \alpha + 1.6 \mu + 2.4 N_\pi - 5.6 )\quad [{\rm meV}],
\end{equation}
where $\eps_ b$ is in meV, $\alpha _d$ is in units of \AA$^3$, and 
$\mu$ is in units of debye. This expression can be viewed as a lowest-order 
Taylor expansion of $\eps_ b$ as a function of the variables $\alpha _d$,
$\mu$, and $N_\pi$.
This fit to the binding energy data is shown in Fig.~\ref{fig:fit}.
Generally, the agreement between the predictions of Eq.~(\ref{eq:fit}) and
the measurements is quite good. The most significant discrepancies are
1-chlorohexane, acetylene (C$_2$H$_2$), ethylene (C$_2$H$_4$), and
2,2-difluoropropane. For all but 2,2-difluoropropane, Eq.~(\ref{eq:fit})
underestimates the binding energy. These molecules probably possess
some additional attraction beyond that described by Eq.~(\ref{eq:fit}).
With the exception
of 1-chlorohexane and 2,2-difluoropropane, the outliers all have double and
triple bonds. The addition of the $N_\pi$ term in Eq.~(\ref{eq:fit}), which
improves agreement for the aromatic species, overestimates $\eb$ for
these smaller molecules. Thus, while a similar effect may well be
operating in small molecules with $\pi $ bonds, its magnitude appears
to be smaller than in the aromatics.

According to Eq.~(\ref{eq:fit}), binding is assured if either
$\alpha _d> 5.4$~\AA$^3$ or $\mu > 3.6$ D. The first of these conditions
is similar to that for positron binding to a model ``alkali atom'',
namely $\alpha _d > 3.5$~\AA$^3$ \cite{MBR99}. The second condition can be
compared with the theoretical critical value $\mu _c= 1.625$~D, beyond which
the dipole supports an infinite series of bound states \cite{Cra67}.
Probably more relevant is a ``practical'' value $\mu _c\approx 2.5$~D
required to obtain binding energies $\geq 1$~meV in negative ions
(i.e., electron-molecule bound states) \cite{AD98}; see  below.
These comparisons show that the threshold values of $\alpha _d$ and
$\mu $ from Eq.~(\ref{eq:fit}) appear to be quite reasonable.

Another feature of Eq.~(\ref{eq:fit}), which is supported by observations
(Table~\ref{tab:bind}), is that it predicts that the binding energies for
fully deuterated hydrocarbons are close to their hydrogenated analogs
(cf. deuterated butane, benzene, and nonane) and that isomers have
similar values of $\eb$ (e.g., pentane vis-\'a-vis isopentane).

For the species studied to date, the largest binding energies for a given 
number of atoms are seen in the aromatic molecules. A study of 
larger species (i.e., polycyclic aromatic hydrocarbons), 
such as the three- and four-ring variants anthracene and pyrene, will 
be of interest. If $\eb $ increases beyond the 
energy of the C-H stretch mode (i.e., the highest-energy mode), the resonant
energy threshold in these molecules becomes negative and the VFR of
the vibrational fundamentals become inaccessible to the positron.
This has already been observed in deuterated naphthalene \cite{YS08a}.
Nevertheless, the annihilation spectra for these molecules may still exhibit
VFRs associated with positronically excited states, or those where the positron
excites higher-lying overtone and combination vibrations.

For molecules with similar chemical composition, $\alpha_d$ increases with
molecular size. However, the linear increase in $\eb$ with $\alpha_d$
predicted by Eq.~(\ref{eq:fit}) is expected to saturate at some point
(e.g., when the positron de Broglie wavelength becomes smaller than the size
of the molecule). The data in Figs.~\ref{fig:bind_alpha}
and \ref{fig:fit} indicate that the experiments have not yet reached
this limit.

The predictions of Eq.~(\ref{eq:fit}) can also be compared with available
calculations for positron binding to molecules. All of the molecules
listed in Table~\ref{tab:fit_cal} are predicted to bind positrons in
agreement with the theoretical calculations. The absolute values of $\eps_ b$
predicted by Eq.~(\ref{eq:fit}) for HCN and formaldehyde (H$_2$CO) agree to
within a factor of 2 with the calculations. For urea and acetone
Eq.~(\ref{eq:fit}) predicts much larger values than the calculated $\eb$. 
This confirms the expectation \cite{Str04} that the binding
energies calculated by \textcite{TBK03} for urea and acetone are
underestimated.

\begin{table}[ht]
\caption{Comparisons of theoretical predictions for positron-molecule binding 
energies with those of Eq.~(\ref{eq:fit}).}
\label{tab:fit_cal}
\begin{ruledtabular}
\begin{tabular}[t]{llcccc}
Name & Formula & $\alpha _d$ & $\mu$ &
$\eb$\footnotemark[1] &
$\eb$\footnotemark[2] \\
 & & (\AA$^3$) & (D) & (meV) & (meV) \\
\hline
Hydr. cyanide & HCN & 2.5 & 3.0 & 21 & 35 \\
Formaldehyde & H$_2$CO & 2.8 & 2.3 & 12 & 19 \\
Urea & (NH$_2$)$_2$CO & 9.7 & 4.6 & 140 & 13  \\
Acetone & (CH$_3$)$_2$CO & 6.4 & 2.9 & 66 & 4 \\
Lith. hydride & LiH & 3.8 & 5.9 & 94 & 1000
\end{tabular}
\end{ruledtabular}
\footnotetext[1]{Predicted from Eq.~(\ref{eq:fit}).}
\footnotetext[2]{Values from quantum-chemistry calculations:
HCN \cite{CS06}, H$_2$CO \cite{Str04}, urea and acetone
\cite{TBK03}, and LiH \cite{MMB00,Str01,BA04}.}
\end{table}

For LiH, the calculated $\eb $ is ten times greater than
the prediction of Eq.~(\ref{eq:fit}). This is also true for other
alkali hydrides, NaH, KH and RbH, where the calculations give
$\eb \gtrsim 1$ eV \cite{BLM05,GFB06}. This discrepancy is due to the fact
that Eq.~(\ref{eq:fit}) is based on binding energies
for molecules with relatively large ionization energies. 
The physical picture of such bound states gained from positron calculations for
atoms \cite{MBR02}, describes them as the positron moving in the attractive
potential of the neutral molecule. In contrast, the ionization energy of
LiH and the other alkali hydrides is close to the binding energy of the
Ps atom (e.g., $E_i - E_{\rm Ps} =1.1$ eV for LiH). In this case, the
calculations indicate that the relevant physical picture is closer to
that of a PsH complex attached to the positively charged metal
ion \cite{BLT06,BLM05}; cf. Fig.~\ref{fig:LiH}. Thus it is not surprising
that there is a discrepancy between the predictions of the Eq.~(\ref{eq:fit})
and calculations of $\eb$ for LiH.

Equation~(\ref{eq:fit}) can be used to identify candidate molecules for further 
theoretical and experimental studies. Theoretical calculations favor
molecules with small numbers of atoms and simple electronic structures.
Experimental studies require species with vapor number densities
$\ge 10^{-7}$~amagat at moderate temperatures (e.g., $T \le 500$ $^\circ $C)
that are stable at this temperature and not deleterious
to the vacuum system. Most convenient for study are small molecules with
relatively large binding energies (e.g., $\eb \geq 20$~meV).

Recently \textcite{DGS10} measured $\eb $ for several
molecules of this type. For carbon disulfide CS$_2$ ($\alpha _d=8.8$~\AA$^3$
and $\mu =0$), the measured value $\eb =75$~meV is a factor of 2 greater than 
40~meV, predicted by Eq.~(\ref{eq:fit}) neglecting $\pi $ bonds. Other
molecules studied, which were chosen for their simplicity and relatively
large dipole moments, include acetaldehyde CH$_3$CHO ($\alpha _d=4.6$~\AA$^3$,
$\mu =2.75$~D, and $\eb =90$~meV), acetone (CH$_3$)$_2$CO
($\alpha _d=6.4$~\AA$^3$, $\mu =2.9$~D, and $\eb =173$~meV), and acetonitrile
CH$_3$CN ($\alpha _d=4.4$~\AA$^3$, $\mu =3.9$~D, and $\eb =180$~meV). All three
have binding energies significantly larger than those predicted by
Eq.~(\ref{eq:fit}). In this previously unexplored regime in
which $\mu >2$~D, the dependence on both $\mu $ and $\alpha _d$ is much
stronger than that given by Eq.~(\ref{eq:fit}). These data provide a
significant new opportunity to make quantitative comparisons between theory
and experiment. It is hoped that they will stimulate new theoretical
calculations. Thus a recent CI calculation for acetonitrile by \textcite{TKB10}
predicted $\eb =135$~meV, which is within 30\% of the experimental value.

Equation~(\ref{eq:fit}) could also be useful in
describing the behavior of low-energy positrons in a wide range of chemical
environments, including biological systems (e.g., in PET analysis) and in
materials science in conjunction with studies of insulators using techniques
such as ACAR and positron annihilation lifetime spectroscopy.

\subsection{Comparison with negative ions of molecules and
clusters}\label{subsec:negions}

Unlike positron-molecule complexes, negative molecular ions have been
studied extensively. They can be separated into two classes.
There is a class of stable molecular anions with large
electron affinities $\sim 1$--3~eV \cite{RTS02}. In these anions, the
excess electron occupies an unfilled orbital in the valence shell of the
molecule, and they have no positron analog.

A second arguably less well-studied class of negative ions
is closely related to the positron-molecule complexes discussed here
\cite{DGS04,ASD02,AD98,DAK94}. In these ``dipole-bound anions,'' the electron
is only weakly bound to the target by a combination of dipolar,
quadrapolar, and polarization forces \cite{GJS98,AD98}. Because of the
small binding energy, the wave function of the excess electron is diffuse,
residing predominantly outside the molecule due to the Pauli exclusion
principle. In the positron-molecule case, the analogous short-range repulsion
is due to the positive charges of atomic cores.
The minimum dipole moment required for the formation of stable anions of
common closed-shell molecules has been determined experimentally to be
2--2.5~D \cite{DAK94}, in qualitative agreement with the critical dipole
moment of 3.6~D, predicted by Eq.~(\ref{eq:fit}).

Table~\ref{tab:neg} lists the measured and calculated binding energies and
relevant molecular parameters for a selection of these dipole-bound
anions \cite{AD98}. This table also lists the positron binding energies
from quantum-chemistry calculations (where available) and estimated
from Eq.~(\ref{eq:fit}). Positron binding energies are generally
greater than their electron counterparts. This could be related to an
electron-positron correlation effect known as virtual Ps formation, which
contributes significantly to the positron-atom attraction \cite{DFG95,GL04}.
One consequence of this is that positrons are predicted to bind to atoms such
as Mg, Zn, and Cd, which do not form stable negative
ions \cite{DFG95,MBR02}.

\begingroup
\renewcommand{\arraystretch}{0.8}
\squeezetable
\begin{table}[ht]
\caption{Measured and calculated electron binding energies and
positron binding energies (in meV).}
\label{tab:neg}
\begin{ruledtabular}
\begin{tabular}[t]{llcccccc}
Molecule & Formula & $\mu$\footnotemark[1] & $\alpha _d$\footnotemark[1] &
\multicolumn{2}{c}{Electron}& \multicolumn{2}{c}{Positron} \\
 & & (D) & (\AA$^3$) & $\eb $\footnotemark[2] & $\eb $\footnotemark[3] &
$\eb $\footnotemark[4] & $\eb $\footnotemark[5] \\
\hline
Formaldehyde & CH$_2$O & 2.33 & 2.8 & $-$  & 0.02--0.05 & 18 & 12\\ 
Acetaldehyde & CH$_3$CHO & 2.75 & 4.6 & 0.65 & 0.95 & $-$ & 42 \\
Acetone & (CH$_3$)$_2$CO & 2.88 & 6.4 & 2.8 & 1.6 & 4 & 66 \\
Hydr. cyanide & HCN & 2.98 & 2.6 & $-$ & 3.3--5 & 35 & 21 \\
Nitrobenzene & C$_6$H$_5$NO$_2$ & 4.2 & 13.5 & 28 & 30.4 & $-$ & 270
\end{tabular}
\end{ruledtabular}
\footnotetext[1]{Values used by \textcite{AD98}.}
\footnotetext[2]{Experimental data \cite{AD98}.}
\footnotetext[3]{Electrostatic model calculations \cite{AD98}.}
\footnotetext[4]{Calculated values, see Table~\ref{tab:fit_cal}.}
\footnotetext[5]{Predictions of Eq.~(\ref{eq:fit}).}
\end{table}
\endgroup

Another interesting electron analog of positron-molecule bound states is
the case of negative ions of small molecular clusters, such as (N$_2$O)$_n$
\cite{WLR99} and (CO$_2$)$_n$ \cite{LBF00}. Using laser-assisted
photoelectron attachment, \textcite{KRH92a,KRH92b,KRH94} have observed
prominent VFRs in the yields of fragment negative ions [i.e.,
(N$_2$O)$_q$O$^-$ with $q<n$, and (CO$_2$)$_q^-$ with $q\leq n$]. These
experiments are particularly relevant to the resonant processes
discussed in this review. First, these VFRs
were clearly identified with individual molecular vibrational 
modes. Second, these resonances displayed downshifts with increasing
cluster size (especially clear for the CO$_2$ clusters), which provided a
measure of the electron-cluster binding energies. Finally, these cluster
anions have the structure of a weakly bound, diffuse excess electron
attached to an essentially unperturbed neutral cluster, similar to the
positron-molecule bound states described here.

For the CO$_2$ clusters, the measured electron binding energies are
comparable to the positron $\eb $ for alkanes with similar numbers
of carbon atoms. They also increase approximately linearly for $n=4$--20, with
some evidence of saturation at larger $n$. This behavior has been
successfully modeled theoretically by combining the $-\alpha _d/2r^4$
polarization outside the cluster with a constant short-range potential
inside \cite{LBF00}. It is likely that including the long-range
$-\alpha _d/2r^4$ potential would also improve the
modeling of positron binding to alkanes (cf. Fig.~\ref{fig:bind}).

A theory has been constructed to describe electron collisions with van der
Waals clusters, such as (CO$_2$)$_n$ \cite{FH05,Fab05}. It allows one to
calculate the attachment cross sections and describe the VFR that are
observed. In this theory the electron interaction with vibrational degrees of
freedom of the CO$_2$ monomers is described in the dipole approximation.
This is analogous to the approach used in Sec.~\ref{subsec:IR}
to describe positron-molecule VFR.

\section{Analysis of annihilation rates measured with thermalized
positrons}\label{sec:therm}

There is an extensive body of experimental data on the annihilation of
thermalized positrons in molecular gases at 300~K
\cite{PS63,HCG82,SPL88,IGM95}. It is useful to examine these results
in light of the more recent, energy-resolved data for
$\Z$ described above \cite{BGS03,BYS06} and the current understanding
of resonant annihilation. In particular, it is now possible to consider the
relationship between the thermal $\Z$ values and those due to VFR 
and IVR. Figure~\ref{fig:alk_norm} shows a
comparison of energy-resolved $\Z$ spectra for alkanes with $n=3$--8 carbons.
For each molecule, its $\Z$ was normalized by $g= \sqrt{\eb/\eps}$,
shifted upward in energy by the binding energy $\eb$, and then normalized to
unity at the C-H stretch peak. In this representation, it is clear that the
spectral shapes of the alkanes change relatively little with molecular size.
As the size of an alkane increases, its $\Z$ spectrum shifts to lower energies.
A mode accessed at 50~meV incident positron energy in the butane
($n=4$) spectrum will be accessed at 5~meV in hexane ($n = 6$). Thus,
the annihilation rates $\Zth $ measured with thermalized positrons at
300~K are intimately related to the corresponding values of the
energy-resolved spectra, albeit shifted downward by the binding energy.

As shown in Fig.~\ref{fig:alk_norm}, the 300~K data (plus signs in circles)
align well with the energy-resolved data when they are normalized by
$g= \sqrt{\eb/\eps _T}$ (and the C-H stretch peak heights) and assigned a
corrected energy of $\eb+\eps _T$, where $\eps _T=\frac{3}{2}k_BT=37.5$~meV
is the average thermal positron energy at 300~K. Also note that, due to their
larger binding energies, alkanes with nine or more carbons begin to sample the
trough in the spectrum between the plateau and the C-H stretch peaks. This
explains the trend, pointed out previously \cite{BGS03}, that the ratio
of $\Zth $ to $\Z$ at the C-H stretch peak decreases by a factor of 2 when
the number of carbon atoms in the alkane is increased from $n\le 9$ to $n>9$.
This analysis connects in a \emph{quantitative} way the thermal data
with the energy-resolved measurements taken at higher positron energies.
Thermal positrons at 300~K annihilate by the same IVR-enhanced resonant
mechanism as higher-energy positrons; the only difference is the specific
vibrational resonances involved.

The values of $\Zth$ measured at 300~K can also be used
to  test Eq.~(\ref{eq:fit}) which estimates the positron
binding energy in terms of molecular parameters \cite{DYS09}. A selection
of molecules and their thermal annihilation rates $\Zth$ from
\textcite{DYS09} are listed in Table~\ref{tab:test}. While these $\Zth$
values do not provide direct evidence of VFR and hence positron binding,
they can indicate whether positrons do or do not bind to the molecular
species \cite{YS08a,Gri00,Gri01,IGG00}.
We take $\Zth>10^3$ (Sec.~\ref{subsec:dir}) to be an indicator of resonant
annihilation and hence positron binding. This is by no means a necessary
condition, and so in the analysis below, we also use the somewhat arbitrary
condition, $\Zth/Z =10$, as the boundary separating the two
groups of molecules. Namely, $\Zth/Z > 10$ for a given
molecule is taken to mean that positrons bind to this target and vice
versa. The predictions of Eq.~(\ref{eq:fit}) are listed in
Table~\ref{tab:test}, together with the molecular parameters used to calculate
them. Generally, molecules with relatively large values of $\Zth/Z $ are
predicted correctly by Eq.~(\ref{eq:fit}) to have positive binding energies. 
One borderline exception is methane which has a value of $\Zth/Z =14$ but
does not bind positrons.

\begingroup
\renewcommand{\arraystretch}{0.8}
\squeezetable
\begin{table}[ht]
\caption{Values of $\Zth/Z$ for a variety of chemical species, the 
predictions of Eq.~(\ref{eq:fit}) for their binding energies, the molecular
parameters relevant to this analysis, and the molecular ionization energies
$E_i$.}
\label{tab:test}
\begin{ruledtabular}
\begin{tabular}{llcccccc}
Name & Formula & $\eb$\footnote{Predicted by Eq.~(\ref{eq:fit}).} 
& $Z$ & $\Zth /Z$\footnote{$\Zth$ is
measured at room temperature with thermalized positrons \cite{IGM95,Iwa97}.}
& $\alpha _d$\footnotemark[3] & $\mu $\footnotemark[3] &
$E_i$\footnotemark[3] \\
 & & (meV) & & & (\AA$^3$) & (D) & (eV) \\
\hline
\multicolumn{8}{c}{Small molecules} \\
Carbon dioxide & CO$_2$ & $-36$ & 22 & 2.5 & 2.7 & 0 & 13.8 \\
Sulfur hexafl. & SF$_6$ & $-14$ & 70 & 1.2 & 4.5 & 0 & 15.3 \\
Water & H$_2$O & $-15$ & 10 & 32 & 1.5 & 1.9 & 12.6 \\
Nitrous oxide & N$_2$O & $-29$ & 22 & 3.5 & 3.0 & 0.2 & 12.9 \\
Nitrogen dioxide & NO$_2$ & $-26$ & 23 & 47 & 3.0 & 0.3 & 9.8 \\ 
Ammonia & NH$_3$ & $-12$ & 10 & 160 & 2.3 & 1.5 & 10.2 \\
Methane & CH$_4$ & $-37$ & 10 & 14 & 2.6 & 0 & 12.7 \\
\multicolumn{8}{c}{Alkenes and alkynes} \\
1-hexene & C$_6$H$_{12}$ & 81 & 48 & 3900 & 11.6 & 0.3 & 9.5 \\
trans 3-hexene & C$_6$H$_{12}$ & 74 & 48 & 4100 & 11.6 & 0 & 8.9 \\
1,3-hexadiene & C$_6$H$_{10}$ & 72\footnotemark[4] & 46 & 8500 & 11.4 & --
& 8.5 \\
1,3,5-hexatriene & C$_6$H$_8$ & 69 & 44 & 9400 & 11.2 & 0 & 8.3 \\
\multicolumn{8}{c}{Perhalogenated alkanes} \\
Carbon tetrafl. & CF$_4$ & $-34$ & 42 & 1.2 & 2.9 & 0 & 16.2 \\
Hexafluoroethane & C$_2$F$_6$ & $-10$ & 66 & 2.3 & 4.8 & 0 & 14.6 \\
Perfluoropropane & C$_3$F$_8$ & 13\footnotemark[4] & 90 & 1.7 & 6.7 & -- & --\\
Perfluorohexane & C$_6$F$_{14}$ & 84 & 162 & 3.3 & 12.4 & 0 & 12.8 \\
Perfluorooctane & C$_8$F$_{18}$ & 131 & 210 & 5.1 & 16.2 & 0 & 12.6 \\
Carbon tetrachl. & CCl$_4$ & 58 & 74 & 130 & 10.3 & 0 & 11.3 \\
Carbon tetrabrom. & CBr$_4$ & 120 & 146 & 270 & 15.3 & 0 & 10.3 \\
Carbon tetraiod. & CI$_4$ & 235 & 218 & 37 & 24.5 & 0 & -- \\
\multicolumn{8}{c}{Partially fluorinated alkanes} \\
Difl.methane & CH$_2$F$_2$ & 0.4 & 26 & 31 & 2.5 & 1.8 & 12.6 \\
Trifluoromethane & CHF$_3$ & $-4$ & 34 & 7.3 & 2.7 & 1.7 & 14.8 \\
Fluoroethane & C$_2$H$_5$F & 21 & 26 & 120 & 4.2 & 2.0 & 12.4 \\
1,1,1-trifl.ethane & C$_2$H$_3$F$_3$ & 29 & 42 & 38 & 4.2 & 2.3 & 13.3 \\
\multicolumn{8}{c}{Oxygen-containing molecules} \\
1-propanol & C$_3$H$_8$O & 50 & 34 & 590 & 7.0 & 1.7 & 10.2 \\
Acetone & C$_3$H$_6$O & 66 & 32 & 3100 & 6.3 & 2.9 & 9.7 \\
\multicolumn{8}{c}{Other aromatics and substituted benzenes} \\
Anthracene & C$_{14}$H$_{10}$ & 422 & 94 & 46000 & 22.8 & 0 & 7.5 \\
Decahydronaphth. & C$_{10}$H$_{18}$ & 151 & 78 & 5000 & 17.7 & 0 & 9.4 \\
o-xylene & C$_8$H$_{10}$ & 208 & 58 & 3100 & 14.1 & 0.6 & 8.6 \\
Toluene & C$_7$H$_8$ & 179 & 50 & 3800 & 12.3 & 0.4 & 8.8 \\
Hexafluorobenzene & C$_6$F$_6$ & 141 & 90 & 13 & 9.8 & 0 & 9.9 \\
Octafluorotoluene & C$_7$F$_8$ & 165\footnotemark[4] & 114 & 11 & 11.7 & --
& 9.9 \\ 
Octafluoronaphth. & C$_{10}$F$_8$ & 272 & 132 & 23 & 15.5 & 0 & 8.9 \\
Nitrobenzene & C$_6$H$_5$NO$_2$ & 254 & 64 & 6700 & 12.1 & 4.2 & 9.9 \\
Chlorobenzene & C$_6$H$_5$Cl & 204 & 58 & 1250 & 12.1 & 1.6 & 9.1 \\
Bromobenzene & C$_6$H$_5$Br & 215 & 76 & 2300 & 13 & 1.7 & 9.0 \\
Fluorobenzene & C$_6$H$_5$F & 176 & 50 & 900 & 10.0 & 1.6 & 9.2 \\
1,2-difluorobenzene & C$_6$H$_4$F$_2$ & 189 & 58 & 570 & 9.8 & 2.4 & 9.3
\end{tabular}
\end{ruledtabular}
\footnotetext[3]{Parameters from \textcite{CRC,Mil90,McC63}.}
\footnotetext[4]{Lacking values of $\mu$
for these species, the predictions of Eq.~(\ref{eq:fit}) are lower bounds on
$\eb$.}
\end{table}
\endgroup

As shown in Table~\ref{tab:test}, the criterion from Eq.~(\ref{eq:fit}) that
molecules will bind for $\alpha_d > 5.4$~\AA$^3$ has the consequence that
most large molecules can bind positrons
whether they have a permanent dipole moment or not. The alkanes are an example
of this, where only methane does not bind. For nonaromatic molecules the
$\alpha _d$ term dominates for all but relatively small molecules with
correspondingly 
small values of $\alpha _d$. A related trend is seen in halogen substitution, 
where $\eb$ rises rapidly as the size (and hence the polarizability) of the 
halogen is increased. For example, CF$_4$ does not bind positrons, while
CBr$_4$ is predicted to have a binding energy in excess of 120 meV.

Restricting comparison to cases where the model predicts $|\eb| \geq 15$~meV
in deference to the likely error bars for the model, negative binding
energies are predicted for almost all small molecules. In 
particular, the diatomic molecules, H$_2$, D$_2$, N$_2$, O$_2$, CO, and NO
(not shown in Table~\ref{tab:test}), with $\Zth/Z$ values ranging from
2.3 to 7.4 (i.e., $< 10$) all have negative binding energies
$\eb < -40$~meV \cite{DYS09}. As shown in Table~\ref{tab:test}, neither 
methane nor carbon tetrafluoride bind, in agreement with the 
interpretation of their $\Zth$ values \cite{IGM95}. As indicated in 
Table~\ref{tab:bind}, molecules for which the binding energies are
``too close to call,'' [i.e., $|\eb| < 15$ meV, based on Eq. (\ref{eq:fit})]
include ethane, propane, cyclopropane, methanol, methyl fluoride,
 and water (cf. Table~\ref{tab:test}), and this is consistent with the
observations \cite{YS08b}. There are some disagreements in cases where the
binding energy
is predicted and/or observed to be reasonably small. From Table~\ref{tab:test},
the only molecule that exceeds the $|\eb| \ge 15$ meV criterion is NO$_2$
with a $\Zth/Z$ ratio of $47$ and yet a predicted binding energy of
$-26$ meV. Like benzene, NO$_2$ has two resonant electronic states involving
$\pi$ bonds, but the geometry of its valence orbitals differs greatly from
that of the aromatics. Additional experiments are needed to better distinguish
the effects of the different types of electronic bonds, especially for the 
smallest molecules.

As shown in Table~\ref{tab:test}, the ratios of $\Zth/Z$ for perfluoroalkane 
molecules are much smaller than those for alkanes. This has been interpreted
as evidence that resonant annihilation is ``switched off'' for them,
possibly due to lack of binding \cite{Gri00}. However, Eq.~(\ref{eq:fit}) 
predicts that perfluorocarbons other than perfluoromethane and 
perfluoroethane can bind positrons. This warrants further investigation.

\section{Other topics}\label{sec:other}

\subsection{Gamma-ray Doppler-broadening measurements}\label{subsec:gamma}

As discussed in Sec.~\ref{subsec:anspec}, positron annihilation
usually results in the production of two, approximately back-to-back
gamma rays, each with an energy $E_\gamma \approx 511$~keV.
However, there are small shifts in the energies of the two gammas that
can be used to obtain microscopic information about
the annihilating pair. According to Eq.~(\ref{eq:Dop}), these energies
are Doppler shifted by the energy $\eps =\pm cP_z/2$,
where $P_z$ is the component of the center-of-mass momentum of the
electron-positron pair along the direction of the gamma rays.
For the low-energy positrons that we deal with here
(i.e., $\eps \leq 0.5$~eV), the center-of-mass momentum $P$ is dominated by 
the momentum distribution of the electron orbitals. As a benchmark, a 4~eV
electron traveling along the direction of the gamma rays produces a 1~keV
Doppler shift.

\begin{figure}[ht]
\includegraphics*[width=7cm]{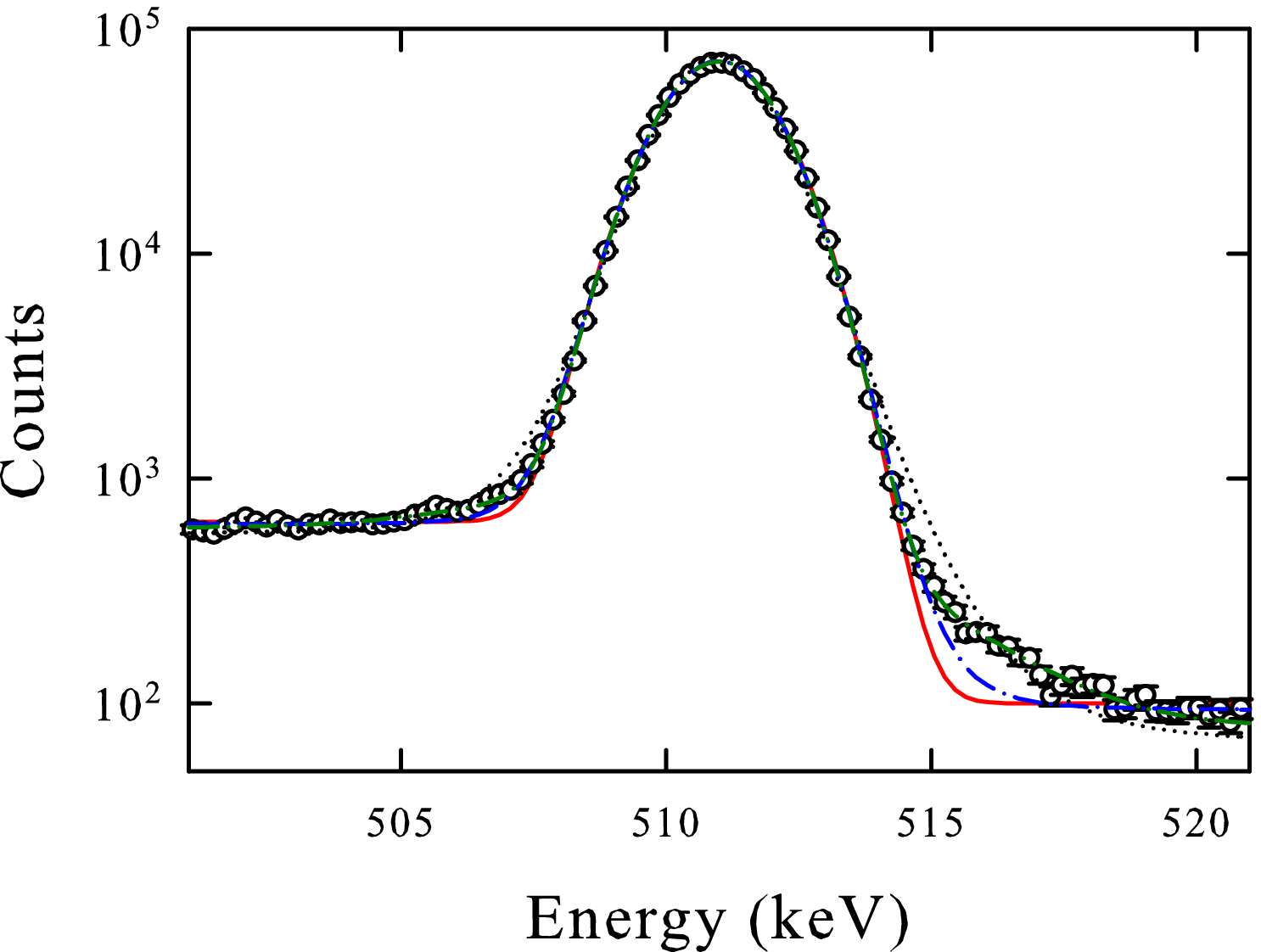}
\caption{$\gamma $-ray spectrum of hexane (C$_6$H$_{14}$):
($\circ $) observed spectrum; solid curve, single-Gaussian fit;
dotted curve, fit using a hydrogenic functional form; dot-dashed curve,
fit using a noninteracting hydrogenic form convolved with a Gaussian;
dashed curve (indistinguishable from the data), two-Gaussian fit.
The statistical error bars are comparable to or smaller than the size
of the data points. See \textcite{IGS97} for details.}
\label{fig:Dop_hex}
\end{figure}

Extensive measurements have been made of the Doppler broadening of 
annihilation gamma rays for the case of a thermal distribution of 
positrons at 300 K interacting with a wide variety of molecules
\cite{IGS97,TTG92}. The apparatus and procedures for these measurements are
described in Sec.~\ref{subsec:gam_meas}. Shown in
Fig.~\ref{fig:Dop_hex} is the spectrum for hexane. The detector response is
modeled by a Gaussian lineshape with a FWHM of 1.16~keV and an error function
to account for the effect of Compton scattering of the gamma rays in the 
detector. The measured gamma spectra could be fit reasonably well 
using a single Gaussian to approximate the Doppler spectrum, convolved
with the detector response function.

A more accurate fit was obtained using two Gaussians to model
the Doppler spectrum. The second Gaussian
was required to describe a smaller,
higher-momentum component, which was typically about 1--3\% of the larger
component. The resulting linewidths (FWHM), with the detector 
resolution de-convolved, ranged from 1.7~keV for H$_2$ to $\sim 3.1$~keV
for fluorocarbons. The smaller, higher-momentum component had widths from 
4 to 9 keV, with most molecules in the narrower range from 5 to 7 keV.

Shown in Table~\ref{tab:gam} are typical values of the linewidths $\Delta E$
(FWHM) using the single Gaussian fit, the annihilation rates $\Z$ for
thermal positrons, and the positron-molecule binding energies $\eps _b$,
for a selection of molecules.
A key conclusion of these Doppler-broadening studies is that the linewidths
are only weakly correlated with the values of either $\eb$ or $\Zth$. 
For example, comparing the ethane and dodecane in Table~\ref{tab:gam}, 
$\eb $ increases from near zero to 220 meV, and $\Zth$ increases by 3
orders of magnitude, while the gamma-ray line width increases by only 5\%. 
Note that there is, however, a significant increase in the linewidth when
a hydrocarbon is partially or fully fluorinated.

\begin{table}[ht]
\caption{Gamma-ray linewidths $\Delta E$ (FWHM, single Gaussian fit),
annihilation rates $\Zth$ for a thermal distribution of positrons at 300 K, and
positron-molecule binding energies $\eps _b$, for selected molecules.}
\label{tab:gam}
\begin{ruledtabular}
\begin{tabular}{llcrc}
Name & Formula & $\Delta E$\footnote{From \textcite{IGS97}.} &
$\Zth $\footnote{See \textcite{IGM95,Iwa97} and references therein.} &
$\eb $\footnote{See \textcite{YS08a,YS08b} and Sec. \ref{sec:bind}.} \\
 & & (keV) & & (meV) \\
\hline
Methane & CH$_4$ & 2.09 & 142 & $<0$ \\
Ethane & C$_2$H$_6$ & 2.18 & 1780 & $>0$ \\
Propane & C$_3$H$_8$ & 2.21 & 3500 & 10 \\
Pentane & C$_5$H$_{12}$ & 2.24 & 40200 & 60 \\
Nonane & C$_7$H$_{16}$ & 2.32 & 643000 & 145 \\
Dodecane & C$_{12}$H$_{26}$ & 2.29 & 1780000 & 220 \\
\hline
Benzene & C$_6$H$_6$ & 2.23 & 15000 & 150 \\
Naphthalene & C$_{10}$H$_8$ & 2.29 & 494000 & $\sim 300$ \\
\hline
1-fluoroethane & C$_2$H$_5$F & 2.62 & 3030 & $>0$ \\
Hexafluoroethane & C$_2$F$_6$ & 3.04 & 149 & $<0$ \\
1-fluorohexane & C$_6$H$_{13}$F & 2.46 & 269000 & 80 \\
\end{tabular}
\end{ruledtabular}
\end{table}

The linewidths $\Delta E$ for alkanes, shown in Table~\ref{tab:gam}, increase
monotonically with increasing molecular size.
This increase has been ascribed to the change in the relative number
of electrons in C-C {\em vs} C-H bonds, since the electrons in these bonds are
expected to be characterized by different Doppler widths. An analysis was made
to determine the fraction of annihilation events involving electrons
from these types of orbitals in alkanes \cite{IGS97}. It showed that there is a
linear increase in the linewidth as a function of the fraction of electrons in
C-C orbitals.
This is consistent with the assumption that the positron annihilates
statistically on any of the valence electrons in the molecule (see below).

\begin{figure}[ht]
\includegraphics*[width=7.5cm]{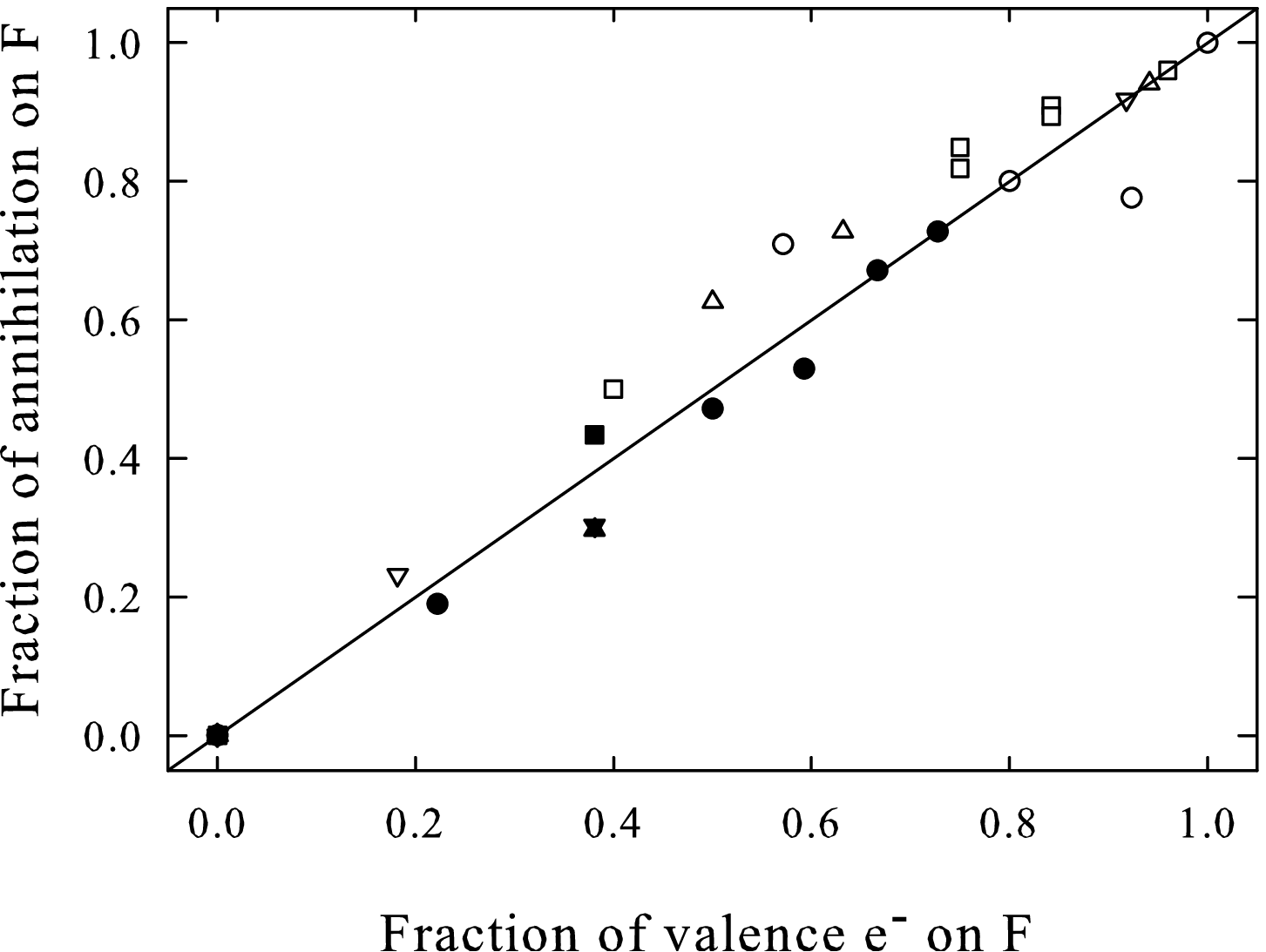}
\caption{Normalized fraction of positrons annihilating on fluorine atoms 
in partially and fully fluorinated alkanes, plotted against the fraction of 
valence electrons in the fluorine atoms. Open symbols are methane (circle), 
ethane (square), propane (triangle), and hexane-based molecules (inverted 
triangle); filled symbols are six-carbon benzene-based molecules:
1,2-difluorobenzene (square), 1,3-difluorobenzene (triangle),
1,4-difluorobenzene (inverted triangle), and other six-carbon benzene-based 
molecules (circle). See text and \textcite{IGS97} for details.}
\label{fig:CHCF}
\end{figure}

A similar study was carried out for the linewidths associated with 
annihilation in partially and fully fluorinated alkanes. In this case, the 
observed line was decomposed into fluorinated and hydrogenated components
by fitting to a sum of two measured line shapes, one for the fully
hydrogenated compound, and one for the fully fluorinated compound \cite{IGS97}. 
As shown in Fig.~\ref{fig:CHCF}, there is
a smooth, linear increase in the linewidth over the full range of the degree
of fluorination, from 0 to 100\%.
This also supports the hypothesis that the positrons annihilate
statistically on any valence electron.

The results of these gamma-ray studies provide confirmation of a model 
by \textcite{Cra94} for positron annihilation on molecules. It was developed
to explain the degree of ion fragmentation that is observed following
positron annihilation on molecules (see Sec.~\ref{subsec:fragm}). Crawford
argued that a low-energy positron should annihilate statistically on any
valence electron. The gamma-ray linewidth measurements shown in
Fig.~\ref{fig:CHCF} are consistent with this hypothesis.
Such uniform spreading of the positron density over the molecule
is related to the fact that at low energies the positron de Broglie
wavelength $\lambda _{\rm deB}$ is larger than the size of the target.
This prevents the positron from being localized strongly on any particular
site in the molecule. This is true both for direct annihilation, for which
$\lambda _{\rm deB}=2\pi /k$ (e.g., $k=0.05$ a.u. for thermal positrons at
300~K), and for resonant annihilation, for which
$\lambda _{\rm deB}\sim 2\pi /\kappa $, where $\kappa =\sqrt{2\eb }$
(see, e.g., Fig.~\ref{fig:c14} for model bound-state wave functions).

In some cases, such as strongly polar molecules, localization of the
positron near specific sites in the molecule might be expected. For example,
in LiH the positron density is strongly localized at the negatively charged
hydrogenic end of the molecule \cite{Str01,Str99}; cf. Fig.~\ref{fig:LiH}.
However, annihilation gamma-ray spectra from this class of molecules have
not yet been investigated.

An analysis is in progress to understand in more detail
the implications of the Doppler-broadening experiments \cite{WSG09}.
It relies on modern quantum-chemistry methods, such as
the density-functional theory B3LYP/TZVP, which give
electron momentum densities in good agreement with experiment \cite{Wan03}.
The linewidths calculated for methane and fluoromethanes, ethane, propane,
butane and benzene, by taking only the electron momentum distribution into
account, are about 30\% greater than the values measured.
The main source of this discrepancy appears to be the neglect of
the Coulomb repulsion between the positron and the atomic cores \cite{WSG09}.
This repulsion suppresses the positron wave function at small positron-nuclear
separations, which effectively reduces the high-Doppler-shift components in the
annihilation spectra.

A more complete theory of the annihilation gamma-ray spectra for molecules
should include the full electron and positron dynamics and account
for the positron-nuclear repulsion and electron-positron correlation effects.
However, the results of \textcite{WSG09} suggest
that these effects in positron annihilation spectra can be modeled 
by a relatively simple scaling factor.
At the next level of analysis, one should examine the high-momentum
component in the spectra [e.g., as described by two-Gaussian fits
\cite{IGG97}], which is likely due to annihilation on inner-shell electrons.
Further work on this topic is in progress.

\subsection{Annihilation-induced fragmentation of molecules}
\label{subsec:fragm}

Two-body collisions between positrons and molecules can produce positive ions
by two mechanisms. For incident positron energies greater than the 
threshold for Ps formation, ionization proceeds via this channel.
If the positron energy is below the Ps formation threshold
(i.e., the principal regime of interest in this review), the positron can
annihilate with a molecular electron, also producing a positive molecular ion. 

The initial studies of annihilation-induced ion formation in molecules below 
the positronium formation threshold were conducted using positrons 
confined in a buffer-gas trap in the presence of low pressure 
gases of alkane molecules \cite{PSL89}. In spite of complications
due to the presence of molecular nitrogen in the trap \cite{GGM94},
these experiments established that sub-Ps-threshold ionization can produce
significant amounts of fragmentation.
Subsequently, extensive studies of
positron-induced ionization of molecules, both below and above the positronium 
formation threshold, were conducted by L. D. Hulett and collaborators
\cite{DHE90,HDX93,XHL93,XHL94,XHL95,HXM96a,HXM96b,XHM97,MSL00}.
The experimental procedures are described in Sec. \ref{subsec:ann_frag}.

\begin{figure}[ht]
\includegraphics*[width=7cm]{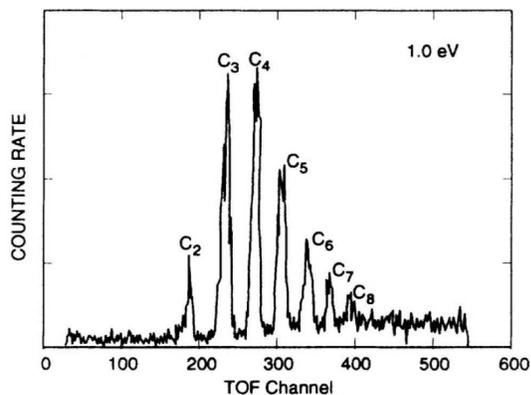}
\caption{Time-of-flight mass spectra of ion fragments from 1-dodecene 
at 1.0~eV, which is 2.1 eV below the Ps
formation threshold. Adapted from \textcite{XHL94}.}
\label{fig:dodecene}
\end{figure}

\begin{figure}[ht]
\includegraphics*[width=8cm]{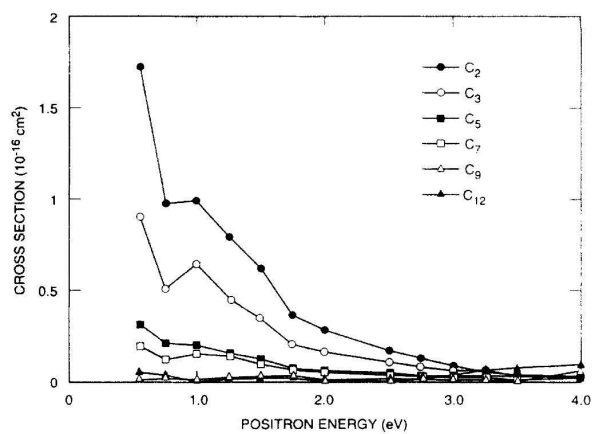}
\caption{Cross sections for fragmentation of 1-dodecene (C$_{12}$H$_{24}$)
by positron annihilation near and below the Ps-formation threshold at 3.1~eV.
From \textcite{XHL94}.}
\label{fig:dode_cs}
\end{figure}

These studies produced several important results. One is that the extent
of fragmentation depends in a nontrivial way on the energy of the
incident positron. For example, fragmentation is a minimum at energies near
the Ps-formation threshold, and increases toward smaller and
greater positron energies. Figure~\ref{fig:dodecene} shows the mass spectrum
of 1-dodecene (C$_{12}$H$_{24}$) at an incident energy of 1.0 eV (i.e.,
$\sim $2~eV below the Ps formation threshold). It illustrates the broad
spectrum of product ions that is produced \cite{XHL94}.
Figure~\ref{fig:dode_cs} shows the cross sections for
producing these fragments as a function of positron energy.

The fact that in many molecules the degree of fragmentation has a minimum
close to the Ps-formation threshold may offer the possibility of using
positrons to advantage in ion mass spectroscopy. 
Another interesting effect is that,
below the Ps-formation threshold, double and triple bonds can stabilize the
species with respect 
to fragmentation. For example, in the series decane
(C$_{10}$H$_{22}$, single bonds only), 1-decene (C$_{10}$H$_{20}$, one double
bond) and 1,9-decadiene (C$_{10}$H$_{18}$, two double bonds), the yield of ion
fragments decreases with each additional double bond \cite{XHL95}. This same
effect was also observed in other molecules [e.g.,
tetravynilsilane {\em vs} tetraethylsilane \cite{HDX93,XHL95}].

\begin{figure}[ht]
\includegraphics*[width=8cm]{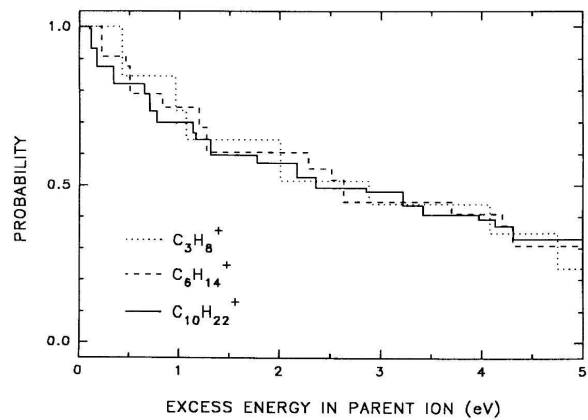}
\caption{Probability $P(E)$ that the energy deposited into a molecular ion
due to positron annihilation with an electron in a valence orbital below the
HOMO is greater than a given excess energy for propane, hexane, and decane.
For details of the calculation, see \textcite{Cra94}.}
\label{fig:ex_en}
\end{figure}

Motivated by these experiments, \textcite{Cra94} constructed a simple 
and insightful model of the fragmentation process. He argued that 
annihilation will occur with comparable probability on electrons in any
valence molecular orbital $i$, not just the highest occupied molecular
orbital (HOMO). As a result, the molecular ion is typically left
in an electronically excited state, with the excitation energy provided by the
energy difference between the HOMO and the orbital $i$. Figure~\ref{fig:ex_en}
shows the probability that this energy exceeds a certain value for propane,
hexane, and decane \cite{Cra94}. Due to vibronic coupling, this
energy flows from the electronic to the vibrational degrees of freedom.
Given time and sufficient energy (e.g., a few electron volts), the molecular
ion can break up in the process known as unimolecular dissociation.

Fragmentation can occur when the incident positron annihilates, either during
direct annihilation or following capture into a VFR. The resonant process
likely dominates in larger polyatomics at low positron energies and is
central to this review. In this case, both the incident positron energy
and the binding energy contribute to the total excitation energy of the
molecular ion. The gamma-ray annihilation studies described above provided
validation of Crawford's model, showing that positrons do indeed annihilate
with approximately equal probabilities on all valence electrons.

The experiments of Hulett and coworkers used a beam with an 
energy resolution $\sim $0.5~eV FWHM. This raises a question as to the 
extent to which ionization could be controlled (e.g., to produce only
parent ions) using the much higher resolution beams that are now
available (e.g., 40~meV FWHM). Another possibility would be to exploit
ionization via the second bound states such as those observed in larger
alkane molecules (cf. Sec.~\ref{subsec:alk}). In this case, the
positron wave function has a nodal plane at the center of the linear carbon 
chain (see Fig.~\ref{fig:c14}), which would be expected to produce a decrease
in annihilation near this location. An experiment to test this might also
provide information as to whether the excess electronic energy
deposited in the molecule in the annihilation process
could diffuse away from the annihilation site and then break a bond.
Investigations to date have only scratched the 
surface of this rich area of matter-antimatter chemistry. We still know 
relatively little about the chemical specificity of annihilation and 
annihilation-induced ion production.

\subsection{Nonlinear dependence of annihilation on molecular density}
\label{subsec:density}

Equations~(\ref{eq:Zefflam}) and (\ref{eq:rate}) imply that the annihilation
rate $\lambda $ is a linear function of the gas density $n$. This was shown to
be correct at low densities, such as those used in
the positron-trap experiments described in Sec.~\ref{subsec:anntrap}.
There are, however, a number of effects that can make $\Z$
density dependent. For example, in dense gases and at lower temperatures,
positrons can cause a local phase transition and become self-trapped in
clusters that gives rise to a strongly nonlinear behavior of the annihilation
rate~\cite{IK82}. Also, in low-density gases one typically observes
$\Z\gg Z$, while at solid or liquid densities the value of
$\Z$ for almost any material are close to the number of valence
electrons \cite{Pom49}.

\begin{figure}[ht]
\includegraphics*[width=8cm]{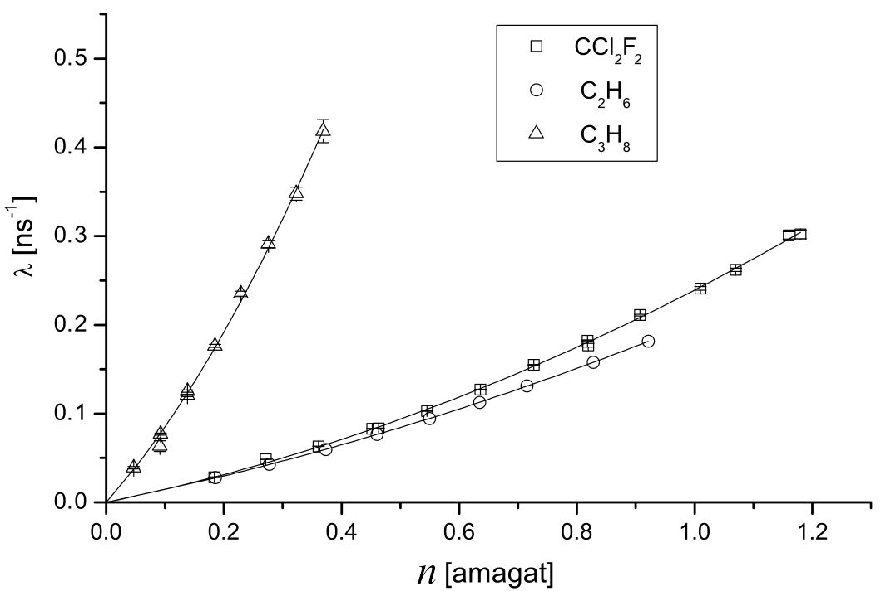}
\caption{Density dependence of the room-temperature positron annihilation
rates for CCl$_2$F$_2$($\square $), C$_2$H$_6$ ($\circ $), and
C$_3$H$_8$ ($\bigtriangleup $). Solid lines are fits of the form
Eq.~(\ref{eq:ab}). Adapted from \textcite{CWL06}.}
\label{fig:nonlin}
\end{figure}

For diatomic molecular gases such as H$_2$, N$_2$, CO and O$_2$
at densities $n\lesssim 50$~amagat, the $\Z$ values remain practically
independent of $n$. In contrast, other species, such as CO$_2$, CH$_4$,
and SF$_6$, display considerable variation with $n$ \cite{HCG82}.
For larger molecules (e.g., C$_2$H$_6$, CCl$_2$F$_2$, C$_3$H$_8$, and
C$_4$H$_{10}$) a strong density dependence is observed even at
$n\sim 1$~amagat~\cite{HCG82}, as shown in Fig.~\ref{fig:nonlin}.
In these cases, the density effect can be expressed as
a quadratic correction to Eq.~(\ref{eq:rate}),
\begin{equation}\label{eq:ab}
\lambda =an+bn^2.
\end{equation}
Here $a=\pi r_0^2c\Z$ represents annihilation in binary collisions, and the
$b$ term accounts for three-body annihilation events involving
a positron and two gas molecules \cite{CWA02,CWL06}. A key finding
is that the coefficient $b$ for molecules with large 
$\Z$, for which resonant annihilation is observed (e.g., C$_2$H$_4$
or C$_3$H$_8$), is approximately proportional to $\Z$. This suggests
that three-body collisions affect the resonant annihilation,
rather than being an additional independent reaction pathway.

There is currently no information about the microscopic nature of
this process, and so one can only mention some possibilities. The annihilation
in the resonant positron-molecule complex could be enhanced through
collisions with other molecules. \textcite{Gri09} recently proposed
that these collisions may stabilize the resonant complexes. If this
stabilization is complete (i.e., preventing the positron from escaping the
molecule), then the corresponding collision rate $\Gamma ^c=\sigma _cv_mn$
must be {\em added} to the annihilation
rate $\Gamma ^a_\nu $ in the numerator of Eq.~(\ref{eq:siga}).
Here $v_m$ is the mean molecular collision velocity and $\sigma _c$ is the
stabilizing collision cross section. The values of $\sigma _c$ required
to explain the observations can be estimated using the experimental values
of $a$ and $b$ \cite{CWL06}, and estimating $\Gamma ^a_\nu $ from
Eqs.~(\ref{eq:gama}) and (\ref{eq:rhoep}) using the binding energies inferred
from the energy-resolved $\Z$ measurements. The values range
from $\sigma _c\sim 1$~a.u. in ethane and ethylene to $\sim $10~a.u. in butane.

There is also the possibility that a collision with a second molecule
enhances the annihilation rate by an amount $p\Gamma^c$, where $p<1$ is the
probability that the collisionally stabilized positron-molecule complex
survives to annihilation rather than autodetaches. In this case, to achieve
the same effect, the cross sections $\sigma_c$ would need to be enhanced by a
factor $p^{-1}$ above that for complete stabilization.

The actual mechanism of collisional stabilization is also unclear. 
It could be due to the transfer of the bound positron from the vibrationally
excited host molecule to a vibrationally colder molecule of the same species.
This process would be facilitated by the fact that the positron binding
energies in the two molecules would be similar, making it a
{\em resonant charge transfer}. The vibrational de-excitation of the
positron-molecule complex could also be due to intermolecular vibrational
energy transfer in a collision with a colder molecule.
However, in both cases, for the species studied,
the thermal vibrational energies at room temperature are larger than the
positron binding energies. Hence these processes might not completely
stabilize the complex.

Finally, if a VFR is excited when two molecules are in close
proximity, the positron binding energy is likely to be greater than that for
a binary positron-molecule interaction. This too might help to explain the
observations. At a minimum, further research is warranted to understand this 
collisional enhancement effect. 

\section{Summary and a look to the future}\label{sec:sum}

This review focuses on the interaction of positrons with molecules in the 
range of energies below the thresholds for electronic excitation and 
Ps formation. Experimental studies of annihilation 
resolved as a function of positron energy show that 
positrons bind to many molecular species. This enables the 
formation of long-lived resonant states (i.e., vibrational Feshbach 
resonances) in two-body positron-molecule collisions, thus altering
the dynamics in a fundamental way for the incident positron energies 
in the range of the molecular vibrations.

These VFR are responsible for the large annihilation rates observed in many 
polyatomic molecules. They exceed by orders of magnitude the contribution of
direct, ``in-flight'' annihilation. This enhancement distinguishes annihilation
from 
conventional scattering processes. For example, the resonant contribution 
would be hard (if at all possible) to observe in elastic scattering, where
much larger contributions come from potential scattering. 

Theoretically, resonant annihilation can be described using a Breit-Wigner
approach. The key quantities required to 
make predictions are energies of the VFR and their decay rates (i.e., the 
annihilation rate, positron capture and autodetachment rates, etc.). Such a 
calculation can be  done for small molecules (e.g., the methyl halides and 
similar molecules with IR-active modes) using a theory which makes use
of the small parameter in the problem, namely, the positron binding energy.
This enables the use of long-range dipole coupling to evaluate the 
positron elastic capture rate $\Gamma ^e$ and the 
(universal) $\sqrt{\eb }$ scaling to estimate the positron annihilation rate 
$\Gamma ^a$. The theory also makes use of the fact that the vibrational
spectra are sufficiently simple and that the IR strengths of the 
modes are known.
The only free 
parameter is the positron-molecule binding energy, which can be determined
experimentally. In the case of methyl halides, the theoretical predictions
are in excellent agreement with the experimentally measured annihilation 
rates as a function of positron energy. Application to the deuterated
methyl halides provides a complete test of the theory with no
adjustable parameters.
The theory has also been successfully extended to include combination and 
overtone vibrations.

A general result of the theory is that the resonances for all
modes with coupling strengths $\Gamma ^e \gg \Gamma ^a$ produce 
annihilation resonances of the same magnitude, modulated only by the 
factor $g=\sqrt{\eb /\eps }$. There are no known cases where dipole coupling 
strengths greater 
than $\Gamma ^a$ fail to produce resonances. On the other hand, the theory
is incomplete for molecules such as ethylene and acetylene, where
explaining the measured annihilation spectra requires the
inclusion of VFR associated with nominally IR-inactive modes
and overtones and combination vibrations. In these cases, the couplings
are more difficult to evaluate.

As the number of atoms $N$ in the molecule is increased, the magnitudes 
of the annihilation resonances exceed (often greatly) those 
explicable on the basis of individual single-mode resonances.
The annihilation rates in these species scale 
as $\Z \propto gN^{4.1}$. While a quantitative explanation for this
$N^4$ scaling is lacking, it is 
likely that the process of intramolecular vibrational energy redistribution 
is responsible for this enhancement. This IVR process couples the single-mode 
resonant
doorway states to baths of dark states, namely, states that are not 
directly coupled to the positron continuum. In this picture, $N$ likely
reflects the number of vibrational degrees of freedom of the molecule. 

For attachment and annihilation in large molecules, there is
a related key piece of the theory that is missing.
If the positrons were able to populate the VFRs associated with all possible
vibrational excitations, $\Z$ would be expected to increase
much faster with molecular size than is observed. While vibrationally
inelastic escape channels could moderate such growth, they appear to be
generally inoperative. Hence, it is not known presently what subset of
multimode dark states and subsequent positron escape channels are accessible.
By inference from the experimental results, detachment from these dark states
likely takes place via a quasielastic escape channel or channels,
including the original and nearby doorway states (e.g., dipole-allowed modes
close in energy to the doorway state through which the positron entered).

The energy-resolved annihilation data have provided measurements of
positron-molecule binding energies $\eb $ for 30 molecular species. The
molecular dipole polarizability plays an important role in fixing the magnitude
of $\eb $. Binding is further enhanced in aromatic molecules by an amount that
increases with the number of electronic $\pi $ bonds. In small molecules, the
permanent dipole moment also increases $\eb $. These results indicate that
most large molecules will bind positrons, and they provide some insight into
promising candidates for further experimental and theoretical study. The goal
of finding molecular species for which the binding energy can be both
calculated and measured is closer to realization with the recent discovery
of small molecules with relatively large binding energies \cite{DGS10}.
Other interesting topics for further research include study of very
large molecules, such as polycyclic aromatic hydrocarbons
that are of astrophysical interest \cite{GJG10,IGS96}, and
cagelike structures, such as C$_{60}$ \cite{GL99}.

Following positron annihilation, VFR-mediated or not, the remaining 
molecular ion is frequently found to fragment. There are many open 
questions in this area, including what determines the degree of 
fragmentation. A practical question of interest is whether positron-induced
annihilation might be a way to produce unfragmented ions from large 
molecular species for applications such as mass spectrometry.

Finally, the resonant enhancement of positron annihilation described here has
an electron analog. In electron-molecule scattering, resonances are known to 
drive processes such as dissociative attachment \cite{Chr84},
a process that involves the slow motion of the heavy atomic nuclei
and that would be weak in a direct electron scattering process. In many 
electron-driven reactions, shape resonances are prominent. While this is not
the case for positrons, both positrons and electrons can 
populate VFR directly. In the positron case, even a weak positron-vibrational
coupling is sufficient to ``turn on'' the resonant annihilation mechanism.
It is possible that a better understanding of positron attachment
to polyatomic molecules, gained through such annihilation studies, can be
useful in understanding similar electron-molecule processes.

\section*{Acknowledgments}

We are indebted to many colleagues for their contributions to the topics
reviewed here. In particular, we acknowledge the collaboration of
L.~D.~Barnes, J.~R.~Danielson, L.~Dunlop, P.~Gill, R.~G.~Greaves, D.~Green,
K.~Iwata, C.~Kurz, C.~M.~R.~Lee, M.~Leventhal, J.~Ludlow, J.~Marler,
T.~J.~Murphy, A.~Passner, and J.~P.~Sullivan. We also gladly acknowledge
helpful conversations with M.~Allan,
M.~Bromley, S.~Buckman, M.~Charlton, P.~Coleman, I.~Fabrikant, F.~Gianturco,
H.~Hotop, and~M. Lima.
G.~F.~G. is grateful to V.~Flambaum for arousing his interest in
the problem of low-energy positron interaction with atoms and molecules
and for much of what he learned during seven years at UNSW (Sydney, Australia).
The experimental work at UCSD was supported by the NSF, Grant No. PHY 07-55809.
The work at UCSD benefited greatly from the technical assistance of
E.~A.~Jerzewski.

\bibliographystyle{apsrmp}
\bibliography{positron}

\end{document}